\documentclass[12pt]{article}
\usepackage{epsfig}
\usepackage{hyperref}
\usepackage{graphicx}
\usepackage{ragged2e}
\usepackage{cite} 

\newcommand{\be}{\begin{equation}}
\newcommand{\ee}{\end{equation}}
\newcommand{\bea}{\begin{eqnarray}}
\newcommand{\eea}{\end{eqnarray}}

\newcommand{\Sigb}{{\overline\Sigma}}
\newcommand{\oot}{\overline {126}}
\newcommand{\boot}{\bf\oot}

\newcommand{\ovl}{\overline}
\newcommand{\nnu}{\nonumber\\}
\def\blfootnote{\xdef\@thefnmark{}\@footnotetext}

\begin{document}
\vspace{4\baselineskip}
\vspace{4cm}
\begin{center}{\Large\bf Baryon Stability on the Higgs Dissolution Edge :
Threshold corrections and suppression of Baryon violation in the
NMSGUT}

\end{center}
\vspace{2cm}
\begin{center}
{\large
 Charanjit S. Aulakh\footnote{ aulakh@pu.ac.in ; http://physics.puchd.ac.in/aulakh/}, Ila Garg and Charanjit K. Khosa
 }
\end{center}
\vspace{0.2cm}
\begin{center}
 {\it
Dept. of Physics, Panjab University,\\ Chandigarh, India.}
\end{center}

\centerline{\Large{\bf{Abstract}}}

\vspace{0.2cm} Superheavy threshold corrections to the matching
condition between matter Yukawa couplings of the effective Minimal
Supersymmetric Standard Model (MSSM) and the New Minimal
Supersymmetric (SO(10)) GUT(NMSGUT)   provide a novel and generic
mechanism for reducing the long standing and generically
problematic operator dimension 5 Baryon decay rates. In suitable
regions of the parameter space strong wave function
renormalization of the effective MSSM Higgs doublets due to the
large number of  heavy fields can take the wave function
renormalization of the MSSM Higgs field close to the dissolution
value ($Z_{H,\overline{H}}=0$). Rescaling to canonical kinetic
terms lowers the SO(10) Yukawas required to match the MSSM fermion
data. Since the same Yukawas determine the dimension 5 B violation
operator coefficients, the associated rates can be suppressed to
levels compatible with current limits. Including these threshold
effects also relaxes the constraint $ y_b-y_\tau\simeq y_s-y_\mu$
operative between $\textbf{10} -\textbf{120} $ plet generated tree
level MSSM matter fermion Yukawas $y_f$. We exhibit accurate fits
of the MSSM fermion mass-mixing data in terms of NMSGUT
superpotential couplings and 5 independent soft Susy breaking
parameters specified at $10^{16.25}\,$ GeV with the claimed
suppression of Baryon decay rates. As before, our s-spectra are of
the mini split supersymmetry type with large
$|A_0|,\mu,m_{H,\overline H} > 100\,\,$ TeV, light gauginos and
normal s-hierarchy. Large $A_0,\mu$ and soft masses allow
significant deviation from the canonical GUT gaugino mass ratios
and ensure vacuum safety. Even without optimization, prominent
candidates for BSM discovery such as the muon magnetic anomaly,
$b\rightarrow s\gamma$ and Lepto-genesis CP violation emerge in
the preferred ball park.

\vskip1pc

\section{ Introduction} \hspace{0.5cm}

Supersymmetric Grand Unification based on the SO(10) gauge
group\cite{fritmin} has received well deserved attention over the
last 3 decades. Models proposed
 fall into two counter posed broad classes. The first consists of just a few models
   which  preserve R-parity down
 to low energies \cite{aulmoh,ckn,abmsv,rparso10,blmdm,nmsgut,rpar1}.
It   uses   Higgs   representations (${\bf{\oot}}$) of
 SO(10) that contain R-parity even SM singlets. The other
 large and diverse class of R-parity violating models
\cite{nRGUTs}   attempts to construct viable models using sets of
small(dimension $d\leq 54, $Index $S_2(d)\leq 12$)  SO(10)
representations even after sacrificing the vital distinction
provided by R-parity between matter and Higgs multiplets in the
first class of models. This issue is only the tip of a sharp wedge
that divides the outlooks of these two schools of supersymmetric
SO(10) unification and discussion of their contrasting attitudes
towards fundamental questions regarding the nature of the UV
completion of the MSSM is unavoidable.

 The defining  feature of R-parity preserving(RPP) GUTs
 \cite{aulmoh,ckn,rparso10,abmsv,blmdm,nmsgut,rpar1} is use of
 a pair of $\mathbf{126 +\oot}$ dimensional irreps which  generate(via
\emph{renormalizable}  B-L/R-parity even vevs) large right handed
neutrino masses(small left handed triplet vevs)   required by Type
I(Type II) seesaw mechanism \cite{seesaw} for light neutrino mass.
Such large irreps \emph{cannot} arise in the massless sector of
known string theory models. Thus this class of models may properly
call itself ``Unstrung GUTs" \cite{csapuri}. Following upon the
proposal of \cite{babmoh} a great deal of attention was paid with
considerable  success,  \cite{allferm} to the issue of fitting the
fermion mass and mixing data using $\mathbf{10\oplus \oot}$ vevs
with \emph{generic} coefficients (rather than derived in terms of
GUT superpotential parameters). However when the realization of
the generic coefficients in terms of actual GUT parameters was
probed it was found that the fits were not
feasible\cite{gmblm,bmsvsnk,blmdm,bert}.

  In direct contrast to RPP GUTs are   R-parity violating(RPV) GUTs\cite{nRGUTs}, which
are typically  ``string inspired'' or ``string compatible'', and
employ \textbf{16} dimensional Higgs irreps(with B-L odd neutral
components) to generate seesaw neutrino masses via ``composite
$\mathbf{\oot}$ '' channels  i.e $d>4$ \emph{non-renormalizable}
operators thought to arise generically in the effective theory
below the Planck/String scale. Out of   the infinite set of
possible $d>4$ operators  these models pick a convenient small
subset and use their coefficients to fit data. The very absence of
\emph{any} calculation of the coefficients of such gravity/string
induced operators from UV theory is taken to justify assuming them
to have convenient values. Suppression of super fast B-decay and
other unpleasant R-parity violating effects is accomplished by
introducing -again with `string inspiration'- suitable   discrete
symmetries. In contrast,  RPP GUTs   use only renormalizable
interactions, avoid invoking \emph{ad hoc} non-gauge symmetries
and claim \emph{parameter counting} -as opposed to field counting-
minimality  as their USP. In this respect they are perhaps closer
in spirit to the original form of String Unification where the
infinite plethora of string excitations was justified by reference
to the single coupling of the stringy TOE just beyond the horizon!
Their neglect of the possibility that all non-renormalizable
operators induced by gravity become strong, in the absence of any
calculation of the coefficients of such operators, while not
provably justified for now,  is at least consistent with their
renormalizable framework and assumptions and provides a fertile
starting point for a self consistent exploration of a very complex
theory which would not be illuminated by induction of an arbitrary
number of new couplings.

Nevertheless the  replies of RPP SO(10) model builders   to
perennial objections (see specially \cite{dixitsher}) bear
repetition and elaboration  since the replies  have evolved
  as the detailed structure of these very well defined and
calculable models   continues to be
excavated\cite{aulmoh,ckn,rparso10,ag1,abmsv,bmsv,fuku04,ag2,gmblm,blmdm,nmsgut,mrs,rpar1}
 due to a   focus  maintained  over 30 years : that  few
 other models have succeeded in inspiring.
Firstly use of the $\mathbf{126 +\oot}$ pair with SO(10) indices
$S_2(126)= 35$ \emph{each} \emph{makes a Landau pole in the SO(10)
gauge coupling inevitable} at a scale $\Lambda$ \emph{within an
order of magnitude above the perturbative unification scale}. We
note that banning   these irreps outright for  large beta function
would also eliminate the only other renormalizable channel for
fermion mass in SO(10) namely the \textbf{120}-plet which has an
index $S_2(120)=28$. This would leave only models with a bunch of
\textbf{10}-plets in consideration before even showing that such
an impoverishment in structural richness is actually  called for.
  SO(10) group theory clearly signals the importance of the $\boot$
dimensional representation for accommodating the most important
mechanism for understanding neutrino mass seamlessly. Secondly  If
perturbative unification is postponed to a higher scale near to or
coinciding with the Planck scale $M_p \sim 10^{18.4}$ GeV then the
neglect of non-renormalizable operators suppressed only  by the
Planck scale is said to be  unjustifiable. We have countered these
objections \cite{trmin,tas,ag2,nmsgut} by arguing that   detailed
calculation of gauge threshold effects
 shows the perturbative unification scale-properly
defined\cite{weinberghall,ag2}- is indeed raised\cite{ag2,nmsgut}
towards the Planck scale. So it makes  inevitable the coincidence
of the SO(10) Landau pole with $M_p$ itself. The unitarity
violation arising in non-renormalizable Fermi theory determined  a
cutoff at the electroweak scale and required   new degrees of
freedom for UV completion and thus led to the discovery of the
Standard Model. Similarly the  Landau pole of RPP GUTs mandated by
the neutrino seesaw dynamics required to assimilate convincingly
the only known BSM dynamics within GUT models points to  a new
physical cutoff and need for a new UV completion. \emph{The
convergence of $\Lambda$ and $M_p$ points to a \emph{origin} for
gravity  in the physics of strongly coupled supersymmetric
SO(10)}. For instance it could arise from that strongly coupled
theory as an induced gravity\cite{adler} with the supersymmetric
strong coupling scale $\Lambda\sim 10^{18}$ GeV setting the Newton
constant much more plausibly and consistently than the
original(inconsistent if non-supersymmetric \cite{david}) proposal
based on an asymptotically free gauge theory. In any case the
existence of a Landau Pole at the Planck scale does not invalidate
the  use of a weakly coupled SO(10) GUT framework \emph{below}
that cutoff scale-where both SO(10) and gravitational couplings
are small- just as it does not invalidate Fermi Theory of Weak
Interactions or Chiral perturbation theory below the appropriate
(internally determined) \emph{physical} cutoffs. In short  the
Landau pole signals  an internally determined physical cutoff of
RPP SO(10) GUTs   and is a potential \emph{addition} to our
physical understanding analogous to the
 information furnished by the breakdown of Chiral at scales $\sim
 1 $ GeV  or Fermi  perturbation theory at scales $\sim 50$ GeV.
 We emphasize that in this wise RPP GUTs are no worse than the plethora of
 RPV SO(10) GUTs which are not only non renormalizable, but typically
 assume without calculation that an infinity of
 operators present by consistency are negligible as also the
 (incalculable?) radiative corrections that should be applied.

 This brings us to the related question of how restrictions to maintain
 perturbativity   should be imposed in  complex QFTs with   many fields and couplings.
   We may keep in mind that well accepted theories like String theory and Kaluza-Klein models would fail this
 test utterly if a naive restriction  like $g^2 <N^{-1}$ (N the number of fields) were adopted.
   As is well   known even the QED perturbation series,  in spite of giving predictions (g-2 of muon etc)
   verifiable to 7   decimal places, does not,
     strictly speaking, converge and is  only `Borel summable'.
     Thus, at the stage when the quantum effects of the RPP
     SO(10)(which \emph{is} at least perturbatively
     renormalizable in contrast to RPV models) have only begun to
     be calculated, demonstration of perturbative consistency
 can   only  proceed, order by order in the loop expansion,
   by demanding that  radiative corrections
to (directly or indirectly) measurable (e.g
$M_X,\alpha(M_X),\alpha_3(M_Z)$\cite{ag2,nmsgut}) or theoretically
central (e.g $Z_{f,\bar f, H,\ovl Z}$ in\cite{gutupend} and  this
paper) quantities should remain    under control and respect basic
consistency requirements such as correct sign(non ghost) kinetic
  terms.  Each such loop corrected element of the theory will obviously
   need to be checked at every loop order reached and there is no
    way of checking this at higher orders before the computation at lower orders.
     This situation is shared with  other
  UV completions. Indeed,   an important implication of our results
  is that theories such as string theory, before claiming
  consistent effective low energy models must check the threshold
  effects involved in specifying the light modes which mix strongly with heavy ones as we have done !
   What was long feared\cite{dixitsher} and we have
  encountered when checking corrections to tree level fits
  found in\cite{nmsgut}  is that \emph{due to the large number of fields} the wave function corrections
 can easily  violate even   basic constraints such as positivity ($Z>0$)  very
   badly.  We therefore imposed positivity of wave function
  renormalization  as  a very  effective   proxy for merely numerical
    guess estimates for the magnitudes of perturbativity limits on
    couplings :  because such a criterion already includes the
    crucial effect of the large number of fields. In fact we continued to find
 good fits  although requiring positivity reduced the magnitude of $Z$  by a factor of
    several hundred, brought  the sign back to the physically
    acceptable one and drastically reduced the magnitude of the SO(10) couplings
    found !
The Baryon decay mechanism we advocate relies on this very (large
N facilitated) limiting value being approached   i.e.  $ Z_{H,\ovl
H}\simeq 0$. Even if
 aggravated large N combinatorics at higher loops
  further restrict the magnitudes of SO(10) couplings   they could
    still -by definition - yield values compatible with positivity.
Finding solutions respecting  $1>>Z\simeq 0$  for  light field
renormalization   has   \emph{improved} our confidence in
  the perturbative status of the couplings so determined.
  We have identified   parameter sets  where the
  achievements of tree level fermion fits and gauge
  unification\cite{nmsgut} are preserved even while
   the magnitudes of the couplings are much smaller
  and positivity of kinetic terms not   violated
  (as the  tree level fits were actually found to do once
   the huge computations we have performed became available).
 This confidence   may well survive higher loop
 corrections as well unless the theory
  has a pathologically ill defined perturbation expansion.
  Our results on threshold corrected gauge unification
  \cite{ag2,blmdm,nmsgut} and fermion fitting\cite{nmsgut} have rather
  lessened this fear by showing that the very complexity of the spectra
  effectively enlarges the possibilities for finding arrangements of parameters for which the
  feared breakdown does not take place.   Our results favour the view that
  there is an intrinsic tendency for a ``Higgs dissolution edge''
  to form when implementing the strange requirement of a fine
  tuned light MSSM Higgs pair to precipitate out of a plethora of
  superheavy MSSM doublets. There is no reason to preclude before computation the
  possibility that higher loop effects may further reduce
  the magnitudes of couplings required to attain the Higgs dissolution edge and thus further  strengthen this
 growing  confidence  that the richness of SO(10)  will dissipate the primordial fears
 of\cite{dixitsher} and similarly render innocuous the threat of the  nearby
 Landau pole. In any event the issue cannot be prejudged. Note that  nothing
 in our interpretations of our  extrapolates the small coupling calculation to a
 region where it is manifestly inapplicable.

 Besides structural
attractions, such as the automatic inclusion of the conjugate
neutrino fields necessary for neutrino mass, SO(10) GUTs
 offer a number of other natural
features. Among these are third generation Yukawa unification
\cite{3Gyukuni,hallrathempf}, automatic embedding of minimal
supersymmetric Left-Right models, natural R-parity preservation
\cite{rparso10} down to the weak scale and consequently natural
LSP WIMP dark matter, economic and explicitly soluble symmetry
breaking at the GUT scale \cite{abmsv}, explicitly calculable
superheavy spectra \cite{ag1,ag2,bmsv,fuku04}, interesting gauge
unification threshold effects \cite{ag2,blmdm,gmblm,nmsgut} which
can lead to a natural elevation of the unification scale to near
the Planck scale \cite{nmsgut}, GUT scale threshold corrections to
the QCD coupling $\alpha_3(M_Z)$ of the required \cite{langpol}
sign and size \cite{precthresh} and a deep interplay between the
scales of Baryon and Lepton number violation as suggested by the
neutrino oscillation measurements and the seesaw formulae
connecting neutrino masses to the B-L breaking scale.

 The fascination of the MSSM RG flow at large $\tan\beta$
 stems from the tendency of third
 generation Yukawa couplings to converge, at the MSSM unification
 scale \cite{3Gyukuni,hallrathempf}, in a manner reminiscent of gauge
unification in the MSSM RG flow \cite{marsenj,amaldimssm}.
 For suitably large
$\tan\beta$ and for close to central input values of SM fermion
couplings at the Susy breaking scale $ M_S \sim M_Z$, third
generation Yukawas actually almost coincide at $M_X$. On the other
hand, in SO(10) theories with only the simplest possible fermion
mass giving (FM) Higgs content (a single \textbf{10}-plet), when
all the complications of threshold effects at $M_X\sim 10^{16}
$GeV (not to speak of those at seesaw scales $M_{\bar\nu}\sim
10^7-10^{12}\,$GeV) are ignored, one does expect to generate
boundary conditions for the gauge and Yukawa couplings that are
unified gauge group wise and (third generation) flavor wise.

However, fitting the rest of the known fermion data (15 more
parameters) definitely requires other Fermion Higgs multiplets
(more \textbf{10}-plets, $\mathbf{120, \oot}$s etc). A principled
position (\emph{monoHiggism}?) with regard to the choice of FM
Higgs irreps   is to introduce  \emph{only one} of each irrep
present in the conjugate of the direct product of fermion
representations. This principle may be motivated by regarding the
different Higgs representations as characteristic ``FM channels''
through which the fermion mass (FM) is transmitted in structurally
distinguishable ways. For example the Georgi-Jarlskog mechanism
distinguishes the \textbf{45} plet Higgs in SU(5) ($\boot$ in
SO(10)) from the $\mathbf{5+{\bar 5}}$ (\textbf{10} in SO(10)) due
to their ability to explain the quark-lepton mass relations in the
second and third generations respectively. Similarly the $\boot$
in SO(10) is peculiarly suitable for implementing the Type I and
Type II seesaw mechanisms for neutrino mass (as well as embedding
the Georgi-Jarlskog mechanism : but the two functions may be
incompatible\cite{blmdm}). If one duplicates the Higgs multiplets
transforming as the \emph{same} gauge group representation, for
example by taking multiple \textbf{10}-plets in SO(10), then one
abandons the quest for a structural explanation of the pattern of
fermion masses in favor of ``just so'' solutions.

In previous work \cite{nmsgut} we have shown that it is possible
to obtain accurate fits of the complete effective MSSM fermion
couplings (including neutrino mass Weinberg operator derived from
Type I and Type II Seesaw masses) from the SO(10) Susy GUT
specified by\cite{aulmoh,ckn} the $\bf{210,10,120,126,\oot}$ Higgs
system. A very notable feature of this fit was that it was
achieved by deducing that threshold corrections at $M_S$ must play
a vital role from the \emph{cul de sac} into which the theory had
apparently\cite{grimu} painted itself by leaving only
$\mathbf{10,120}$ -plets to fit charged fermion masses. The use of
threshold corrections to evade the no-go of too small $d,s$ masses
found in\cite{grimu} then led to the remarkable prediction, well
ahead of the discovery of Higgs mass at 126 GeV, that the
shierachy is \emph{normal} i.e \emph{stops are heavy}   and
supersymmetry is in the decoupling regime ($M_A>>
M_Z$)\cite{decoupsusy} and mini-split\cite{minisplit} :
$A_0,\mu,m_{3/2}$ are multi or tens of TeV. All these were
anathema to Susy orthodoxy in those years : now they are accepted
to be required by Susy and 126 GeV light Higgs ! However this
notable success was faced with the uncomfortable fact that the
parameters found implied \cite{nmsgut} proton decay lifetimes
$\sim 10^{28} $ yrs i.e. at least six orders of magnitude greater
than current limits.

To tackle this situation we proposed \cite{gutupend} that in
Minimal renormalizable SO(10) theories \cite{aulmoh,ckn,abmsv} due
to the large number of heavy fields running within the light field
propagators entering the fermion Yukawa vertices a strong wave
function renormalization is possible even in the perturbative
regime. This can then radically modify the MSSM-GUT Yukawa
matching conditions by suppressing the SO(10) Yukawas required to
match the MSSM fermion `data'. An  preliminary calculation- with
some  defects- of the threshold corrections to the matching
condition between MSSM and GUT determined Yukawas was used to
argue that the B-decay rate in renormalizable Susy SO(10) could be
strongly suppressed. In this paper we present a complete
calculation of the threshold corrections to the matter fermion and
MSSM Higgs vertices. We also found realistic fits of the earlier
type \cite{nmsgut} but now fully viable inasmuch as the
$d=5,~\Delta B\neq 0$ lifetimes can be $ 10^{34}$ yrs or more. We
note that superheavy threshold corrections also relax the
stringent constraint $y_b-y_\tau\simeq y_s-y_\mu$ that we found
\cite{core,msgreb,bs} operative at $M_X$ in SO(10) models with a
\textbf{10-120} FM Higgs system.

A detailed discussion of the historical developments, motivations
and phenomenological issues related to the present work can be
found in the preliminary survey in \cite{gutupend}. Other
calculable quantities include Quark and Lepton Flavor violation
rates, muon $g-2$ anomaly, candidate loop corrected  Susy
discovery spectra, Lepto-genesis parameters  and  NMSGUT based
inflection point inflation(with inflaton scale set by the Type I
Seesaw mass scale\cite{SSI}). In this paper we mainly focus  on
resolving the major issue of $d=5,\Delta B \neq 0$ rates.

In Section \textbf{2} we briefly review the structure of the
NMSGUT \cite{blmdm,nmsgut} to establish the notation for
 presentation of our results on threshold effects in
Subsection \textbf{3.1} and Appendix \textbf{A}. In
 Subsection \textbf{3.2} we present illustrative examples to underline the
 significance of the GUT scale threshold effects and the need to include
 them. In Section \textbf{4} we discuss various aspects of our
 fitting criteria together with threshold effects, and give a
 description of the Tables in Appendix \textbf{B} in
 Subsection \textbf{4.1}. In Section \textbf{5} we
 discuss   exotic observables and specially
 the  acceptable $d=5$ operator Baryon violation rates we have found.
 In Section \textbf{6}
 we summarize our conclusions and discuss   which  improvements
 in the fitting, RG flows and searches are urgently called
 for. Appendix \textbf{A} contains details of the calculation of threshold
 effects at $M_X$. In Appendix \textbf{B} we give two example solutions of NMSGUT parameters
 which fit fermion mass-mixing data and are compatible with B decay limits.

\section{NMSGUT recapitulated }

 The NMSGUT \cite{nmsgut} is a
 renormalizable globally supersymmetric $SO(10)$ GUT
 whose Higgs chiral supermultiplets consist of AM (Adjoint Multiplet) type totally
 antisymmetric tensors: $
{\bf{210}}(\Phi_{ijkl})$, $
{\bf{\overline{126}}}({\bf{\Sigb}}_{ijklm}),$
 ${\bf{126}} ({\bf\Sigma}_{ijklm})(i,j=1...10)$ which break the $SO(10)$ symmetry
 to the MSSM, together with Fermion mass (FM)
 Higgs {\bf{10}}(${\bf{H}}_i$) and ${\bf{120}}$($\Theta_{ijk}$).
 The SO(10) anti-self dual  ${\bf{\overline{126}}}$ plays a dual or AM-FM
role since it also enables the generation of realistic charged
fermion and neutrino masses and mixings (via the Type I and/or
Type II Seesaw mechanisms); three {\bf{16}}-plets
${\bf{\Psi}_A}(A=1,2,3)$ contain the matter including the three
conjugate neutrinos (${\bar\nu_L^A}$).
 The superpotential (see \cite{abmsv,ag1,ag2,blmdm,bmsv,nmsgut} for
 comprehensive details ) contains the mass parameters
 \bea
 m: {\bf{210}}^{\bf{2}} \quad ;\quad M : {\bf{126\cdot{\overline {126}}}}
\qquad ;\qquad M_H : {\bf{10}}^{\bf{2}}\qquad;\qquad m_{\Theta} :{\bf{120}}^{\bf{2}}
\eea

and trilinear couplings corresponding to the superfield chiral
invariants indicated :
 \bea
 \lambda &:& {\bf{210}}^{\bf{3}} \qquad ; \qquad \eta :
 {\bf{210\cdot 126\cdot{\overline {126}}}}
 \qquad;\qquad \rho :{\bf{120\cdot 120 \cdot{ { 210}}}}
\nnu k &:& {\bf{ 10\cdot 120\cdot{ {210}}}} \qquad;\qquad \gamma \oplus
{\bar\gamma} : {\bf{10 \cdot 210}\cdot(126 \oplus{\overline
{126}}}) \nnu \zeta \oplus {\bar\zeta} &:& {\bf{120 \cdot
210}\cdot(126 \oplus {\overline {126}}})
 \eea

In addition one has two symmetric matrices $h_{AB},f_{AB}$ of
Yukawa couplings of the $\mathbf{10,\oot}$ Higgs multiplets to the
$\mathbf{16_A .16_B} $ matter bilinears and one antisymmetric
matrix $g_{AB}$ for the coupling of the ${\bf{120}}$ to
 $\mathbf{16_A .16_B} $. One of the complex symmetric matrices can be
made real and diagonal by a choice of SO(10) flavor basis. Thus
 initially complex FM Yukawas contain 3 real and 9 complex
parameters. Five overall phases (one for each Higgs), say those of
 $m,M, \lambda ,\gamma,\bar\gamma$ can be set by fixing phase conventions.
 One (complex parameter) out of the rest of the superpotential
parameters i.e
 $m,M_H,M,m_{\Theta},\lambda,\eta,\rho,k,\gamma,\bar\gamma,\zeta,\bar\zeta$ , say
 $M_H$, can be fixed by the fine tuning condition to keep two doublets
 light so that the effective theory is the MSSM.
 After removing un-physical phases this leaves 23 magnitudes and
15 phases as parameters : still in the lead out of any theories
aspiring to do as much \cite{abmsv}. As explained in \cite{abmsv,ag1,ag2}
 the fine tuning fixes the Higgs fractions i.e the composition
 of the massless electroweak doublets in terms of the (6 pairs of suitable)
 doublet fields in the GUT.

The GUT scale vevs and therefore the mass spectrum are all
expressible \cite{abmsv,ag2,bmsv} in terms of a single complex
parameter $x$ which is a solution of the cubic equation

\be 8 x^3 - 15 x^2 + 14 x -3 +\xi (1-x)^2=0 \label{cubic} \ee
where $\xi ={{ \lambda M}\over {\eta m}} $.

In our programs we find it convenient to scan over the ``three for
a buck" \cite{abmsv,bmsv06,blmdm} parameter $x$ and determine
$\xi$ therefrom. Then the phase of $\lambda$ is adjusted to be
that implied by $x$ and the relation $\xi ={{ \lambda M}\over
{\eta m}} $ and is not itself scanned over independently.
 It is a convenient fact that the $\bf{592}$
fields in the Higgs sector fall into precisely $\bf{26}$ different types
of SM gauge representations which can hence be naturally labelled
by the $\bf{26}$ letters of the English alphabet \cite{ag2}.
 The decomposition of SO(10) in terms of the labels of its ``Pati-Salam''
maximal subgroup $SU(4)\times SU(2)_R\times SU(2)_L$ provided
\cite{ag1} a translation manual from SO(10) to unitary group
labels. The complete GUT scale spectrum and couplings of this
theory have been given in \cite{ag2,nmsgut}.
 The MSSM fermion Yukawa couplings and
neutrino mass(Weinberg) operator  of the effective  MSSM arising
from this GUT after fine tuning(but before application of   GUT
scale threshold corrections), along with the implementation of
loop corrected electroweak symmetry breaking based on a fixed
value of $\tan\beta, M_Z$ and the run down values of $M^2_{H,\ovl
H}$ and the threshold corrections to the matching conditions
between MSSM and SM   fermion Yukawa are given in
\cite{nmsgut}(Appendix C).

In the NMSGUT, to enhance the light neutrino Type I seesaw
masses\cite{blmdm,nmsgut}, the conjugate (i.e ``right handed'')
neutrino Majorana masses are 4 or more orders of magnitude smaller
than the GUT scale due to very small $\boot$ couplings. Therefore
for purposes of calculating the threshold corrections to the
Yukawa couplings at $M_X$ we can consistently treat the conjugate
neutrinos as light particles on the same footing as the other 15
fermions of each SM family. These fermion mass formulae, after
correcting for threshold effects, are to be confronted with the
fermion Yukawa couplings and Weinberg neutrino mass operator
(RG-extrapolated from $Q=M_Z $ to $Q=M_X^0=10^{16.25}$ GeV). The
calculation of the change in the unification scale exponent
($\Delta_X$) also fixes\cite{ag2} the scale $ {m}$ of the high
scale symmetry breaking \cite{blmdm,nmsgut}. The simultaneous
requirements of a common origin for the unification-seesaw scale,
gauge unification, with the right high scale and Susy breaking
scale, RG threshold corrections to shift the GUT prediction of
$\alpha_3(M_Z)$ down to acceptable values \cite{precthresh} and to
lower the down and strange fermion Yukawas to a level achievable
in this type of GUT\cite{nmsgut}, are very stringent. They are
effective in singling out characteristic and suggestive GUT
parameters (including Susy breaking parameters at $M_X$) which
realize a fully realistic effective theory with distinctive
signatures \emph{derived} from its UV completion. We now show how
the NMSGUT can successfully bypass the remaining roadblock of
rapid dimension 5 proton decay which is generic to Susy GUTs.

\section{ GUT scale Yukawa threshold corrections}
\subsection{One loop threshold correction formulae}
The technique of \cite{wright} for calculating high scale
threshold corrections to Yukawa couplings, generalizes the
Weinberg-Hall \cite{weinberghall} method for calculating threshold
corrections to gauge couplings, and has long been available but
has not been exploited much; possibly due to the assumption that
such effects are always negligible. In supersymmetric theories the
superpotential parameters are renormalized only due to wave
function correction and this is easy -if tedious- to calculate for
the large number of heavy fields which couple to the light
fermions and MSSM Higgs at SO(10) Yukawa and gauge vertices. The
calculation involves going to a basis in which the heavy field
supermultiplet mass matrices are diagonal. This basis is easily
computable given the complete set of mass matrices and trilinear
coupling decompositions given in \cite{ag1,ag2,nmsgut}. For a
generic heavy field type $\Phi$ (conjugate $\overline \Phi$) the
mass terms in the superpotential diagonalize as :
 \bea{\overline{\Phi }}= U^{\Phi}{\overline{\Phi' }} \quad ;
 \qquad {\Phi } = V^{\Phi}\Phi'\quad \Rightarrow \quad
 {\overline{\Phi}}^T M \Phi ={\overline{\Phi'}}^T M_{Diag}
 \Phi'\eea

The circulation of heavy supermultiplets within the one loop
insertions on each of the 3 chiral superfield lines
 ($f_c=\bar f,f,H_f=H,\ovl{H}$) entering the matter Yukawa vertices :
\bea {\cal L} = [f_c^T Y_f f H_f]_F + H.c. +....\eea implies
\cite{wright} a finite wave function renormalization in the
Kinetic terms : \bea {\cal L}=[\sum_{A,B}( {\bar f}_A^\dagger
(Z_{\bar f})_A^B {\bar f}_B +{f}_A^\dagger (Z_{f})_A^B {f}_B ) +
H^\dagger Z_H H + {\ovl H}^\dagger Z_{\ovl H} {\ovl H}]_D +..\eea
where $A,B=1,2,3$ run over matter generations and $H,\ovl H$ are
the light Higgs doublets of the MSSM. The light Higgs superfields
are actually mixtures of 6 Higgs doublets $h_i,\bar h_i, i=1...6$
from the GUT multiplets:\bea H=\sum_i \alpha_i^* h_i \qquad
;\qquad
 \ovl H=\sum_i \bar \alpha_i^* \bar h_i \eea where the Higgs
fractions $ \alpha_i,\bar \alpha_i$ are components of the null
eigenvectors of the Higgs doublet mass matrix ${\cal H}$
\cite{abmsv,ag1,ag2,nmsgut}.

Let $U_{Z_f},{\ovl U}_{Z_{\bar f}}$ be the unitary matrices that
diagonalize ($U^\dagger Z U= \Lambda_Z$) $Z_{f,\bar f}$ to
positive definite form $\Lambda_{Z_f,Z_{\bar f}}$. We define a new
basis to put the Kinetic terms of the light matter and Higgs
fields in canonical form :\bea f &=& U_{Z_f}
\Lambda^{-\frac{1}{2}}_{Z_f} \tilde f =\tilde U_{Z_f} \tilde
f\qquad ; \qquad \bar f= U_{Z_{\bar f}}
\Lambda^{-\frac{1}{2}}_{Z_{\bar f}} \tilde{\bar f}=\tilde
U_{Z_{\bar {f}}} \tilde {\bar f}\nnu H&=&\frac{\widetilde
H}{\sqrt{Z_H}}\qquad ; \qquad \ovl H=\frac{\widetilde {\ovl
H}}{\sqrt{Z_{\ovl H}}}\label{f2ftilde}\eea then \bea {\cal
L}&=&[\sum_A( {\tilde{\bar f}}_A^\dagger \tilde{\bar f}_A +
{\tilde{f}}_A^\dagger \tilde{f}_A ) + \widetilde{H}^\dagger
\widetilde{H} + \widetilde{{\ovl H}}^{\dagger} \widetilde{{\ovl
H}}]_D + [\tilde{\bar f} ^T \tilde{Y}_{f} \tilde{f} \widetilde{H}_
{f}]_F + H.c. +..\nonumber \eea \be\tilde Y_f= \Lambda_{Z_{\bar
f}}^{-\frac{1}{2}} U_{Z_{\bar f}}^T {\frac{Y_f}{\sqrt{Z_{H_f}}}}
U_{Z_f} \Lambda_{Z_f}^{-\frac{1}{2}} = \tilde{U}_{Z_{\bar f}}^T
{\frac{Y_f}{\sqrt{Z_{H_f}}}} \tilde{U}_{Z_f} \label{Ynutilde}\ee
Thus when matching to the effective MSSM it is $\tilde Y_f$ and
not the original $Y_f$ obtained \cite{abmsv,ag1,ag2} from the
SO(10) Yukawas that must equal the value of the MSSM Yukawa at the
matching scale.

For any light Chiral field $\Phi_i$ the corrections have generic
form ($Z= 1 -{\cal K}$) :\bea{\cal K}_i^j=- {\frac{g_{10}^2}{8
\pi^2}} \sum_{\alpha,k}{Q^\alpha_{ik}}^* {Q^\alpha_{kj}}
F(m_\alpha,m_k) +{\frac{1}{32 \pi^2}}\sum_{kl} Y_{ikl} Y_{jkl}^*
F(m_k,m_l) \eea

where ${\cal L}= g_{10} \, Q^\alpha_{ik} \psi^\dagger_i
{\gamma^\mu A_\mu}^\alpha \psi_k$ describes the generic gauge
coupling of the (fermion component $\psi_i$ of) $\Phi_i$ to a
generic SO(10) \emph{heavy} gauge boson $A^\alpha$ and charge
$Q^\alpha$ ($g_{10}= g_5/\sqrt{2}$  and $g_5$ are  the   SO(10)
and  $SU(5)$ gauge  couplings). The generic Yukawa couplings are
defined by the superpotential $W={\frac{1}{6}} Y_{ijk}\Phi_i
\Phi_j \Phi_k$.

 When both the fields running in the loop are heavy
 fields the  symmetric Passarino-Veltman function  $F(m_1,m_2)$
 should be taken to be
\bea F_{12}(M_A,M_B,Q)={1\over {(M_A^2- M_B^2)}}( M_A^2\ln
{M_A^2\over Q^2} -M_B^2\ln    {M_B^2\over Q^2} )- 1 \eea which
reduces to just \bea F_{11}(M_A,Q)=F_{12}(M_A,0,Q)=  \ln
{M_A^2\over Q^2} - 1 \eea when one field is light ($M_B\rightarrow
0)$.
 When one of the heavy fields in the loop has MSSM doublet type
 $G_{321}$ quantum numbers $[1,2,\pm 1]$ (so that one eigenvalue
 is light while the other \emph{five} \cite{nmsgut} are heavy)
 care should be taken to avoid summing over light-light loops: since
 that calculation belongs to the MSSM radiative corrections.

The crucial point to notice is that the SO(10) Yukawa couplings
$(h,f,g)_{AB}$ also enter into the coefficients
$L_{ABCD},R_{ABCD}$ of the $d=5 $ baryon decay operators in the
effective superpotential  obtained by integrating out the heavy
chiral supermultiplets that mediate baryon decay(see
\cite{ag1,ag2,nmsgut} for discussion of the contributing Higgsino
modes and derivation of expressions) : \bea
 W_{eff}^{\Delta B\neq 0} = -{ L}_{ABCD} ({1\over 2}\epsilon
{ Q}_A { Q}_B { Q}_C { L}_D) -{ R}_{ABCD} (\epsilon {\bar{ e}}_A
{\bar{ u}}_B { \bar{ u}}_C {\bar{ d}}_D) \eea After the
redefinition (\ref{f2ftilde}) to the tilde basis to make the
kinetic terms canonical, $\tilde Y_f$ must be diagonalized to mass
basis (denoted by primes) using bi-Unitary $U_{\bar f}(N_g)\times
U_{ f}(N_g)$ kinetic term redefinitions via the unitary matrices
$(U_f^{L,R})$ made up of the left and right eigenvectors of
$\tilde Y_f$ with phases fixed by the requirement that $(U^L_f)^T
{\tilde Y}_f U^R_f =\Lambda_f$ yields positive definite
$\Lambda_f$ : \bea W &=& (\bar f')^T \Lambda_f f' {\tilde H_f}\nnu
 f&=& \tilde U_{Z_f} U^R_f f'= {\tilde U_{f}}' f'\nnu
 \bar{f}&=& \tilde U_{Z_{\bar{f}}} U^L_f \bar{f}'= {\tilde U_{{\bar{f}}}}' \bar{f}'\eea
As a result the $d=5, \Delta B =\pm 1$ decay operator coefficients
in terms of the Yukawa eigenstate basis become \bea L_{ABCD}'
&&=\sum_{a,b,c,d} L_{abcd} (\tilde U_Q')_{aA} (\tilde
U_Q')_{bB}(\tilde U_Q')_{cC} (\tilde U_L')_{dD} \nnu R_{ABCD}'
&&=\sum_{a,b,c,d} R_{abcd} (\tilde U_{\bar e}')_{aA} (\tilde
U_{\bar u}')_{bB}(\tilde U_{\bar u}')_{cC} (\tilde U_{\bar
d}')_{dD} \eea

When we search for a fit of the MSSM Yukawas in terms of the
SO(10) parameters under the constraint that $L'_{ABCD}, R'_{ABCD}$
be sufficiently suppressed (i.e yielding proton lifetime $\tau_p >
10^{34}$ yrs) we find that the search is guided ineluctably
towards those regions of SO(10) parameter space where $Z_{H,\ovl
H} <<1$. As a result the SO(10) Yukawa couplings required to match
the MSSM become much smaller than they would be if these threshold
corrections are ignored. The same $SO(10)$ Yukawa couplings enter
$L'_{ABCD},R'_{ABCD}$ but  there is no boost derived from wave
function renormalization because $d=5$ operators have no external
Higgs line. This  mechanism is generically available in realistic
multi-Higgs theories. It remains to be checked what is the effect
on $d=6$ B violation operators with one external Higgs line.
However those operators are severely suppressed to begin with.

 The decomposition of SO(10) invariant terms in the superpotential
and gauge terms yields \cite{ag1,ag2,nmsgut} a large number ($\sim
1100$ ) of relevant Light-Heavy-Heavy/Light SO(10) vertices. It
then requires a tedious but straightforward calculation to
determine the threshold corrections. The explicit expressions are
given in Appendix {\bf{A}}.

Heretofore such threshold corrections have mostly been argued to
be negligible ($< 1\%$) although at least one paper \cite{baer}
faced with the difficulties of literal third generation Yukawa
unification has considered the possibility, without any explicit
model which permitted calculation, that the third generation
Yukawa unification relations must necessarily be subject to
threshold corrections of up to $50\%$. In which case it was found
that the various  stratagems invoked to permit precise 3
generation Yukawa unification could become redundant. We shall see
that  the calculation of the GUT scale 1-loop Yukawa threshold
effects in the NMSGUT can actually change the naive (i.e pure
\textbf{10}-plet) unification relations $y_t=y_b=y_\tau$
significantly.

Furthermore the $ \mathbf{10-120}$ plet fermion fits have been
shown ( in the absence of GUT scale threshold effects) to require
a close equality $ | y_b-y_\tau/ ( y_s- y_\mu)| \approx 1 $ at
$M_X $ which is very constricting when searching for fits. The
fits we exhibited in \cite{nmsgut} were all of this type. However
in the present case the fits we obtain can deviate significantly
from $ {\frac{y_b-y_\tau}{y_s- y_\mu}}\simeq 1 $. Of course one
should study the higher loop threshold corrections to see if the
1-loop results we find are stable. At present this task seems
computationally prohibitive. However   we have  calculated  the
complete SO(10) two loop beta functions\cite{lettgutupend} using
the fact that the beta functions are determined by  anomalous
dimensions alone. Since the two loop threshold corrections will
also rely upon essentially the same type of anomalous dimensions,
its may be possible to convolute the GUT scale mass spectra with
our SO(10) loop sums to determine the two loop threshold
corrections as well.  In any case our one loop results are a
necessary first step for higher loop studies. As noted  before
the restriction   $Z>0$ also leads to smaller couplings and to
heavy spectra that are  significantly less spread out than in our
previous solutions.

The effect of the wavefunction renormalization on the relation
between  other GUT and MSSM parameters is also interesting and
illuminating. The MSSM superpotential  $\mu$ parameter is larger
than the the GUT $\mu $ parameter  by the factor $(Z_H Z_{\bar
H})^{-1/2}$ and the same goes for the soft Susy breaking parameter
parameter $B$. On the other hand the matter sfermion soft masses
are enhanced only by $Z_f^{-1}$ which will be very close to 1. The
soft Higgs masses will however be boosted by $Z_{H/{\overline
H}}^{-1}$. It is the boosted parameters we determine in our fits
and it is interesting to note(see Appendix \textbf{B} and
\cite{nmsgut}) that we typically find $\mu,A_0,|m_{H/{\overline
H}}|>>m_{\tilde f/\tilde {\bar f}}>>M_{1/2} $ ! However  the $A_0$
parameter does not change since the wave function enhancements are
absorbed by the Yukawa coupling in terms of which it is defined
($A=A_0 \tilde Y$).

Finally the right handed neutrino masses $(M_{\bar{\nu}})_{AB}\sim
f_{AB}<\bar \sigma>$ will also change due to finite corrections to
the SO(10) breaking induced mass term
  due to heavy field loops.
 However since the vev $<\bar \sigma>$ is protected by the non-renormalization theorem
 i.e is fixed in terms of the parameters $m,\lambda,M,\eta$, and the
 corresponding field fluctuation  is \emph{not} a part of the low
 energy effective theory,    the heavy loops  will
  redefine   $ f_{AB}\rightarrow
\tilde{f}_{AB} =({\widetilde U_{\bar \nu}}^T f{\widetilde U_{\bar
\nu}})_{AB}  $    along with  $Y^{ \nu}_{AB}\rightarrow
\tilde{Y}^{ \nu}_{AB}  $(eqn(\ref{Ynutilde})). As a result when
the  right handed neutrinos $\bar\nu$ are integrated out the
factors $\tilde{U}_{\bar\nu}$
  actually cancel  out of the Type I seesaw formula leaving only
  $  \tilde{U}_{\nu},Z_H^{-1/2}$ to dress the formula obtained without threshold corrections.
Since $Z_{\bar \nu}$ is rather close to unity the effect on
neutrino masses is likely to be small.  We have included these
factors in our calculations. This discussion also shows that we
have given a complete calculation of the germane  1-loop GUT scale
threshold corrections to  the relation between observable  gauge,
Yukawa, Seesaw and B-decay couplings and GUT scale parameters.

\subsection{Necessity of including threshold
effects}

To appreciate the importance of the threshold corrections at $M_X$
for the matter fermion Yukawas it is sufficient to consider
 what one obtains for $Z_{f,{\bar{f}},H,\ovl H}$ using
parameters from the examples of tree level fits (found ignoring
GUT scale threshold corrections) given in \cite{nmsgut}.

\begin{table}
 $$
 \begin{array}{|c|c|c|c|}
\hline
 \multicolumn{4}{|c|}{\mbox{SOLUTION 1}} \\
 \hline
 \mbox{Eigenvalues}(Z_{\bar u})& 0.928326& 0.930946& 1.031795\\
 \mbox{Eigenvalues}(Z_{\bar d})& 0.915317& 0.917464& 0.979132\\
 \mbox{Eigenvalues}(Z_{\bar \nu})& 0.870911& 0.873470& 0.975019\\
 \mbox{Eigenvalues}(Z_{\bar e})& 0.904179& 0.908973& 0.971322\\
 \mbox{Eigenvalues}(Z_{Q})& 0.942772& 0.946127& 1.027745\\
 \mbox{Eigenvalues}(Z_{L})& 0.911375& 0.916329& 0.997229\\
 Z_{\bar H},Z_{H}& -109.367 & -193.755 & \\
\hline \multicolumn{4}{|c|}{\mbox{SOLUTION 2}}\\
 \hline
 \mbox{Eigenvalues}(Z_{\bar u})& -7.526729& -7.416343& 1.192789\\
 \mbox{Eigenvalues}(Z_{\bar d})& -7.845885& -7.738424& 1.191023\\
 \mbox{Eigenvalues}(Z_{\bar \nu})& -8.830309& -8.681419& 1.234923\\
 \mbox{Eigenvalues}(Z_{\bar e})& -7.880892& -7.716853& 1.238144\\
 \mbox{Eigenvalues}(Z_{Q})& -9.203739& -9.109832& 1.171956\\
 \mbox{Eigenvalues}(Z_{L})& -9.797736& -9.698265& 1.217620\\
 Z_{\bar H},Z_{H}& -264.776 & -386.534 & \\
 \hline
 \end{array}
 $$
 \caption{\small{
 Eigenvalues of the wavefunction
 renormalization matrices $Z_f$ for fermion lines and for
MSSM Higgs ($Z_{H,\overline H}$) for solutions found in
\cite{nmsgut}.  \label{table1} }}
 \end{table}

It is clear from Table 1 that neglect of the
 wave function corrections would be a serious error since they are
 easily so large as to change the sign of the effective  kinetic terms !
 In the case of Solution 2, not only the Higgs but even the
fermion line corrections can be large enough to do this ! This
seems to put the solutions found in \cite{nmsgut} (as well as all
previous GUTs with a Higgs structure rich enough to account for
the observed charged fermion and neutrino data) in a dubious
light. However we shall see that the disease contains its own
cure: when the wave function corrections are correctly accounted
for, and searches mounted while maintaining $Z>0$ for all fields
we are led to regions of the parameter space where not only the
matter Yukawa couplings but also the other super potential
parameters are significantly lowered in magnitude: \emph{inter
alia} improving the status of the model vis a vis perturbativity.
Since accounting for the effects of threshold corrections also
allows us to lower the $d=5$ operator mediated B-violation rate,
it is clear that a central result of our work is that henceforth
close attention must be paid to the consequences of the fact that
MSSM Higgs multiplets derive from multiple GUT sources. Analyses
of GUT models that neglect the multiple GUT level parentage of
MSSM Higgs and the consequent drastic threshold corrections to
tree level effective MSSM couplings should no longer be
countenanced  uncritically. Of course this warning traces back to
\cite{dixitsher}, but our emphasis\cite{ag2,nmsgut} has been to
exploit the richness of the Quantum effects rather than a
pessimistic one.

\section{ Realistic fits with threshold corrections included}
\subsection{Description of search strategy and conditions}
We follow the procedure described in \cite{nmsgut} to find sets of
GUT superpotential and GUT compatible soft Susy breaking
parameters which allow accurate fits of \emph{all} the fermion
masses and mixing angles. The new features are that
\begin{itemize}
\item{ } We use our search programs to find fits after including
the threshold effects at $M_X^0$ \item{ } We include the effects
of Susy thresholds on the gauge unification parameters
($\Delta_{X,G,3}$) which we earlier neglected but should not have
since the sparticle spectrum we found is decoupled (large
$A_0,\mu,m_0>>M_Z$) and quite spread out. \end{itemize}

 We impose strict
unitarity in the sense that the wave function renormalization must
remain  positive i.e \bea Z_{f,\bar f,H,\bar H}
>0 \label{strictpert}\eea The search programs \cite{nmsgut} do
find solutions (quite far from the examples of \cite{nmsgut} in
that many couplings, such as most noticeably $\eta$ undergo major
changes,  being driven  towards smaller values) which satisfy this
constraint and still provide consistent unification and accurate
fits of the fermion mass data. Unless  higher loop effects could
 somehow  overcome and forbid the tendency of Z to be reduced
below 1 that we found by calculating 1-loop effects, it is likely
that even smaller values of the couplings will make $Z_H\simeq 0$
achievable. Then the suppression of proton decay may become even
easier.

 Moreover the effectiveness of our  mechanism for reducing the size of the $d=5$ B decay operators
is verified. When we conduct searches while demanding that these
coefficients be suppressed strongly the search program
incorporating threshold corrections succeeds in finding solutions
: whereas earlier proton decay lifetimes greater than about
$10^{28}$ yrs could not be achieved. Specifically, without the
threshold corrections the generic values of the maximal absolute
magnitude $Max(O^{(4)})$ of the LLLL and RRRR coefficients in the
$d=5,\Delta B\neq 0$ effective superpotential was found to be
typically of order $10^{-17} \,$  GeV${}^{-1} $ corresponding to
fast baryon decay rates $\sim 10^{-27}\,$ yr${}^{-1}$. Our quick
fix to the problem of limiting the B-decay rate while searching
for accurate fermion fits is to limit ($\tilde{O}$ is the
dimensionless operator in units of $|m/\lambda|$)
$Max(\tilde{O}^{(4)})< 10^{-5 }$(in dimensionful terms $Max(
{O}^{(4)})< 10^{-22 } GeV^{-1}$).
  This produces fits with proton lifetimes
  above $10^{34}$ yrs, so we  work with a penalty
 for violating :  $Max(\tilde{O}^{(4)})< 10^{-5}$.
 These fits are  always in regions where
  $Z_{H,\ovl H}$ approach zero (from above) while $Z_{f,\bar f}$
  suffer only minor  corrections since the $\mathbf{16-}$plet Yukawas
 are now suppressed. In addition to the penalty for rapid
 proton decay we also imposed the following conditions for
 acceptable fits :

 \begin{enumerate}

   \item As already explained in detail in \cite{nmsgut} the gauge
   unification RG flow is constrained so that perturbation theory in the
 gauge coupling at unification remains valid, the unification
  scale is less than $M_{Planck}$  and the GUT threshold
  contributions to $\alpha_3(M_Z)$ (together with the corrections from the
  rather high value of the superpartner masses : see below)
  are in the right range\cite{nmsgut,langpol,precthresh}.  :
\bea
-22.0\leq \Delta_G &\equiv&  \Delta  (\alpha_G^{-1}(M_X))  \leq 25 \nonumber \\
3.0 \geq  \Delta_X &\equiv &\Delta (Log_{10}{M_X}) \geq - 0.03\nonumber \\
-.0126< \Delta_{3} &\equiv & \Delta\alpha_3(M_Z)  <
-.0122\label{criteria} \eea

\item  We constrain  the $|\mu(M_Z)|, | A_0(M_Z) | $ parameters to
be smaller than  150 TeV.Two loop RGE flow from $M_X$ to $M_Z$,
\emph{ignoring generation mixing}, was   used to determine these
soft Susy parameters (by imposing consistency with   Susy
  threshold effects required for fitting $y_{d,s,b}$) since only the diagonal
threshold correction formulae are available at present. This is
justified in  view of our limited expectations of overall
 accuracy of sfermion spectra which are so far uncorrected by loops.
  Typically these parameters emerge in the
range $\sim 50-150 $ TeV while the gaugino masses $M_i$  are
driven to the lower limits imposed (since it is the ratios $
\mu(M_Z)/M_i(M_Z), A_0(M_Z) /M_i(M_Z) $ which control the efficacy
of the large tan$\beta$ corrections required for our
purposes\cite{nmsgut}(the search selects very small gaugino masses
at $M_X$ compatible with $M_i(M_X)\simeq 0$, since in any case the
two loop running of gaugino masses, specially with large $A_0$, is
enough to generate adequate gaugino masses)). Sfermion masses lie
 in the $1-50$ TeV range   though a few (notably the
  Right chiral smuon)  can be lighter than a  TeV. This
is the price one must pay to correct the fermion Yukawas to
achievable values in the NMSGUT. Large values of $A_0$ are often
feared   to lead to  charge and color breaking (CCB) minimae
\cite{CCB} or unbounded from below (UFB) potentials \cite{UFB}.
However it is also established \cite{kuslangseg} that the
metastable standard vacua that we are considering (with all mass
squared parameters of charged or colored or  sneutrino scalar
fields \emph{positive} i.e at a local minimum which preserves
color, charge and R-parity) can well be stable on time scales
   of order the age of the universe($\sim 10 $ Giga-years),
provided   $|A_0|,\mu$  are above about 5 TeV : as found in our
fits. This is natural for the decoupled/Mini Split Susy
s-spectra\cite{decoupsusy,minisplit} we have always found since
2008.

\item In accordance with experimental constraints \cite{pdg2012}
we also constrain lightest chargino (essentially wino $\tilde
W^\pm$ ) masses to be greater than 110 GeV. All the charged
sfermions  as well as the charged
 Higgs are constrained to lie above 110 GeV and
 the uncharged loop corrected Higgs($h^0$) mass
 to be in the measured range   $124\, GeV < m_{h^0} < 126 \, GeV$.
 The Higgs masses were calculated using the 1-loop
 corrected electroweak symmetry breaking
 conditions and  1-loop effective potential using a subroutine \cite{porod}
 based on \cite{loophiggs}. The large values of $A_0, \mu$
 (and thus $X_{t }=A_{t }-\mu \tan \beta,X_{b }=A_{b }-\mu \cot \beta$)
 favor  large masses for the light
 Higgs through loop corrections. It is a matter of
 gratification for the NMSGUT that
 it selected such values in 2008 : long before the Higgs discovery in 2012
 which abruptly promoted  large $A_0$ values(even if not the NMSGUT!)  from eccentric
 to  fashionable and rigorous.

 \item  The LHC Susy searches have now arrived \cite{susy2013}
 at a fairly model independent lower limits of about $ 1200$ GeV for the Gluino mass.
 In models with very large $A_0$ and Non Universal Higgs masses like ours
 the correlation between  gaugino masses at low
 scales  can deviate substantially from the standard $1:2:7$ ratio
 common to GUT models with universal gaugino masses at the
 Unification scale. However the scales are still grouped together
 so the characteristic spectrum associated with the NMSGUT finds a
 useful anchor in the LHC  gluino limit($M_{\tilde G} > 1 $ TeV) which we implement via a
 penalty. This   has the effect of not allowing LSP(Bino)  masses
 lower than about 200 GeV so that the LHC limit may be regarded as
 signalling also the inability of the NMSGUT to provide a very light
 LSP. The friability  of the standard  gaugino   mass ratio
 is also remarkable.  For small $A_0$ this ratio is almost
 fixed in stone by one loop RGE and GUT
mandated  gaugino mass universality at $M_X$. However, invocation
of   gaugino masses generated by SO(10) variant F terms which is
sometimes   advocated\cite{ananthpandita} seems a
 too much  to pay for such a freedom. Inasmuch as it is
  assumes   hidden  Supersymmetry breaking
  involving SO(10) Higgs multiplets can be consistently
  sequestered without proof, such a scenario is orthogonal to
  the motivation of our work. We find that the SO(10) GUT is rich
  enough to allow generation of variant gaugino mass ratios
  via    $A_0\sim 100$ TeV consistently with other demands of our model.

\item An   improvement concerning the treatment of Susy threshold
effects on gauge unification parameters
$\alpha_3(M_Z),M_X,\alpha(M_X)$ is  introduced   to account for
the spread out  spectrum of supersymmetric masses. A   weighted
sum over all the Susy particles ($M_{Susy}$) is used in
$\Delta^{Susy}_{\alpha_s}$ as given in \cite{langpol}. \bea
\Delta^{Susy}_{\alpha_s} & \approx & \frac{-19\alpha_s^2}{28\pi}
\ln\frac{M_{Susy}}{M_Z}\nnu M_{Susy} &=&  \prod_i m_i^{-{\frac{
5}{38}} (4 b_i^1 -9.6 b_i^2 +5.6 b_i^3)} \eea

\be \Delta_X^{Susy} = \frac{1}{11.2\pi } \sum_i (b_1 -
b_2)Log_{10}\frac{m_i}{M_Z} \ee

\be \Delta_G^{Susy} = \frac{1}{11.2\pi}\sum_i (6.6\, b_2 - b_1)
\ln\frac{m_i}{M_Z}\ee

Here $b_1$, $b_2,b_3$ are the 1-loop $\beta$ function coefficient
of U(1), SU(2), SU(3) in the MSSM  respectively.
$\Delta^{Susy}_{\alpha_s}$ can
 be significant  so  it changes the allowed range at GUT scale.
  We considered the following limits for $\Delta^{Susy}_{\alpha_s}$ in the search program.
\be -.0146< \Delta^{Susy}_{\alpha_s}  <-0.0102  \ee

   \end{enumerate}

\subsection{Description of Tables}
In Tables 2-13 in Appendix B we have shown  two example fits of
fermion mass mixing parameters in terms of NMSGUT parameters. In
Tables   2,8  we give the complete set
 of NMSGUT parameters defined at the one loop unification scale
 $M_X^0=10^{16.25}\,$ GeV-which we always use as the GUT-MSSM matching scale-
 together with the values of the soft Susy breaking parameters ($m_0,m_{1/2},A_0,B,
  M^2_{H,\bar H}$) and  the   superpotential parameter $\mu $.
  The values of $\mu(M_X), B(M_X)$ are determined by RG evolution
   from $M_Z$ to $M_X$ of the values determined by the
loop corrected electro-weak symmetry breaking
conditions\cite{nmsgut,piercebagger}.

  Our soft supersymmetry breaking parameters are thus
 those of a N=1 Supergravity GUT compatible  scenario with
 different soft scalar masses allowed for different SO(10)
irreps. As a result  Non Universal soft  Higgs Masses (NUHM) for
the light Higgs of the MSSM are  justified
 since the light doublets are a mixture of doublets from several sources
  in different SO(10) irreps each of which is free to have its own soft mass.
  Our solutions always   find negative values for these soft masses which can readily
  arise only if the soft masses of at least some of the
  originating representations are themselves negative. Another point
  to be noted is that $|m_{1/2}|$ is quite small (0-500 GeV) compared to other soft parameters.
 Besides these parameter values of the  SUGRY-NUHM NMSGUT
 we also give the mass spectrum of superheavy fields
 including the right handed neutrinos. We also report Type I
 and Type II neutrino seesaw masses as well as the changes($\Delta_{X,G,3}^{GUT/Susy}$) in
 gauge unification parameters from their 1-loop MSSM values due to GUT scale
 and Susy breaking  scale threshold corrections. The benefit of imposing $1>>Z
>0$ i.e.  that it guides the Nelder-Mead search amoeba\cite{pressteukolsky}  to regions of the parameter  space with a
smaller spread in superheavy masses and smaller values for the
non-matter  superpotential couplings couplings as well (making the
spectrum and perturbation theory in the superpotential parameters
more trustworthy) can be appreciated by comparing the values in
Tables 2 and 8 with those in the corresponding tables of
\cite{nmsgut}.

In Tables 3,9  we give  the values of the target fermion
parameters (i.e two loop RGE extrapolated, Susy threshold
corrected MSSM Yukawas, mixing angles, neutrino mass squared
differences and neutrino mixing angles). Their uncertainties are
estimated as in\cite{antuschspinrath}, together with the achieved
values and pulls.  We obtain excellent fits with typical
fractional errors O(0.1\%).   We also give the eigenvalues of  the
GUT scale Yukawa vertex threshold correction factors $Z_{f,\bar f,
H,\ovl H}$  and ``Higgs fractions'' \cite{abmsv,ag2,nmsgut}
  $\alpha_i, {\bar \alpha}_i$ crucial for determining the
 fermion mass formulae \cite{ag2,blmdm,gmblm,nmsgut}.
 These parameters are determined as a consequence of the GUT scale
symmetry breaking and the fine tuning to preserve a light pair of
MSSM Higgs doublets. They  distill the influence of the SO(10) GUT
on the low energy fermion physics. The reader may use them
together with the formulae given in \cite{nmsgut} to check the
fits even without entering into the details of our GUT scale mass
spectra. We note that the values of the $\alpha_1, {\bar
\alpha}_1$ quoted were chosen real by convention (see Appendix
\textbf{C} in the arXiv version of \cite{nmsgut} where full
expressions are given) but the phases of
$V^H_{i1}\sim\alpha_i,U^H_{i1}\sim \bar\alpha_i$  used in the
threshold correction formulae were fixed by demanding
semi-positive eigenvalues for the Higgs mass matrix. Since the
overall phase of the $\alpha, \bar{\alpha}$ nowhere enters our
physical parameters we have let the discrepancy stand. Tables 2,8
show the reduction in magnitude of SO(10) matter Yukawas. As a
result universal corrections dominate and make the GUT scale
threshold corrections to all three generations small and almost
equal.

In Tables  4,10 values of the SM masses at $M_Z$ are compared with
those of masses from the run down Yukawas achieved in the NMSGUT
both before and after large $\tan\beta$ driven radiative
corrections. Note that due to the inclusion of Susy threshold
corrections  the current experimental central value of
$m_b(M_Z)=2.9$ GeV can become  acceptable (see Solution 2, Tables
9,10) in contrast to small $A_0$ scenarios where the need for
$m_b(M_Z)>3.1$ GeV, i.e. more than one standard deviation away,
has been a source of tension   for small $A_0$ models \cite{king}.

 In Tables  5,11 we give  values of the soft supersymmetry
 breaking parameters which are a  crucial and remarkable
  output of this study since they tie the survival of the NMSGUT to a
  distinctive  type of soft Susy spectrum with large
  $\mu,A_0,B > 100$ TeV and third generation sfermion masses   in
 the 10-50 TeV range. Remarkably, and in sharp
  contrast to received (small $A_0,M^2_{H,{\bar H}}$) wisdom,
 the third s-generation is much \emph{heavier} than the first two
 sgenerations, which however are themselves not very light \emph{except}
possibly for the \emph{right chiral} sfermions particularly the
smuon (see Solution 1)  which can descend close to their
experimental lower limits. Light smuon solutions are  very
interesting since they permit a significant supersymmetric
contribution to the muon $g-2$ anomaly.  They can also contribute
to the effectiveness  of the pure Bino LSP (and pure Wino lightest
chargino and  next to lightest Neutralinos) as candidate
  dark matter by providing co-annihilation channels of the sort
 a light $stau$ is often enlisted for
 in standard Susy scenarios.

 Tables 6,12  give Susy particle masses
 determined using two loop RGEs and without generation mixing
switched on while in Tables 7,13 give  the masses with generation
mixing.  They are so similar as to justify the use of the diagonal
values for estimating the Susy threshold
 corrections. For the case of the lightest sfermions however
 the corrections are sometimes as large as 10-30\%.
 This again sounds a note of caution regarding the exact
 numerical values of the (tree level)  lighter sfermion masses we obtain.

In Table 14,  we collect values of B-decay rates for our example
solutions.  In Table 15 we give the  values of the $b\rightarrow s
\gamma$ branching ratio, the contribution to the muon $g-2$
anomaly,  the variation in the Standard model $\rho$ parameter,
and the value of the CP violation parameter
$\epsilon$\cite{buchplumach} in the Leptonic sector which is
relevant for Lepto-genesis :

\begin{equation}
 \epsilon\simeq - \frac{3 M_1 }{8 \pi M_2 } \frac{Im[(Y^\dag_\nu Y_\nu)^2]_{12}}{(Y^\dag_\nu Y_\nu)_{11}}
 \end{equation}

 We have not yet optimized our
solutions with respect to flavor violation observables and limits.
The overlap of the  range of values seen with the  range allowed
by experimental constraints  implies   that a successful
optimization is possible and highly constraining once
Supersymmetric particles are observed.

\section{Discussion of exotic observables}

 Baryon decay via $ d=5$ operators is,  as usual \cite{lucasraby,gotonihei},
 dominated by the chargino mediated channels. The heavy
 sfermions help with suppressing B-decay. The dominant
channels are $ Baryon \rightarrow Meson + neutrino $. We emphasize
that the flavor violation required by $d=5$ B violation is
supplied entirely by the rundown values of the (off diagonal)
SuperCKM values determined by the fitting of the fermion Yukawas
at $M_X$ by the SO(10)  light fermion Yukawa formulae
\cite{abmsv,ag1,ag2,blmdm,nmsgut}. We calculated the proton decay
rates in the   dominant channels   using the formulae for the
dimension 5 operators obtained  in \cite{nmsgut}, after running
them down to $M_Z$ using 1-loop RG equations,  adapting the
formalism of \cite{lucasraby,gotonihei}.

 Table 14 shows that    we have been able to suppress the B decay
rates to lie comfortably within the current limits. Thus the
search criteria may even be loosened without conflict with
experiment. Given enough computational resources, we could also
conduct fine grained searches where B-decay rates are calculated
for every trial parameter set. We note that our programs can
already calculate the rates in other channels driven by Gluino,
Neutralino, Higgsino etc exchange. However we defer a presentation
of the results for the subdominant channels till the various
corrections and improvements still needed   have been implemented.
Our aim was to show that the NMSGUT is quite compatible with the
stability of the proton to the degree it has been tested, and even
beyond. Firm predictions will ensue only once the Susy spectrum is
anchored in reality by a discovery of a supersymmetric particle.

We plugged our soft Susy parameters at $M_Z$ into the  SPHENO
\cite{porod} routines to obtain the ``flavor'' violation
contributions shown in Table 15. The very heavy third sgeneration
masses  imply   acceptable rates    $BR( b\rightarrow s \gamma)$
which are  uniform over  the fits. These branching ratio values
are right in the center of the region $(3-4\times 10^{-4})\pm
15\%$ determined by measurements at CLEO, BaBar and Belle
\cite{pdg2012,cleo,babar,belle}. The Susy contribution to muon
anomalous magnetic moment $\Delta a_\mu=\Delta(g-2)_\mu/2$ ranges
from negligible to significant depending on the smuon mass.
  The current difference between experiment and theory for the muon magnetic
moment anomaly is $\Delta a_\mu=287(63)(49)\times
10^{-11}$\cite{pdg2012}. Thus our light smuon solutions give
$a_\mu$ in the right range. The $\rho $ parameter $\Delta\rho$ is
also found to be severely suppressed by the decoupled spectrum of
sfermions. The predicted change in the $\rho$ parameter is so
small as to be insignificant compared with the experimental
uncertainties $\sim .001$\cite{pdg2012}. Finally the  values  of
the Leptonic  CP violaton parameters $\epsilon,\delta_{PMNS}$ seem
to be somewhat small relative to estimates\cite{buchplumach} in
the literature but may well increase upon optimization since CP
violation parameters which arise from phases  are notoriously
fickle. The values in Table 15 are thus  in the right ball park
and we may well begin to use the value of $\Delta a_\mu$ to
discriminate between different models provided one is confident
that all instabilities in the parameter determination process have
been controlled by adequate attention to loop and threshold
effects. At the moment however we simply note that there is no
gross conflict.

 The unification scale tends to be raised
above $M_X^0$ in the NMSGUT i.e. $\Delta_X>0$ . This is especially
true once we demand that $d=5$ operators mediating proton decay be
suppressed. In fact in  fits of  \cite{nmsgut}  the values of
$\Delta_X$  are $-0.30, 2.15$  while with threshold corrections we
get (Tables 2,8)  $ 0.67, 0.80$. Thus we see that the unification
scale (defined as the mass of the B-violating gauginos of type $
X[3,2,\pm{5\over 3}]$) is typically raised one order of magnitude
to $\sim 10^{17} -10^{17.5}$ GeV. On the other hand the correction
to the inverse value of the fine structure constant ($\Delta_G$)
at  the unification scale tends to make the gauge coupling at
Unification quite large($\alpha_G\sim 0.2$). Both these tendencies
together with the well known UV Landau pole in the SO(10) gauge RG
flow due to the large gauge beta functions of the large SO(10)
irreps used again point to the existence of a physical cutoff
lying around $10^{17.5}$ GeV. This is  close to the Planck scale
where gravity is expected to become strong. Solutions with smaller
$\alpha_G$ can at most improve the coincidence of the two scales.
  An ideal scenario\cite{trmin,tas} is  that the theory is
  still weakly coupled enough to be well  described by
   perturbative SO(10) at the threshold
corrected unification scale $M_X
 \sim  10^{17.5}$ GeV,  but that thereafter the Susy GUT becomes strongly
coupled simultaneously with gravity. In that case the Planck scale
may be identified as a physical cutoff for the Susy NMSGUT where
it condenses as strongly coupled Supersymmetric gauge theory
described by an appropriate SO(10) singlet supersymmetric  sigma
model. We envisaged \cite{tas} the possibility that gravity arises
dynamically as an induced effect of the quantum fluctuations of
the Susy GUT calculated in a coordinate independent framework.
This may be realized as a path integral over a background metric
that begins to propagate only at low energies leading to the near
canonical N=1 Supergravity perturbative NMSGUT as the effective
theory below $M_{Planck}$: as we assume in this work.

  \section{Conclusions and Outlook}
This paper is the second of a series \cite{nmsgut} devoted to
evaluating the ability of the NMSGUT to fit all the known fermion
mass and mixing data and be consistent with known constraints on
exotic BSM processes. The ultimate aim is to develop  the
NMSO(10)GUT into complete and calculable theory of particle
physics and particle cosmology \cite{SSI} at scales below the
Planck scale. In earlier papers, after developing a translation
manual to rewrite field theories invariant under orthogonal groups
in terms of labels of their unitary subgroups \cite{ag1} as a
basic enabling technique,  we showed \cite{ag2,blmdm,gmblm,nmsgut}
that the theory is sufficiently simple as to allow explicit
calculation of the spontaneous symmetry breaking, mass spectra and
eigenstates. It allows  computation of the RG flow in terms of the
fundamental GUT parameters to the point where one can attempt to
actually fit the low energy data, i.e the SM parameters together
with the neutrino mixing data, in its entirety.   However,
although successful in fitting the fermion mass data \cite{nmsgut}
and yielding distinctive and falsifiable signals regarding the
required Susy spectra, the fits gave  $d=5$ operator mediated
proton decay rates that are at least 6 orders of magnitude larger
than the current experimental limits\cite{pdg2012}.

Faced with an apparent nullification of the previous successes
 we re-examined our treatment of the relation
between the Higgs doublets of effective and High scale theories
\cite{gutupend}. Our approximate treatment\cite{gutupend} of
threshold corrections  immediately showed that superheavy
corrections to Higgs (and matter) kinetic terms and thus to the
Yukawa couplings would inevitably play a critical role due to the
large number of fields involved in dressing each line entering the
effective MSSM vertices. In fact
  care must be taken to maintain positivity of the kinetic terms
  after renormalization which is otherwise generically badly violated : in
particular by the   fits found earlier. In this paper we have
completed and corrected the approximate treatment of
\cite{gutupend} while maintaining positive kinetic terms. As a
result we find that searches incorporating threshold corrected
Yukawa couplings, and a constraint to respect B-decay limits,
naturally flow to region of parameter space that have weak Yukawa
couplings and $Z_{H,\ovl H}$ close to zero and hence imply strong
lowering of the required SO(10)  matter Yukawa couplings.  The
mechanism that we have demonstrated is likely \cite{lettgutupend}
to work in \emph{any} realistic GUT since the features required
are so generic and the necessity of implementation of threshold
corrections while maintaining unitarity undeniable. Since its
success depends on $Z_H$ approaching zero while remaining positive
rather than fine tuning to some specific parameter values our
mechanism  is likely to be robust against 2 and higher loop
corrections.  Moreover the large wave function renormalization
driven threshold/matching effects can also have notable influence
on  soft supersymmetrty breaking parameters, enhancing
$\mu,M^2_{H,\ovl H}$  relative to their GUT scale values
consistent with the patterns found in our fitshere and before.
  As such our  paper yet again confirms\cite{dixitsher,ag2,nmsgut}
 that the calculation of threshold effects should be a \emph{sine
qua non} of serious work on Grand Unified models.

In this paper our focus has been to report only the details of the
calculation of the complete threshold corrections for the NMSGUT
and exhibit successful fits that also respect Baryon decay limits.
We have also exhibited the values of the most prominent monitors
of BSM viability such as estimates of $a_\mu,\Gamma(b \rightarrow
s\gamma)$ and found that the $d=5,\Delta B\neq 0 $ problem is
essentially solved but there is room for optimization of other BSM
parameters in future searches.

Since our  theory   claims to be a realistic UV completion of the
MSSM a  host of phenomenological issues arises. Serious
consideration of these requires implementation of improvements
such as using loop corrected sparticle masses, implementation of
heavy neutrino thresholds, detailed and generic analysis of the RG
flows in novel Susy parameter region indicated by the NMSGUT,
incorporation of generation mixing flows in the soft sector,
issues of safety as regards Color and Charge breaking minima,
detailed BSM phenomenology, calculation of Leptogenesis using the
calculable Leptonic CP violation,  Dark matter constraints,
Inflationary scenarios\cite{SSI}  and so on. These will be
reported in the sequels.

In summary, by solving the conundrum of fast dimension 5 operator
mediated B decay  the NMSGUT has passed another formidable barrier
to its development into a complete, calculable and falsifiable
theory providing consistent UV completion to Particle Physics and
Cosmology.

\section{Acknowledgments}
 \vspace{ .5 true cm}

 It is a pleasure for C.S.A  to acknowledge
the hospitality of the High Energy Theory Group ICTP,Trieste and
in particular Goran  Senjanovic at various times over the long
gestation period of this paper. He is also indebted to  Kaladi
Babu,  Jogesh Pati,   Barbara Szczerbinska  for invitations to
visit stimulating  meetings(CETUP)  at Lead(SD) and Asilomar(CA)
during 2012-2013. He also thanks Kaladi Babu,
 Jogesh Pati,  Borut Bajc,  Zhurab Tavartkiladze, for useful
conversations and encouragement,  and W.Porod for permissions and
help with SPHENO. I.G and C.K.K would like to thank the CSIR
(Council of Scientific and Industrial Research of India ) and  the
UGC (University Grants Commission of India) for financial support
through Senior Research Fellowships during the period of this
work.

\section*{Appendix A}

We give below our results for the threshold corrections to the
Yukawa couplings of the matter fields due to  heavy fields running
in a self energy loop on a line leading into the Yukawa vertex.
The calculation is quite tedious but we applied various
consistency checks to ensure that we had included contributions
from all members of multiplets.

 The corrections to the matter field lines are given by using the
trilinear invariants of the matter fields in the \textbf{16}-plet
to the Higgs in the $\mathbf{10,120,\oot}$ irreps, and gauge
fields in the \textbf{45}-plet, decomposed into MSSM irreps
\cite{ag1,ag2,blmdm,nmsgut}. With $Z=1-{\cal K}$ in the notation
of eqn(10), $K^f_\Phi$ refers to the loop corrections on the
matter ($f$) line in which the loop contains the heavy multiplet
$\Phi$. The corrections on the Higgs lines ${\cal K}_{H,\overline
H}$ are described below. Using the formulae in Section \textbf{3 }
leads straightforwardly to

\bea \justifying ({ {16\pi^2}}) {{\cal K}}^{\bar{u}}&=&
K^{\bar{u}}_{\bar{T}}+K^{\bar{u}}_{T}+2
K^{\bar{u}}_{H}+\frac{16}{3}K^{\bar{u}}_{C}
+2K^{\bar{u}}_{D}+K^{\bar{u}}_{J}+4K^{\bar{u}}_{L}+4K^{\bar{u}}_{\bar
K}+16 K^{\bar{u}}_{M}\nnu &&- 2 {g_{10}^2} (0.05
F_{11}(m_{\lambda_G},Q) + F_{11}(m_{\lambda_J},Q) +
F_{11}(m_{\lambda_F},Q) +4 F_{11}(m_{\lambda_X},Q) \nnu
&&+2 F_{11}(m_{\lambda_E},Q)) \\
K^{\bar{u}}_{\bar{T}}&=&2\sum_ {a =
1}^{\mbox{7}}\biggr(\bar{h}U^T_{1a}-2\bar{f}U^T_{2a}-\sqrt{2} i
\bar{g}U^T_{7a}\biggr)^*
\biggr(\bar{h}U^T_{1a}-2\bar{f}U^T_{2a}\nnu &&+\sqrt{2} i
\bar{g}U^T_{7a}\biggr) F_{11}(m^T_{a}, Q)\\
K^{\bar{u}}_{T}&=&\sum_ {a =
1}^{\mbox{7}}\biggr(\bar{h}V^T_{1a}-2\bar{f}V^T_{2a}-2\sqrt{2} i
\bar{f}V^T_{4a}-\sqrt{2}\bar{g}V^T_{6a}+\sqrt{2} i
\bar{g}V^T_{7a}\biggr)^*
\biggr(\bar{h}V^T_{1a}-2\bar{f}V^T_{2a}\nnu &&-2\sqrt{2} i
\bar{f}V^T_{4a}+\sqrt{2}\bar{g}V^T_{6a}-\sqrt{2} i
\bar{g}V^T_{7a}\biggr) F_{11}(m^T_{a}, Q)\\ K^{\bar{u}}_{H}&=&
\sum_ {a =
2}^{\mbox{6}}\biggr(\bar{h}V^H_{1a}-\frac{2i}{\sqrt{3}}\bar{f}V^H_{2a}-
 \bar{g}V^H_{5a}+\frac{i}{\sqrt{3}} \bar{g}V^H_{6a}\biggr)^*
\biggr(\bar{h}V^H_{1a}-\frac{2i}{\sqrt{3}}\bar{f}V^H_{2a}+
 \bar{g}V^H_{5a}\nnu &&-\frac{i}{\sqrt{3}} \bar{g}V^H_{6a}\biggr)
F_{11}(m^H_{a}, Q)\\ K^{\bar{u}}_{C}&=&\sum_ {a =
1}^{\mbox{3}}\biggr(-4\bar{f}V^C_{2a}+
 2\bar{g}V^C_{3a}\biggr)^*\biggr(-4\bar{f}V^C_{2a}-
2\bar{g}V^C_{3a}\biggr) F_{11}(m^C_{a}, Q)\\
K^{\bar{u}}_{D}&=&\sum_ {a =
1}^{\mbox{3}}\biggr(-4\bar{f}V^D_{1a}+
 2\bar{g}V^D_{3a}\biggr)^*\biggr(-4\bar{f}V^D_{1a}-
2\bar{g}V^D_{3a}\biggr) F_{11}(m^D_{a},Q)\\
K^{\bar{u}}_{J}&=&\sum_ {a = 1,a\neq
4}^{\mbox{5}}\biggr(-4i\bar{f}V^J_{1a}+
 2i\bar{g}V^J_{5a}\biggr)^*\biggr(-4i\bar{f}V^J_{1a}-
 2i\bar{g}V^J_{5a}\biggr) F_{11}(m^J_{a}, Q)\\
K^{\bar{u}}_{L}&=&\sum_ {a =
1}^{\mbox{2}}\biggr(-2\sqrt{2}i\bar{f}V^L_{1a}-
 \sqrt{2}\bar{g}V^L_{2a}\biggr)^*\biggr(-2\sqrt{2}i\bar{f}V^L_{1a}+
 \sqrt{2}\bar{g}V^L_{2a}\biggr) F_{11}(m^L_{a}, Q)\\
K^{\bar{u}}_{\bar{K}}&=&2\sum_ {a = 1}^{\mbox{2}}
(-i\bar{g})^*(i\bar{g}) |U^K_{2a}|^2F_{11}(m^K_{a}, Q)\\
K^{\bar{u}}_{M}& =& (2i\bar{f})^*(2i\bar{f}) F_{11}(m^M, Q) \eea

\bea({ {16\pi^2}}) {\cal K}^{\bar{d}}&=&
K^{\bar{d}}_{\bar{T}}+K^{\bar{d}}_{T}+2K^{\bar{d}}_{\bar
H}+\frac{16}{3}K^{\bar{d}}_{\bar C}+2K^{\bar{d}}_{E}+K^{\bar{d}}_{
K}+4K^{\bar{d}}_{L}+4K^{\bar{d}}_{\bar J}+16K^{\bar{d}}_{N}\nnu
&&-2 {g_{10}^2} (0.45 F_{11}(m_{\lambda_G},Q) +
F_{11}(m_{\lambda_J},Q) + F_{11}(m_{\lambda_F},Q) +
2F_{11}(m_{\lambda_X},Q) \nnu &&+4F_{11}(m_{\lambda_E},Q)) \\
K^{\bar{d}}_{\bar{T}}&=&2\sum_ {a =
1}^{\mbox{7}}\biggr(\bar{h}U^T_{1a}-2\bar{f}U^T_{2a}+\sqrt{2} i
\bar{g}U^T_{7a}\biggr)^*
\biggr(\bar{h}U^T_{1a}-2\bar{f}U^T_{2a}\nnu &&-\sqrt{2} i
\bar{g}U^T_{7a}\biggr) F_{11}(m^T_{a}, Q)\\
K^{\bar{d}}_{T}&=&\sum_ {a =
1}^{\mbox{7}}\biggr(-\bar{h}V^T_{1a}+2\bar{f}V^T_{2a}-2\sqrt{2} i
\bar{f}V^T_{4a}+\sqrt{2}\bar{g}V^T_{6a}+\sqrt{2} i
\bar{g}V^T_{7a}\biggr)^*
\biggr(-\bar{h}V^T_{1a}+2\bar{f}V^T_{2a}\nnu &&-2\sqrt{2} i
\bar{f}V^T_{4a}-\sqrt{2}\bar{g}V^T_{6a}-\sqrt{2} i
\bar{g}V^T_{7a}\biggr) F_{11}(m^T_{a}, Q)\\
K^{\bar{d}}_{\bar{H}}&=&\sum_ {a =
2}^{\mbox{6}}\biggr(-\bar{h}U^H_{1a}+\frac{2i}{\sqrt{3}}\bar{f}U^H_{2a}+
 \bar{g}U^H_{5a}-\frac{i}{\sqrt{3}} \bar{g}U^H_{6a}\biggr)^*
\biggr(-\bar{h}U^H_{1a}+\frac{2i}{\sqrt{3}}\bar{f}U^H_{2a}-
 \bar{g}U^H_{5a}\nnu &&+\frac{i}{\sqrt{3}} \bar{g}U^H_{6a}\biggr)
F_{11}(m^H_{a}, Q)\\ K^{\bar{d}}_{\bar{C}}&=&\sum_ {a =
1}^{\mbox{3}}\biggr(4\bar{f}U^C_{1a}-
 2\bar{g}U^C_{3a}\biggr)^*\biggr(4\bar{f}U^C_{1a}+
2\bar{g}U^C_{3a}\biggr) F_{11}(m^C_{a}, Q)\\
K^{\bar{d}}_{E}&=&\sum_ {a = 1,a \neq
5}^{\mbox{6}}\biggr(4\bar{f}V^E_{1a}-
 2\bar{g}V^E_{6a}\biggr)^*\biggr(4\bar{f}V^E_{1a}+
 2\bar{g}V^E_{6a}\biggr) F_{11}(m^E_{a},Q)\\
K^{\bar{d}}_{K}&=&\sum_ {a =
1}^{\mbox{2}}\biggr(4i\bar{f}V^K_{1a}-
 2i\bar{g}V^K_{2a}\biggr)^*\biggr(4i\bar{f}V^K_{1a}+
 2i\bar{g}V^K_{2a}\biggr) F_{11}(m^K_{a}, Q)\\
 K^{\bar{d}}_{L}&=&\sum_ {a =
1}^{\mbox{2}}\biggr(-2\sqrt{2}i\bar{f}V^L_{1a}+
 \sqrt{2}\bar{g}V^L_{2a}\biggr)^*\biggr(-2\sqrt{2}i\bar{f}V^L_{1a}-
 \sqrt{2}\bar{g}V^L_{2a}\biggr) F_{11}(m^L_{a}, Q)\\
K^{\bar{d}}_{\bar{J}}&=&2\sum_ {a = 1,a \neq 4}^{\mbox{5}}
(i\bar{g})^*(-i\bar{g}) |U^J_{5a}|^2F_{11}(m^J_{a}, Q)\\
K^{\bar{d}}_{N} &=& (2 i\bar{f})^*(2i\bar{f}) F_{11}(m^N, Q) \eea

\bea ({ {16\pi^2}}) {\cal K}^{\bar e}&=& 3
K^{\bar{e}}_{T}+2K^{\bar{e}}_{\bar H}+K^{\bar{e}}_{\bar
F}+6K^{\bar{e}}_{\bar D}+3K^{\bar{e}}_{K}+4K^{\bar{e}}_{\bar{A}}-
2 {g_{10}^2} (0.05 F_{11}(m_{\lambda_G},Q) \nnu &&
+3F_{11}(m_{\lambda_J},Q) +F_{11}(m_{\lambda_F},Q)
+6F_{11}(m_{\lambda_X},Q))\\ K^{\bar{e}}_{T}&=&\sum_ {a =
1}^{\mbox{7}}\biggr(\bar{h}V^T_{1a}-2\bar{f}V^T_{2a}-2\sqrt{2} i
\bar{f}V^T_{4a}+\sqrt{2}\bar{g}V^T_{6a}-\sqrt{2} i
\bar{g}V^T_{7a}\biggr)^*
\biggr(\bar{h}V^T_{1a}-2\bar{f}V^T_{2a}\nnu &&-2\sqrt{2} i
\bar{f}V^T_{4a}-\sqrt{2}\bar{g}V^T_{6a}+\sqrt{2} i
\bar{g}V^T_{7a}\biggr) F_{11}(m^T_{a}, Q)\\
K^{\bar{e}}_{\bar{H}}&=&\sum_ {a =
2}^{\mbox{6}}\biggr(-\bar{h}U^H_{1a}-2\sqrt{3}i\bar{f}U^H_{2a}+
 \bar{g}U^H_{5a}+\sqrt{3} i \bar{g}U^H_{6a}\biggr)^*
\biggr(-\bar{h}U^H_{1a}-2\sqrt{3}i\bar{f}U^H_{2a}\nnu &&-
 \bar{g}U^H_{5a}-\sqrt{3} i \bar{g}U^H_{6a}\biggr)
F_{11}(m^H_{a}, Q)\\ K^{\bar{e}}_{\bar{F}}&=&\sum_ {a = 1,a\neq
3}^{\mbox{4}}\biggr(4i\bar{f}U^F_{1a}-
 2\bar{g}U^F_{4a}\biggr)^*\biggr(4i\bar{f}U^F_{1a}+
 2\bar{g}U^F_{4a}\biggr) F_{11}(m^F_{a}, Q)\\
K^{\bar{e}}_{\bar D}&=&\sum_ {a =
1}^{\mbox{3}}\biggr(4\bar{f}U^D_{2a}-
 2\bar{g}U^D_{3a}\biggr)^*\biggr(4\bar{f}U^D_{2a}+
 2\bar{g}U^D_{3a}\biggr) F_{11}(m^D_{a},Q)\\
K^{\bar{e}}_{K}&=&\sum_ {a =
1}^{\mbox{2}}\biggr(4i\bar{f}V^K_{1a}+
 2i\bar{g}V^K_{2a}\biggr)^*\biggr(4i\bar{f}V^K_{1a}-
 2i\bar{g}V^K_{2a}\biggr) F_{11}(m^K_{a}, Q)\\
K^{\bar{e}}_{\bar{A}}&=& (2\sqrt{2}i\bar{f})^*(2\sqrt{2}i\bar{f})
F_{11}(m^A, Q)\eea

\bea ({ {16\pi^2}}) {\cal K}^{\bar \nu}&=&
3K^{\bar{\nu}}_{T}+2K^{\bar{\nu}}_{ H}+K^{\bar{\nu}}_{\bar
F}+6K^{\bar{\nu}}_{\bar
E}+3K^{\bar{\nu}}_{J}+4K^{\bar{\nu}}_{\bar{G}}- 2 {g_{10}^2} (1.25
F_{11}(m_{\lambda_G},Q) \nnu && +3F_{11}(m_{\lambda_J},Q)
+F_{11}(m_{\lambda_F},Q) +6F_{11}(m_{\lambda_E},Q))\\
K^{\bar{\nu}}_{T}&=&\sum_ {a =
1}^{\mbox{7}}\biggr(-\bar{h}V^T_{1a}+2\bar{f}V^T_{2a}-2\sqrt{2} i
\bar{f}V^T_{4a}-\sqrt{2}\bar{g}V^T_{6a}-\sqrt{2} i
\bar{g}V^T_{7a}\biggr)^*
\biggr(-\bar{h}V^T_{1a}+2\bar{f}V^T_{2a}\nnu &&-2\sqrt{2} i
\bar{f}V^T_{4a}+\sqrt{2}\bar{g}V^T_{6a}+\sqrt{2} i
\bar{g}V^T_{7a}\biggr) F_{11}(m^T_{a}, Q)\\
K^{\bar{\nu}}_{H}&=&\sum_ {a =
2}^{\mbox{6}}\biggr(\bar{h}V^H_{1a}+2\sqrt{3}i\bar{f}V^H_{2a}-
 \bar{g}V^H_{5a}-\sqrt{3} i \bar{g}V^H_{6a}\biggr)^*
\biggr(\bar{h}V^H_{1a}+2\sqrt{3}i\bar{f}V^H_{2a}+
 \bar{g}V^H_{5a}\nnu &&+\sqrt{3} i \bar{g}V^H_{6a}\biggr)
F_{11}(m^H_{a}, Q)\\ K^{\bar{\nu}}_{\bar{F}}&=&\sum_ {a = 1,a \neq
3}^{\mbox{4}}\biggr(-4i\bar{f}U^F_{1a}+
 2\bar{g}U^F_{4a}\biggr)^*\biggr(-4i\bar{f}U^F_{1a}-
 2\bar{g}U^F_{4a}\biggr) F_{11}(m^F_{a}, Q)\\
K^{\bar{\nu}}_{\bar E}&=&\sum_ {a = 1,a \neq
5}^{\mbox{6}}\biggr(-4\bar{f}U^E_{2a}+
 2\bar{g}U^E_{6a}\biggr)^*\biggr(-4\bar{f}U^E_{2a}-
 2\bar{g}U^E_{6a}\biggr) F_{11}(m^E_{a},Q)\\
K^{\bar{\nu}}_{J}&=&\sum_ {a = 1,a\neq
4}^{\mbox{5}}\biggr(-4i\bar{f}V^J_{1a}-
 2i\bar{g}V^J_{5a}\biggr)^*\biggr(-4i\bar{f}V^J_{1a}+
 2i\bar{g}V^J_{5a}\biggr) F_{11}(m^J_{a}, Q)\\
K^{\bar{\nu}}_{\bar{G}}&=& \sum_ {a = 1}^{\mbox{5}}
(-2\sqrt{2}i\bar{f})^*(-2\sqrt{2}i\bar{f}) |U^G_{5a}|^2
F_{11}(m^G_{a}, Q) \eea

\bea ({ {16\pi^2}}) {\cal K}^u &=&K^{u}_{\bar{T}}+K^{u}_{T}+K^{u}_{\bar
H}+K^{u}_{H}+\frac{8}{3}K^{u}_{C}+\frac{8}{3}K^{u}_{\bar
C}+K^{u}_{\bar{E}}+K^{u}_{\bar D}+3K^{u}_{\bar P}+12K^{u}_{
P}+48K^{\bar{u}}_{W}+4K^{\bar{u}}_{\bar L}\nonumber\\&&
 - 2 {g_{10}^2} (0.05F_{11}(m_{\lambda_G},Q)
+F_{11}(m_{\lambda_J},Q) +3F_{11}(m_{\lambda_X},Q) \nnu
&&+3F_{11}(m_{\lambda_E},Q))= ({{16\pi^2}}) {\cal K}^d \\
K^{u}_{\bar{T}}&=&\sum_ {a =
1}^{\mbox{7}}\biggr(-\bar{h}U^T_{1a}-2\bar{f}U^T_{2a}+\sqrt{2}
 \bar{g}U^T_{6a}\biggr)^*
\biggr(-\bar{h}U^T_{1a}-2\bar{f}U^T_{2a}-\sqrt{2}
 \bar{g}U^T_{6a}\biggr) F_{11}(m^T_{a}, Q)\\
K^{u}_{T}&=&2\sum_ {a =
1}^{\mbox{7}}\biggr(\bar{h}V^T_{1a}+2\bar{f}V^T_{2a}\biggr)^*
\biggr(\bar{h}V^T_{1a}+2\bar{f}V^T_{2a}\biggr) F_{11}(m^T_{a},
Q)\\ K^{u}_{\bar{H}}&=&\sum_ {a =
2}^{\mbox{6}}\biggr(-\bar{h}U^H_{1a}+\frac{2i}{\sqrt{3}}\bar{f}U^H_{2a}-
 \bar{g}U^H_{5a}+\frac{i}{\sqrt{3}} \bar{g}U^H_{6a}\biggr)^*
\biggr(-\bar{h}U^H_{1a}+\frac{2i}{\sqrt{3}}\bar{f}U^H_{2a}+
 \bar{g}U^H_{5a}\nnu &&-\frac{i}{\sqrt{3}} \bar{g}U^H_{6a}\biggr)
F_{11}(m^H_{a}, Q)\\
 K^{u}_{H}&=&\sum_ {a =
2}^{\mbox{6}}\biggr(\bar{h}V^H_{1a}-\frac{2i}{\sqrt{3}}\bar{f}V^H_{2a}+\bar{g}V^H_{5a}-
\frac{i\bar{g}}{\sqrt{3}}V^H_{6a}\biggr)^*\biggr(\bar{h}V^H_{1a}
-\frac{2i}{\sqrt{3}}\bar{f}V^H_{2a}-\bar{g}V^H_{5a}\nnu
&&+\frac{i\bar{g}}{\sqrt{3}}V^H_{6a}\biggr)
F_{11}(m^H_{a}, Q)\\
 K^{u}_{C}&=&\sum_ {a =
1}^{\mbox{3}}\biggr(-4\bar{f}V^C_{2a}-
 2\bar{g}V^C_{3a}\biggr)^*\biggr(-4\bar{f}V^C_{2a}+
 2\bar{g}V^C_{3a}\biggr) F_{11}(m^C_{a}, Q) \\ K^{u}_{\bar{C}}&=&\sum_ {a =
1}^{\mbox{3}}\biggr(4\bar{f}U^C_{1a}+
 2\bar{g}U^C_{3a}\biggr)^*\biggr(4\bar{f}U^C_{1a}-
 2\bar{g}U^C_{3a}\biggr) F_{11}(m^C_{a}, Q)\\
K^{u}_{\bar E}&=&\sum_ {a = 1,a\neq 5}^{\mbox{6}}
\biggr(-4\bar{f}U^E_{2a}-
 2\bar{g}U^E_{6a}\biggr)^*\biggr(-4\bar{f}U^E_{2a}+
 2\bar{g}U^E_{6a}\biggr) F_{11}(m^E_{a},Q)\\
K^{u}_{\bar D}&=&\sum_ {a = 1}^{\mbox{3}}\biggr(4\bar{f}U^D_{2a}+
 2 \bar{g}U^D_{3a}\biggr)^*\biggr(4\bar{f}U^D_{2a}-
 2 \bar{g}U^D_{3a}\biggr) F_{11}(m^D_{a}, Q) \\
 K^{u}_{\bar P}&=&\sum_ {a =
1}^{\mbox{2}}\biggr(2 \sqrt{2}\bar{f}U^P_{1a}-
 \sqrt{2}\bar{g}U^P_{2a}\biggr)^*\biggr(2 \sqrt{2}\bar{f}U^P_{1a}+
 \sqrt{2}\bar{g}U^P_{2a}\biggr) F_{11}(m^P_{a}, Q) \\
 \nnu
K^{u}_{P}&=&2\sum_ {a = 1}^{\mbox{2}}
\biggr(-\frac{\bar{g}}{\sqrt{2}}\biggr)^*\biggr(\frac{\bar{g}}{\sqrt{2}}\biggr)
|V^P_{2a}|^2F_{11}(m^P_{a}, Q)\\ K^{u}_{W} &=&
(\sqrt{2}\bar{f})^*(\sqrt{2}\bar{f})
F_{11}(m^W, Q)\\
 K^{u}_{\bar L}&=&\sum_ {a = 1}^{\mbox{2}} (-\sqrt{2}\bar{g})^*(\sqrt{2}\bar{g})
|U^L_{2a}|^2F_{11}(m^L_{a}, Q)\eea

\bea ({ {16\pi^2}}) {\cal K}^{e}&=& 3K^{e}_{\bar T}+K^{e}_{\bar
H}+K^{e}_{ H}+3K^{e}_{ D}+3K^{e}_{E}+ 9 K^{e}_{\bar
P}+K^{e}_{F}+12K^{e}_{\bar O}- 2 {g_{10}^2} (0.45
F_{11}(m_{\lambda_G},Q) \nnu && +3F_{11}(m_{\lambda_J},Q)
+3F_{11}(m_{\lambda_X},Q) +3F_{11}(m_{\lambda_E},Q))=({ {16\pi^2}}) {\cal K}^{\nu}\\
K^{e}_{\bar T}&=&\sum_ {a =
1}^{\mbox{7}}\biggr(-\bar{h}U^T_{1a}-2\bar{f}U^T_{2a}-\sqrt{2}\bar{g}U^T_{6a}\biggr)^*
\biggr(-\bar{h}U^T_{1a}-2\bar{f}U^T_{2a}+\sqrt{2}\bar{g}U^T_{6a}\biggr)\nnu &&
F_{11}(m^T_{a}, Q)\\ K^{e}_{\bar{H}}&=&\sum_ {a =
2}^{\mbox{6}}\biggr(\bar{h}U^H_{1a}+2\sqrt{3}i\bar{f}U^H_{2a}+
 \bar{g}U^H_{5a}+\sqrt{3} i \bar{g}U^H_{6a}\biggr)^*
\biggr(\bar{h}U^H_{1a}+2\sqrt{3}i\bar{f}U^H_{2a}-
 \bar{g}U^H_{5a}\nnu &&-\sqrt{3} i \bar{g}U^H_{6a}\biggr)
F_{11}(m^H_{a}, Q)\\ K^{e}_{H}&=&\sum_ {a =
2}^{\mbox{6}}\biggr(-\bar{h}V^H_{1a}-
 2 \sqrt{3}i\bar{f}V^H_{2a}- \bar{g} V^H_{5a}-i
\sqrt{3} \bar{g}V^H_{6a}\biggr)^*\biggr(-\bar{h}V^H_{1a}-
 2 \sqrt{3}i\bar{f}V^H_{2a}+ \bar{g} V^H_{5a}\nnu &&+i
\sqrt{3} \bar{g}V^H_{6a}\biggr) F_{11}(m^F_{a}, Q)\\
 K^{e}_{ D}&=&\sum_ {a =
1}^{\mbox{3}}\biggr(4\bar{f}V^D_{1a}+
 2\bar{g}V^D_{3a}\biggr)^*\biggr(4\bar{f}V^D_{1a}-
 2\bar{g}V^D_{3a}\biggr) F_{11}(m^D_{a},Q)\\
 K^{e}_{E}&=&\sum_ {a =
1,a \neq 5}^{\mbox{6}}\biggr(-4\bar{f}V^E_{1a}-
 2\bar{g}V^E_{6a}\biggr)^*\biggr(-4\bar{f}V^E_{1a}+
 2\bar{g}V^E_{6a}\biggr) F_{11}(m^E_{a}, Q)\\
 K^{e}_{\bar P}&=&\sum_ {a =
1}^{\mbox{2}}\biggr(2 \sqrt{2}\bar{f}U^P_{1a}+
 \sqrt{2}\bar{g}U^P_{2a}\biggr)^*\biggr(2 \sqrt{2}\bar{f}U^P_{1a}-
 \sqrt{2}\bar{g}U^P_{2a}\biggr) F_{11}(m^P_{a}, Q)\\
 K^{e}_{F}&=& \sum_{a=1,a\neq 3}^{\mbox{4}}(-2\bar{g})^*(2\bar{g})|V^F_{4a}|^2 F_{11}(m^F_a,
Q)\\K^{e}_{\bar O}&= &(2i\bar{f})^*(2i\bar{f}) F_{11}(m^O, Q) \eea
Here $g_{10}$ is the SO(10) gauge coupling    and
\[ \bar h= 2 \sqrt{2} h \quad ;  \quad \bar g= 2 \sqrt{2} g   \quad ;  \quad \bar f= 2 \sqrt{2} f  \]
The calculation for the corrections to the light Higgs doublet
lines  $H,{\overline H}$ is  much more complicated than the matter
lines  since these are mixtures of pairs of doublets from the
$\mathbf{10,120}$(2 pairs)$\mathbf{ \oot,126,210}$ SO(10) Higgs
multiplets($ H= (V^H)^\dagger h, {\overline H}=  (U^H)^\dagger
{\overline h}$). The couplings of the GUT field doublets
$h_a,{\bar h}_a, =1,2...6 $ (see \cite{nmsgut} for conventions) to
various \emph{pairs} of the 26 different MSSM irrep-types
(labelled conveniently by the letters of the  alphabet :
see\cite{ag2,nmsgut}) that occur in this theory can be easily -if
tediously- worked out using the technology\cite{ag1} of SO(10)
decomposition via the Pati-Salam group. Amusingly there are again
precisely 26 different combinations of GUT multiplets (labelled by
the letter pairs for irreps which can combine to give operators
that can form singlets with the MSSM $H [1,2,1]$ and 26 with
${\overline H}[1,2,-1]$). Then we get\bea (16 \pi^2){\cal K}_{H}
&=&8K_{R\bar C}+
 3K_{J\bar D}+3K_{E\bar J}
 + 9K_{X \bar P} +3K_{X\bar T}
  + 9K_{P\bar E}+
 3K_{T\bar E}
 + 6K_{Y\bar L}
 + K_{VF}
 + 8K_{C\bar Z}\nonumber\\&&
 + 3K_{D\bar I}+
 24K_{Q\bar C}
 + 9K_{E\bar U}
 + 9K_{U\bar D}
 + 6K_{L\bar B}
 + 3K_{K \bar X}
 + 6K_{B\bar M}
 + 18K_{W\bar B}\nonumber\\&&
 + 18K_{Y\bar W}
 + 3K_{V\bar O}
 + 6K_{N\bar Y}
 + K_{\bar V\bar A}
 + 3K_{HO}
 + 3K_{S\bar H}
 + K_{H\bar F}+ K_{G\bar H}\eea

Similarly for $\overline H$ we get the conjugated pairs running in
the loop (unless it is a real irrep)

\bea
  K_{R\bar C}
  & = &\sum_ {a = 1}^{\mbox{d(R)}}\sum_ {a' =
  1}^{\mbox{d(C)}} \biggr| \biggr(
  \frac {i\kappa} {\sqrt {2}} V^R_{2a} U^C_{3a'}-\gamma V^R_{1a} U^C_{2a'} + \frac {\gamma } {\sqrt {2}}V^R_{2a} U^C_{2a'} -
  \bar\gamma V^R_{1a}U^C_{1a'} - \frac {\bar\gamma } {\sqrt{2}}V^R_{2a}U^C_{1a'} \biggr) V^H_{11} \nonumber \\ && +\biggr(\frac {2\eta } {\sqrt{3}} V^R_{1a}U^C_{2a'}- \sqrt{\frac{2}{3}}\eta V^R_{2a} U^C_{2a'} + \frac {i \bar\zeta} {\sqrt{6}} V^R_{2a} U^C_{3a'}\biggr) V^H_{21} +\biggr(\frac {2 \eta} {\sqrt{3}} V^R_{1a}U^C_{1a'} +\sqrt{\frac{2}{3}}\eta V^R_{2a} U^C_{1a'} \nonumber \\ &&+ \frac {i\zeta}{\sqrt{6}} V^R_{2a} U^C_{3a'}\biggr) V^H_{31} +\biggr( \frac {\zeta} {\sqrt{2}} V^R_{2a} U^C_{2a'}-\frac {i \rho} {3\sqrt {2}} V^R_{2a} U^C_{3a'} +
   \frac {\bar\zeta}{\sqrt{2}} V^R_{2a} U^C_{1a'}\biggr) V^H_{51} -\biggr(\frac {i\zeta} {\sqrt {6}} V^R_{2a} U^C_{2a'}\nonumber \\ && + \frac {i\bar\zeta} {\sqrt{6}} V^R_{2a} U^C_{1a'} -
  \frac {\rho} {3\sqrt{3}}V^R_{1a} U^C_{3a'}\biggr) V^H_{61} \biggr| ^2 F_{12}(m^R_{a}, m^C_{a'}, Q)\eea

\bea
 K_{J\bar D}
& = & \sum_ {a =1}^{\mbox{d(J)}}\sum_ {a' =1}^{\mbox{d(D)}}
\biggr| \biggr(\gamma V^J_{2a} U^D_{1a'} -\frac {\gamma}{\sqrt
{2}}
  V^J_{3a}U^D_{1a'} +
 \bar\gamma V^J_{2a} U^D_{2a'} +
 \frac { \bar\gamma} {\sqrt {2}} V^J_{3a}U^D_{2a'} - \frac {i\kappa}{\sqrt {2}} V^J_{3a} U^D_{3a'}\biggr) V^H_{11}
 \nonumber \\ && + \biggr( \frac {2\eta } {\sqrt {3}}V^J_{2a} U^D_{1a'} - \sqrt {6}\eta  V^J_{3a} U^D_{1a'} -
  \frac {2 i\bar\zeta } {\sqrt {3}}V^J_{2a} U^D_{3a'} + \sqrt{\frac
{3} {2}}i\bar\zeta V^J_{3a} U^D_{3a'}\biggr) V^H_{21} +
\biggr(\frac{-i}{\sqrt{6}} \zeta V^J_{3a} U^D_{3a'}\nonumber \\
&&-\frac {2i\zeta} {\sqrt{3}}V^J_{2a}U^D_{3a'} +
  \frac{2\eta}{\sqrt{3}} V^J_{2a} U^D_{2a'} - \sqrt{\frac{2}{3}}\eta V^J_{3a} U^D_{2a'}\biggr) V^H_{31} -\biggr(\frac{i \rho}{3} V^J_{5a} U^D_{3a'} +
  4\eta V^J_{1a} U^D_{1a'} \nonumber \\ &&+
  2 i\bar\zeta V^J_{1a} U^D_{3a'} +
  2\bar\zeta V^J_{5a} U^D_{2a'}\biggr) V^H_{41} +\biggr(\frac{i \rho}{3\sqrt{2}} V^J_{3a} U^D_{3a'} -
  \frac{\bar\zeta}{ \sqrt{2}} V^J_{3a} U^D_{2a'} -
  \frac {\zeta}{\sqrt {2}} V^J_{3a} U^D_{1a'}\biggr) V^H_{51} \nonumber \\ && +\biggr( \frac{2i\zeta}{\sqrt{3}} V^J_{2a} U^D_{1a'}-\sqrt{\frac{3}{2}}i\zeta V^J_{3a} U^D_{1a'}
    +
    \frac{i\bar\zeta}{\sqrt{6}}V^J_{3a} U^D_{2a'}+
   \frac{2i\bar\zeta}{ \sqrt{3}} V^J_{2a} U^D_{2a'} +
   \frac {\rho}{3\sqrt{3}}V^J_{2a} U^D_{3a'} \nonumber \\ &&- \frac { \sqrt{2}\rho}{3\sqrt{3}} V^J_{3a} U^D_{3a'}\biggr) V^H_{61} \biggr| ^2 F_{12}(m^J_a, m^D_{a'},
   Q) \nonumber \\ &&-2{g_{10}^2} \biggr|\frac{-2i}{\sqrt{3}}\biggr(V^{D*}_{1a'}V^H_{21}+V^{D*}_{2a'}V^H_{31}+V^{D*}_{3a'}V^H_{61}\biggr)\biggr|^2 F_{12}(m_{\lambda_J}, m^D_{a'},
   Q)\eea

\bea K_{E \bar J} & = & \sum_ {a =1}^{\mbox{d(E)}}\sum_ {a'
=1}^{\mbox{d(J)}}\biggr| \biggr(\gamma V^E_{2a} U^J_{2a'} +
\sqrt{2} \gamma i V^E_{3a} U^J_{1a'} - \frac{\gamma}{\sqrt{2}}
V^E_{2a} U^J_{3a'} + \bar\gamma V^E_{1a} U^J_{2a'} +
\frac{\bar\gamma}{\sqrt{2}}V^E_{1a}U^J_{3a'}\nonumber \\&& +
   i \kappa V^E_{4a} U^J_{5a'} - \frac{i \kappa}{\sqrt{2}}V^E_{6a} U^J_{3a'}\biggr) V^H_{11}
    +\biggr( \frac{2 \eta}{\sqrt{3}} V^E_{2a} U^J_{2a'}+
   2 \sqrt{\frac{2}{3}}i\eta V^E_{3a} U^J_{1a'} +
   \sqrt{\frac{2}{3}}\eta V^E_{2a} U^J_{3a'}\nonumber \\ && + \frac{4i \eta }{\sqrt{3}} V^E_{4a} U^J_{1a'}
   -\frac{i\bar\zeta}{\sqrt{6}} V^E_{6a} U^J_{3a'} + \frac {2 i\bar\zeta }{\sqrt{3}} V^E_{6a} U^J_{2a'} -\frac {i\bar\zeta }{\sqrt{3}}V^E_{4a} U^J_{5a'}\biggr) V^H_{21} +\biggr( \frac {2i \zeta}{\sqrt{3} } V^E_{6a} U^J_{2a'}\nonumber \\ && +\sqrt{\frac{3 }{2}}i \zeta V^E_{6a} U^J_{3a'} -
   \frac {2\sqrt{2}i \zeta}{\sqrt{3} } V^E_{3a} U^J_{5a'} + \frac {i \zeta}{\sqrt{3}} V^E_{4a} U^J_{5a'}
   + \sqrt{6}\eta V^E_{1a} U^J_{3a'} + \frac {2\eta}{\sqrt{3}} V^E_{1a} U^J_{2a'}\biggr) V^H_{31}
    \nonumber \\ && + \sqrt{2}i \lambda \biggr(2 V^E_{3a} U^J_{2a'} -
   V^E_{4a} U^J_{3a'} - \sqrt{2} V^E_{3a} U^J_{3a'}\biggr) V^H_{41}
   +\biggr(\frac{i \rho }{3\sqrt{2} }V^E_{6a} U^J_{3a'} +
   \frac {i \rho}{3} V^E_{4a} U^J_{5a'}\nonumber \\ && - \frac{\zeta }{\sqrt{2}}V^E_{2a} U^J_{3a'}
   -\sqrt{2} i\zeta V^E_{3a} U^J_{1a'} - \frac {\bar\zeta}{\sqrt{2} } V^E_{1a} U^J_{3a'}\biggr) V^H_{51} +\biggr( \frac{\sqrt{2} \rho}{3 \sqrt{3}} V^E_{6a} U^J_{3a'} -\frac{\sqrt{2}\rho}{3\sqrt{3}} V^E_{3a} U^J_{5a'}\nonumber \\ && - \frac{\rho }{3\sqrt{3}}V^E_{4a} U^J_{5a'} + \frac{\rho }{3 \sqrt{3}}V^E_{6a} U^J_{2a'} + \frac {2 \zeta}{\sqrt{3} } V^E_{4a} U^J_{1a'} -
   \frac {2i \zeta}{\sqrt{3} } V^E_{2a} U^J_{2a'}+ \frac {\sqrt{2} \zeta}{\sqrt{3} } V^E_{3a} U^J_{1a'}\nonumber \\ && +
    \frac {i\zeta }{\sqrt{6} }V^E_{2a} U^J_{3a'} -
   \sqrt{\frac{3}{2}}i\bar\zeta V^E_{1a} U^J_{3a'} - \frac{2i\bar\zeta}{\sqrt{3}} V^E_{1a} U^J_{2a'}\biggr) V^H_{61} \biggr| ^2 F_{12} (m^E_a, m^J_{a'},
   Q)\nonumber \\ &&-2{g_{10}^2} \biggr|\frac{2i}{\sqrt{3}}U^{E*}_{2a}V^H_{21}+\frac{2i}{\sqrt{3}}
   U^{E*}_{1a}V^H_{31}-\sqrt{2}U^{E*}_{3a}V^H_{41}+\frac{2i}{\sqrt{3}}U^{E*}_{6a}V^H_{61}\biggr|^2
   F_{12}(m^E _a, m_{\lambda_J},Q)\nonumber \\ &&
   -2{g_{10}^2} \biggr|\frac{2}{\sqrt{3}}V^{J*}_{1a'}V^H_{21}-iV^{J*}_{2a'}V^H_{41}
   +\frac{i}{\sqrt{2}}V^{J*}_{3a'}V^H_{41}+iV^{J*}_{5a'}V^H_{51}
   -\frac{1}{\sqrt{3}}V^{J*}_{5a'}V^H_{61}\biggr|^2\nonumber \\ && F_{12} ( m_{\lambda_E},m^J_{a'}, Q)\eea

\bea K_{X\bar P} &= &\sum_ {a = 1}^{\mbox{d(X)}} \sum_ {a' =
  1}^{\mbox{d(P)}} \biggr|\biggr(\bar\gamma V^X_{1a} U^P_{1a'} - \frac{\kappa}{\sqrt{2}} V^X_{2a} U^P_{2a'}\biggr) V^H_{11} -
  \biggr(\frac{2 \bar\zeta}{\sqrt{3}} V^X_{1a} U^P_{2a'} + \frac{\bar\zeta}{\sqrt{6}} V^X_{2a} U^P_{2a'}\biggr) V^H_{21}\nonumber\\&&
  + \biggr(\frac{\zeta}{\sqrt{6}} V^X_{2a} U^P_{2a'} +
\frac{2 \eta}{\sqrt{3}} V^X_{1a} U^P_{1a'} - \frac{2 \sqrt{2}
\eta}{\sqrt{3}} V^X_{2a} U^P_{1a'}\biggr) V^H_{31}+
\biggr(\frac{\rho}{3\sqrt{2}} V^X_{2a} U^P_{2a'} + \bar\zeta
V^X_{1a} U^P_{1a'}\biggr) V^H_{51} \nonumber\\&&+
  \frac{i}{\sqrt{3}} \biggr(\sqrt{2}\bar\zeta V^X_{2a} U^P_{1a'} - \bar\zeta V^X_{1a}
  U^P_{1a'}
  + \frac{\rho}{3} V^X_{1a} U^P_{2a'}- \frac{\rho}{3 \sqrt{2}} V^X_{2a} U^P_{2a'}\biggr) V^H_{61} \biggr| ^2 F_{12} (m^X_a, m^P_{a'},
  Q)\nonumber\\&&
  -2{g_{10}^2} \biggr|i\sqrt{\frac{2}{3}}V^{P*}_{1a'}V^H_{31}-\frac{V^{P*}_{2a'}}{\sqrt{2}}V^H_{51}
  +\frac{i}{\sqrt{6}}V^{P*}_{2a'}V^H_{61}\biggr|^2 F_{12}(m_{\lambda_X}, m^P_{a'}, Q)
\eea

\bea K_{X\bar T} &= & \sum_{ a = 1}^{ \mbox{d(X)}} \sum_{ a' =
1}^{
 \mbox{d(T)}} \biggr|\biggr( \kappa V^X_{1a} U^T_{6a'}-\gamma V^X_{2a} U^T_{3a'} -
  i\gamma V^X_{1a} U^T_{4a'} - \bar\gamma V^X_{2a} U^T_{2a'}  -
  \frac{i\kappa}{ \sqrt{2}} V^X_{2a} U^T_{7a'}\biggr) V^H_{11}\nonumber\\&&+ \biggr( \frac {\sqrt{2} i \bar\gamma}{\sqrt{3}} V^X_{1a} U^T_{1a'} -\frac {\bar\gamma}{\sqrt{3}}
  V^X_{2a} U^T_{1a'}- 2\sqrt{\frac{2}{3}}\eta V^X_{1a} U^T_{3a'}- \frac{2i\eta}{\sqrt{3}} V^X_{1a} U^T_{4a'}-
  2\sqrt{\frac{2}{3}}i\eta V^X_{2a} U^T_{4a'}\nonumber\\&& -\frac {\bar\zeta}{\sqrt{3}}V^X_{1a} U^T_{6a'}  - \frac {\sqrt{2}\bar\zeta}{\sqrt{3}} V^X_{2a} U^T_{6a'} +
 \frac{ i\bar\zeta}{\sqrt{6} } V^X_{2a} U^T_{7a'}\biggr) V^H_{21} + \biggr(2 \sqrt{\frac{2}{3}}\eta V^X_{1a} U^T_{2a'} -\frac {\zeta}{\sqrt{3}}V^X_{1a}
 U^T_{6a'} \nonumber\\&&+ \frac {\sqrt{2}\zeta}{\sqrt{3}} V^X_{2a} U^T_{6a'} +\frac {2i\zeta}{\sqrt{3}} V^X_{1a} U^T_{7a'}
- \frac{i\zeta}{\sqrt{6}} V^X_{2a} U^T_{7a'} + \frac
{\sqrt{2}i\gamma }{\sqrt{3}} V^X_{1a} U^T_{1a'} +
  \frac
{\gamma}{\sqrt{3}} V^X_{2a} U^T_{1a'}\biggr) V^H_{31}
\nonumber\\&&-2 i \lambda \biggr(V^X_{2a} U^T_{5a'} + \sqrt{2}
V^X_{1a} U^T_{5a'}\biggr) V^H_{41}  + \biggr( i \zeta V^X_{1a}
U^T_{4a'}-\kappa V^X_{2a} U^T_{1a'} - \frac {i \rho}{3\sqrt{2}}
V^X_{2a} U^T_{7a'} \biggr) V^H_{51}\nonumber\\&& + \biggr( \frac {
\sqrt{2} i \zeta}{\sqrt{3}} V^X_{1a} U^T_{3a'} - \frac
{\sqrt{2}\zeta}{\sqrt{3}} V^X_{2a} U^T_{4a'} + \frac
{i\zeta}{\sqrt{3}} V^X_{2a} U^T_{3a'} -
  \frac
{\zeta}{\sqrt{3}} V^X_{1a} U^T_{4a'}
 -\frac{\sqrt{2} i}{ \sqrt{3}} \bar\zeta V^X_{1a}U^T_{2a'}\nonumber\\&& +\frac{ i \bar\zeta}{ \sqrt{3}}
  V^X_{2a}U^T_{2a'} + \frac{i \rho}{3\sqrt{3}} V^X_{1a} U^T_{6a'} +\frac{ \rho}{3\sqrt{3}} V^X_{1a} U^T_{7a'} +
  \frac { \rho }{3\sqrt{6}}V^X_{2a} U^T_{7a'}-
\sqrt{\frac{2}{3}}i\kappa V^X_{1a} U^T_{1a'} \biggr)\nonumber\\&&
V^H_{61} \biggr| ^2 F_{12} (m^X_{a}, m^T_{a'}, Q)
  -2{g_{10}^2} \biggr|-V^{T*}_{1a'}V^H_{11}-\biggr(\frac{i}{\sqrt{3}}V^{T*}_{2a'}+\sqrt{\frac{2}{3}}V^{T*}_{4a'}\biggr)V^H_{21}
  -\frac{i}{\sqrt{3}}V^{T*}_{3a'}V^H_{31}\nonumber\\&&+i V^{T*}_{5a'}V^H_{41}-\frac{i}{\sqrt{2}}V^{T*}_{7a'}V^H_{51}
  +\biggr(i\sqrt{\frac{2}{3}}V^{T*}_{6a'}+\frac{V^{T*}_{7a'}}{\sqrt{6}} \biggr)V^H_{61}\biggr|^2F_{12}(m_{\lambda_X}, m^T_{a'}, Q) \eea

\bea K_{P \bar E} &=& \sum_ {a = 1}^{\mbox{d(P)}}\sum_ {a' =
  1}^{\mbox{d(E)}} \biggr| \biggr(\gamma V^P_{1a} U^E_{3a'} - \frac {\kappa}{\sqrt{2}} V^P_{2a} U^E_{4a'}\biggr) V^H_{11} +
  \biggr( \frac{2\eta}{\sqrt{3}}V^P_{1a}( U^E_{3a'} -
\sqrt{2} U^E_{4a'})
  + \frac {\bar\zeta}{\sqrt{6}} V^P_{2a} U^E_{4a'}\biggr) V^H_{21}\nonumber\\&& - \frac{\zeta}{\sqrt{3}}\biggr(2
  V^P_{2a}U^E_{3a'}
  + \frac{V^P_{2a} }{\sqrt{2}} U^E_{4a'} \biggr) V^H_{31} + \biggr(2 \sqrt{2}\eta i V^P_{1a} U^E_{2a'} -
  \frac{\rho V^P_{2a}}{3\sqrt{2}} U^E_{6a'}- \sqrt{2} \zeta V^P_{1a} U^E_{6a'} \nonumber\\&& + \sqrt{2} i \zeta V^P_{2a} U^E_{1a'}\biggr) V^H_{41}
 + \biggr(\frac{\rho}{3\sqrt{2}} V^P_{2a} U^E_{4a'} +
\zeta V^P_{1a} U^E_{3a'}\biggr) V^H_{51} +
 \biggr(\frac {i\zeta }{\sqrt{3}}
  V^P_{1a} U^E_{3a'}- \frac {\sqrt{2}i\zeta }{\sqrt{3}} V^P_{1a} U^E_{4a'}
 \nonumber\\&& - \frac {i\rho}{3\sqrt{3}}V^P_{2a} U^E_{3a'} +
  \frac {i\rho}{3\sqrt{6}}V^P_{2a}U^E_{4a'}\biggr) V^H_{61}\biggr| ^2 F_{12} (m^P_a, m^E_{a'}, Q)
  -2{g_{10}^2} \biggr|-i \sqrt{\frac{2}{3}}U^{P*}_{1a} V^H_{21}\nonumber\\&&-\frac{U^{P*}_{2a}}{\sqrt{2}}V^H_{51}
  -\frac{iU^{P*}_{2a}}{\sqrt{6}}V^H_{61}\biggr|^2 F_{12}(m^P_a, m_{\lambda_{E}}, Q) \quad \quad\eea

 \bea K_{T\bar E} & = &\sum_ {a =1}^{\mbox{d(T)}}\sum_ {a' =1}^{\mbox{d(E)}} \biggr| \biggr(\gamma
   V^T_{5a} U^E_{1a'} -\gamma V^T_{3a} U^E_{4a'} -\bar\gamma V^T_{2a} U^E_{4a'} -\bar\gamma
   V^T_{5a} U^E_{2a'} -i\bar\gamma V^T_{4a} U^E_{3a'}
    +\kappa V^T_{6a} U^E_{3a'} \nonumber \\ &&+ i\kappa V^T_{5a} U^E_{6a'}-
   \frac{i\kappa }{\sqrt{2}} V^T_{7a} U^E_{4a'}\biggr) V^H_{11} +\biggr( 2\sqrt{\frac{2}{3}}\eta V^T_{3a} U^E_{3a'} +
   2 \sqrt{3}\eta V^T_{5a} U^E_{1a'}+
   i\bar\gamma \sqrt{\frac{2}{3}} V^T_{1a} U^E_{3a'}\nonumber \\ &&+
   \frac {\bar\gamma}{\sqrt{3 } } V^T_{1a} U^E_{4a'} - \frac {\bar\zeta}{\sqrt{3 } }V^T_{6a} U^E_{3a'}  -i \sqrt{3}\bar\zeta V^T_{5a} U^E_{6a'} +
   \frac{2 i\bar\zeta }{\sqrt{3}} V^T_{7a} U^E_{3a'}-
   \frac{i\bar\zeta }{\sqrt{6}} V^T_{7a} U^E_{4a'} \nonumber \\ && +
   \sqrt{\frac{2}{3}}\bar\zeta V^T_{6a} U^E_{4a'}\biggr) V^H_{21} +\biggr( \frac{i\zeta}{\sqrt{6}} V^T_{7a} U^E_{4a'}-\frac {\zeta}{\sqrt{3 } } V^T_{6a}U^E_{3a'} -
   \sqrt{\frac{2}{3}}\zeta V^T_{6a} U^E_{4a'} +
   \frac{i}{\sqrt{3}} \zeta V^T_{5a} U^E_{6a'}\nonumber \\ &&
    + \sqrt{\frac{2}{3}}i\gamma V^T_{1a} U^E_{3a'}-
   \frac{\gamma}{\sqrt{3}} V^T_{1a} U^E_{4a'} -2 \sqrt{\frac{2}{3}}\eta V^T_{2a} U^E_{3a'} -
   \frac{2 i\eta}{\sqrt{3}} V^T_{4a} U^E_{3a'}  +
   \frac{2\eta}{\sqrt{3}} V^T_{5a} U^E_{2a'} \nonumber \\ &&-
   2  \sqrt{\frac{2}{3}}i\eta V^T_{4a} U^E_{4a'}\biggr) V^H_{31} +\biggr( 2 \eta i V^T_{3a} U^E_{2a'}- 2\eta i V^T_{2a} U^E_{1a'}
    -
   2 \sqrt{2}\eta V^T_{4a} U^E_{1a'} - \gamma V^T_{1a} U^E_{1a'} \nonumber \\ &&- \bar\gamma V^T_{1a} U^E_{2a'} -
    \kappa V^T_{1a} U^E_{6a'}-
    \frac {\sqrt{2}\rho}{ 3} V^T_{6a} U^E_{6a'} -\frac{i\rho }{3\sqrt{2}} V^T_{7a}
 U^E_{6a'} + \sqrt{2}\zeta i V^T_{6a} U^E_{1a'} -\zeta
   V^T_{3a} U^E_{6a'} \nonumber \\ &&+
   \sqrt{2}i\bar\zeta V^T_{6a} U^E_{2a'} +\bar\zeta
   V^T_{2a} U^E_{6a'} -
   \sqrt{2} i\bar\zeta V^T_{4a} U^E_{6a'}
   - \sqrt{2}\bar\zeta V^T_{7a} U^E_{2a'}\biggr) V^H_{41} +\biggr( \zeta V^T_{5a} U^E_{1a'}\nonumber \\ && -\kappa V^T_{1a} U^E_{4a'} - \frac{i\rho}{3 }V^T_{5a} U^E_{6a'}
   -\frac {i \rho }{3 \sqrt{2}}V^T_{7a} U^E_{4a'} + i \bar\zeta V^T_{4a} U^E_{3a'} + \bar\zeta V^T_{5a} U^E_{2a'}\biggr) V^H_{51} \nonumber \\ && +\biggr(\sqrt{\frac{2}{3}} i \kappa V^T_{1a} U^E_{3a'} +  \sqrt{\frac{2}{3}}i\zeta V^T_{3a} U^E_{3a'} +
   \sqrt{3}i\zeta V^T_{5a} U^E_{1a'} -
    \frac {i \zeta}{\sqrt{3}}V^T_{3a} U^E_{4a'}
-\sqrt{\frac{2}{3}}i \bar\zeta V^T_{2a} U^E_{3a'}\nonumber \\ &&-
   \frac{i \bar\zeta }{\sqrt{3}}V^T_{5a} U^E_{2a'} +
   \sqrt{\frac{2}{3}} \bar\zeta V^T_{4a} U^E_{4a'} -
    \frac{ i \bar\zeta}{\sqrt{3}} V^T_{2a} U^E_{4a'}
   +\frac{ \bar\zeta }{ \sqrt{3}}V^T_{4a} U^E_{3a'} -\frac {i\rho}{3\sqrt{3}} V^T_{6a} U^E_{3a'}
   \nonumber \\ &&- \frac {\rho }{3\sqrt{3}} V^T_{7a} U^E_{3a'} - \frac {\rho }{3\sqrt{6}}V^T_{7a} U^E_{4a'}
   + \frac {2\rho }{3\sqrt{3}} V^T_{5a} U^E_{6a'}\biggr) V^H_{61} \biggr| ^2 F_{12} (m^T_a,
   m^E_{a'}, Q) \nonumber \\ && -2{g_{10}^2} \biggr|-U^{T*}_{1a}V^H_{11}+\frac{i}{\sqrt{3}}U^{T*}_{2a}V^H_{21}
  +\biggr(\frac{i}{\sqrt{3}}U^{T*}_{3a}+\sqrt{\frac{2}{3}}U^{T*}_{4a}\biggr)V^H_{31}-\frac{i}{\sqrt{2}}U^{T*}_{7a}V^H_{51}\nonumber \\ &&-\biggr(\frac{U^{T*}_{7a}}{\sqrt{6}}+i \sqrt{\frac{2}{3}}U^{T*}_{6a}\biggr) V^H_{61}\biggr|^2 F_{12}(m^T_a,
   m_{\lambda_{E}}, Q)\eea

\bea
  K_{Y\bar L} & = &\sum_{a=1}^{\mbox{d(L)}}
  \biggr|\biggr( k U^L_{2a}-i\gamma U^L_{1a} \biggr) V^H_{11} +\biggr(\frac {2 i\eta} {\sqrt{3}}  \ U^L_{1a} + \frac {\bar\zeta}
{\sqrt{3}} U^L_{2a}\biggr)
 V^H_{21} +\frac {\zeta }{\sqrt{3}}U^L_{2a}
  V^H_{31} +i \zeta U^L_{1a}
 V^H_{51} \nonumber \\ && +\biggr(\frac {\zeta }{\sqrt{3}}U^L_{1a} -
   \frac {i \rho}{3\sqrt{3}} U^L_{2a}\biggr) V^H_{61}\biggr|^2 F_{12}(m^Y, m^L _a,
   Q)
  \eea

\bea K_{VF} &= &\sum_{a=1}^{\mbox{d(F)}}
 \biggr|\biggr( \kappa V^F_{4a}-i \gamma V^F_{1a} \biggr)
  V^H_{11}- \biggr(2 \sqrt{3}i \eta V^F_{1a}
  +\sqrt{3}\bar\zeta V^F_{4a}\biggr)
  V^H_{21} -\sqrt{3} \zeta V^F_{4a}
  V^H_{31} \nonumber\\&&- 2\sqrt{3} \lambda V^F_{2a}
  V^H_{41} + i \zeta V^F_{1a}
  V^H_{51} -\biggr(\sqrt{3} \zeta V^F_{1a} -
  \frac{i\rho}{\sqrt{3} } V^F_{4a}\biggr) V^H_{61}
\biggr|^2 F_{12}(m^V, m^F_a, Q)\nonumber\\&&-2{g_{10}^2}
\biggr|iV^H_{41} \biggr|^2F_{12}(m^V, m_{\lambda_F}, Q) \quad
\quad \eea

\bea K_{C\bar Z} &=& \sum_ {a =
  1}^{\mbox{d(C)}} \biggr| \biggr( \bar\gamma V^C_{2a}-\gamma V^C_{1a} -
  i\kappa V^C_{3a}\biggr) V^H_{11} + \biggr(\frac {2\eta}{ \sqrt{3}} V^C_{1a} - \frac
{i\bar\zeta }{ \sqrt{3}} V^C_{3a}\biggr) V^H_{21} - \biggr(\frac
{i\zeta }{ \sqrt{3}}  V^C_{3a} + \frac {2 \eta}{ \sqrt{3}}
V^C_{2a}\biggr) V^H_{3 1}\nonumber\\&& + \biggr(\frac {i\rho }{ 3}
V^C_{3a} - \zeta V^C_{1a} - \bar\zeta V^C_{2a}\biggr) V^H_{51} +
\biggr( \frac{i\zeta }{\sqrt{3}} V^C_{1a} +
  \frac{i\bar\zeta }{\sqrt{3}} V^C_{2a}\biggr) V^H_{61} \biggr| ^2 F_{12}(m^Z, m^C_a, Q) \eea

\bea K_{D\bar I} &=& \sum_ {a =
  1}^{\mbox{d(D)}} \biggr| \biggr(\gamma V^D_{2a}- \bar\gamma V^D_{1a} +
  i\kappa V^D_{3a}\biggr) V^H_{11} + \biggr( \frac{i \bar\zeta }{\sqrt{3}} V^D_{3a}-\frac{2 \eta}{\sqrt{3}} V^D_{2a} \biggr)
V^H_{21} + \biggr(-i\zeta \sqrt{3} V^D_{3a} - 2\sqrt{3}\eta
V^D_{1a}\biggr) V^H_{31}\nonumber\\&& + \biggr( \zeta
V^D_{2a}-\frac{i\rho }{3} V^D_{3a} + \bar\zeta V^D_{1a}\biggr)
V^H_{51} -\frac{1}{\sqrt{3}} \biggr(i\zeta
 V^D_{2a} -3 i\bar\zeta V^D_{1a}
+\frac {2 \rho }{3} V^D_{3a}\biggr) V^H_{61} \biggr| ^
 2 F_{12}(m^I, m^D_a, Q)
\quad \quad \eea

\bea K_{Q\bar C} &=& \sum_ {a =
  1}^{\mbox{d(C)}} \biggr| \biggr( \frac{i\bar\gamma }{\sqrt{2}} U^C_{1a}-\frac{i\gamma }{\sqrt{2}} U^C_{2a} -\frac{\kappa}{\sqrt{2}} U^C_{3a}\biggr) V^H_{11} + \biggr(\sqrt{\frac{2}{3}} i \eta U^C_{2a} -
\frac{\bar\zeta}{\sqrt{6}} U^C_{3a}\biggr) V^H_{21} -
\biggr(\frac{\zeta}{\sqrt{6}} U^C_{3a}+ \sqrt{\frac{2}{3}} i \eta
U^C_{1a}\biggr) V^H_{31} \nonumber\\&&+ \biggr( \frac {i\zeta
}{\sqrt{2}} U^C_{2a}-\frac {\rho}{3 \sqrt{2}} U^C_{3a} + \frac
{i\bar\zeta }{\sqrt{2}} U^C_{1a}\biggr) V^H_{51} + \biggr(\frac
{\zeta }{\sqrt{6}} U^C_{2a} + \frac {\bar\zeta }{\sqrt{6}}
U^C_{1a}\biggr) V^H_{61} \biggr| ^2 F_{12}(m^Q, m^C_a, Q) \eea

\bea
 K_{E\bar U} &= &\sum_ {a =
  1}^{\mbox{d(E)}} \biggr| \biggr(\frac{ i\gamma}{\sqrt{2}} V^E_{2a} - \frac{i\bar\gamma }{\sqrt{2}} V^E_{1a} +
  \frac{\kappa}{\sqrt{2}} V^E_{6a}\biggr) V^H_{11} + \biggr(\sqrt{6} i\eta V^E_{2a} -
  \sqrt{\frac{3}{2}} \bar\zeta V^E_{6a}\biggr) V^H_{21}+ \biggr(
  \sqrt{\frac{2}{3}} i \eta V^E_{1a}\nonumber\\&& +\frac{\zeta}{\sqrt{6}} V^E_{6a} \biggr) V^H_{31} + \biggr( \sqrt{2} \lambda V^E_{4a} -2 \lambda  V^E_{3a}\biggr) V^H_{41}  +
  \biggr(\frac {\rho}{3 \sqrt{2}} V^E_{6a} - \frac
{i\zeta }{\sqrt{2}} V^E_{2a}- \frac{i\bar\zeta }{\sqrt{2}}
V^E_{1a}\biggr) V^H_{51} + \biggr( \sqrt{\frac{3}{2}}\zeta
V^E_{2a}\nonumber\\&& - \frac{\bar\zeta}{\sqrt{6}} V^E_{1a}+
   \frac{\sqrt{2}i \rho}{3\sqrt{3}} V^E_{6a}\biggr) V^H_{61}\biggr | ^2 F_{12} (m^U, m^E_a, Q)
  -2{g_{10}^2} \biggr|\frac{-V^H_{41}}{\sqrt{2}}\biggr|^2 F_{12}(m^U, m_{\lambda_{E}}, Q) \eea

   \bea
  K_{U \bar D} &=& \sum_ {a =
  1}^{\mbox{d(D)}} \biggr| \biggr(\frac{i\gamma }{\sqrt{2}} U^D_{1a} - \frac{i \bar\gamma }{\sqrt{2}} U^D_{2a} +
  \frac{\kappa}{\sqrt{2}} U^D_{3a}\biggr) V^H_{11} + \biggr( \frac{\bar\zeta}{\sqrt{6}} U^D
_{3a}- \sqrt{\frac{2}{3}}i\eta U^D_{1a} \biggr) V^H_{21}
\nonumber\\&&+ \biggr(-\sqrt{\frac{3}{2}} \zeta U^D_{3a} -i
\sqrt{6} \eta U^D_{2a}\biggr) V^H_{31} + \biggr(\frac{\rho}{ 3
\sqrt{2}} U^D _{3a} - \frac{i\zeta }{\sqrt{2}} U^D_{1a} -
\frac{i\bar\zeta }{\sqrt{2}} U^D_{2a}\biggr) V^H_{51}
\nonumber\\&&+ \biggr( \sqrt{\frac{3}{2}} \bar\zeta U^D_{2a}-\frac
{\zeta}{\sqrt{6}} U^D_{1a} -
   \frac{\sqrt{2} i\rho}{3 \sqrt{3}}U^D_{3a}\biggr) V^H_{61}\biggr|^2 F_{12} (m^U,
  m^D_a, Q)
\eea

\bea
 K_{L\bar B} &=& \sum_ {a =
  1}^{\mbox{d(L)}} \biggr| \biggr( \kappa V^L_{2a}-i \bar\gamma V^L_{1a} \biggr) V^H_{11} + \frac{\bar\zeta}{\sqrt{3} } V^L_{2a} V^H_{21}
  + \biggr(\frac{ \zeta}{\sqrt{3}} V^L_{2a} + \frac
{2 i\eta }{\sqrt{3}} V^L_{1a}\biggr) V^H_{31} + i\bar\zeta
V^L_{1a} V^H_{51} \nonumber\\&&+ \biggr(
  \frac{i \rho}{3\sqrt{3}} V^L_{2a}-\frac {\bar\zeta }{\sqrt{3}} V^L_{1a} \biggr) V^H_{61}\biggr|^2 F_{12}(m^B, m^L_a, Q)) \eea

\bea K_{K \bar X} &=& \sum_ {a = 1}^{\mbox{d(K)}} \sum_ {a' =
  1}^{\mbox{d(X)}}\biggr| \biggr(\sqrt{2} i\bar\gamma V^K_{1a} U^X_{1a'} +
  i\kappa V^K_{2a} U^X_{2a'}\biggr) V^H_{11} + \biggr(
  \frac{i\bar\zeta }{\sqrt{3}} V^K_{2a} U^X_{2a'}-2\sqrt{\frac {2}{3}} i\bar\zeta V^K_{2a} U^X_{1a'} \biggr) V^H_{21} \nonumber\\&&+ \biggr( 2 \sqrt{\frac{2}{3}} i\eta V^K_{1a} U^X_{1a'}- \frac{ i\zeta}{\sqrt{3} }
  V^K_{2a}
  U^X_{2a'}
  +
 \frac{ 4 i\eta}{\sqrt{3}} V^K_{1a} U^X_{2a'}\biggr) V^H_{31} + \biggr( \frac{i\rho }{3} V^K_{2a} U^X_{2a'}-
  \sqrt{2}i \bar\zeta V^K_{1a} U^X_{1a'}\biggr) V^H_{51} \nonumber\\&& + \biggr( \frac{\rho}{3} \sqrt{\frac{2}{3}}
  V^K_{2a}
  U^X_{1a'}
+ \frac{\rho}{3 \sqrt{3} } V^K_{2a} U^X_{2a'}-\frac{2
\bar\zeta}{\sqrt{3}} V^K_{1a} U^X_{2a'} -
  \sqrt{\frac{2}{3}} \bar\zeta V^K_{1a} U^X_{1a'} \biggr) V^H_{61} \biggr| ^2
F_{12} (m^K_a,
 m^X_{a'},
 Q)\nonumber\\&&-2{g_{10}^2} \biggr|-\frac{2 U^{K*}_{1a}}{\sqrt{3}}V^H_{31}+iU^{K*}_{2a}V^H_{51}
  +\frac{U^{K*}_{2a}}{\sqrt{3}}V^H_{61}\biggr|^2F_{12}(m^K_a,m_{\lambda_X},Q)
\eea

\bea K_{B \bar M} &=& \biggr|\sqrt{2} i \gamma V^H_{11} -
2\sqrt{\frac{2}{3}} i \eta V^H_{21} - \sqrt{2} i \zeta V^H_{51} -
\sqrt{\frac{2}{3}} \zeta V^H_{61}\biggr | ^2 F_{12}(m^B, m^M,Q)
\eea

\bea K_{W\bar B} &=& \biggr| \gamma V^H_{11} - \frac{2
\eta}{\sqrt{3}}
 V^H_{21} + \zeta V^H_{51} -
 \frac{i\zeta }{\sqrt{3}} V^H_{61} \biggr| ^2 F_{12}(m^W,m^B, Q)
\eea

\bea K_{Y\bar W} &=& \biggr| \bar\gamma V^H_{11} - \frac{2
\eta}{\sqrt{3}} V^H_{31} + \bar\zeta V^H_{51} +
 \frac{i \bar\zeta}{\sqrt{3}}  V^H_{61} \biggr| ^2 F_{12} (m^Y, m^W, Q) \eea

\bea K_{V\bar O} &=& \biggr| \bar\gamma V^H_{11} + 2 \sqrt{3} \eta
V^H_{31} + \bar\zeta V^H_{51} -
 \sqrt{3}i\bar\zeta V^H_{61} \biggr| ^2 F_{12}(m^V,
 m^O, Q)
\eea

\bea K_{N \bar Y} &=& \biggr| \sqrt{2} i \bar\gamma V^H_{11} -
  2 \sqrt{\frac{2}{3}}i \eta  V^H_{31} - \sqrt{2} i \bar\zeta V^H_{51} + \sqrt{\frac{2}{3}} \bar\zeta V^H_{61}\biggr | ^2 F_{12}(m^N,
 m^Y, Q)
\eea

\bea K_{\bar V \bar A} &=& \biggr|\sqrt{2} i \bar\gamma V^H_{11} +
  2 \sqrt{6} i \eta V^H_{31} - \sqrt{2} i \bar\zeta V^H_{51} - \sqrt{6} \bar\zeta V^H_{61} \biggr| ^2 F_{12}(m^V,
m^A, Q) \eea

\bea K_{HO} & = & \sum_{a = 2}^{\mbox{d(H)}} \biggr|
 \gamma V^H_{4a}  V^H_{11} + 2\sqrt{3}\eta V^H_{4a} V^H_{21} +\biggr(\gamma V^H_{1a} +
 2\sqrt{3}\eta  V^H_{2a} + \zeta  V^H_{5a}+ \sqrt{3} i \zeta  V^H_{6a}\biggr)  V^H_{41} \nonumber\\&&+ \zeta  V^H_{4a}  V^H_{51} + \sqrt{3} i \zeta  V^H_{4a}  V^H_{61} \biggr| ^ 2 F_{12} (m^O, m^H_a, Q)\nonumber\\&&
  +\biggr| 2\gamma  V^H_{41}  V^H_{11} +
  4\sqrt{3}\eta  V^H_{41}  V^H_{21} +2 \zeta  V^H_{51}  V^H_{41} +
  2  \sqrt{3} i \zeta  V^H_{61}  V^H_{41}\biggr| ^ 2 F_{11} (m^O, Q) \eea

\bea K_{S\bar H} & = &\sum_ {a =
  2}^{\mbox{d(H)}}\biggr | \biggr( \frac{i\bar\gamma }{\sqrt{2} } U^H_{2a} -\frac{i\gamma }{\sqrt{2} }  U^H_{3a} -
  \frac{\kappa }{\sqrt{2}}  U^H_{6a}\biggr)  V^H_{11} - \biggr(2  \sqrt{\frac{2}{3}}i  \eta U^H_{3a} -
  \sqrt{\frac{2}{3}} \bar\zeta U^H_{6a} + \frac{i\bar\zeta }{\sqrt{2}} U^H_{5a} + \frac{ i\bar\gamma}{\sqrt{2} }  U^H_{1a}\biggr)
   V^H_{21} \nonumber\\&&+ \biggr(\sqrt{\frac{2}{3}} \zeta U^H_{6a} -  \frac {i\zeta  }{\sqrt{2} } U^H_{5a} + \frac{i\gamma }{\sqrt{2} } U^H_{1a} +
  2  \sqrt{\frac{2}{3}} i \eta U^H_{2a}\biggr)  V^H_{31} - \sqrt{6 } i \lambda  U^H_{4a}  V^H_{41} -
  \biggr(\frac{\rho }{3 \sqrt{2}}  U^H_{6a} -
\frac{i\bar\zeta }{\sqrt{2} } U^H_{2a} \nonumber\\&&- \frac{i
\zeta }{\sqrt{2} } U^H_{3a}\biggr)  V^H_{51} +
\biggr(\frac{\kappa}{\sqrt{2}} U^H_{1a} - \sqrt{\frac{2}{3}}
\bar\zeta U^H_{2a} - \sqrt{\frac{2}{3}} \zeta U^H_{3a} +
 \frac{\rho}{3\sqrt{2 }} U^H_{5a}\biggr)  V^H_{61} \biggr| ^2 F_{12} (m^S, m^H_a, Q)\nonumber\\&& +\biggr| \biggr(\frac{i \bar\gamma }{\sqrt{2}
}  U^H_{21} - \frac{i\gamma }{\sqrt{2} }  U^H_{31} -
\frac{\kappa}{\sqrt{2}} U^H_{61}\biggr)  V^H_{11} - \biggr(2
\sqrt{\frac{2}{3}} i \eta U^H_{31}- \sqrt{\frac{2}{3}} \bar\zeta
U^H_{61} + \frac{i\bar\zeta }{\sqrt{2} } U^H_{51} + \frac{i
\bar\gamma }{\sqrt{2} } U^H_{11}\biggr) V^H_{21}\nonumber\\&& +
\biggr(\sqrt{\frac{2}{3}} \zeta U^H_{61} - \frac{i \zeta
}{\sqrt{2} } U^H_{51} + \frac{i\gamma }{\sqrt{2} } U^H_{11} +
  2  \sqrt{\frac{2}{3}} i \eta U^H_{21}\biggr)  V^H_{31} - \sqrt{6}i  \lambda  U^H_{41} V^H_{41} -
  \biggr(\frac{\rho}{3 \sqrt{2}} U^H_{61} - \frac{i\bar\zeta }
{ \sqrt{2}}  U^H_{21}\nonumber\\&&- \frac{i\zeta }{\sqrt{2}}
U^H_{31}\biggr)  V^H_{51} + \biggr( \frac{\kappa}{\sqrt{2} }
U^H_{11} - \sqrt{\frac{2}{3}} \bar\zeta U^H_{21} -
\sqrt{\frac{2}{3}} \zeta U^H_{31} +
  \frac{\rho}{3\sqrt{2 }} U^H_{51}\biggr)  V^H_{61} \biggr| ^2 F_{11} (m^S, Q) \eea

\bea K_{H\bar F} &= &\sum_ {a = 2}^{\mbox{d(H)}}\sum_ {a' =
  1}^{\mbox{d(F)}} \biggr| \biggr( \bar\gamma U^F_{2a'}  V^H_{2a} -
  i \bar\gamma U^F_{1a'} V^H_{4a} -\gamma U^F_{2a'}  V^H_{3a} + \kappa U^F_{4a'} V^H_{4a} -
  i \kappa U^F_{2a'}  V^H_{6a}\biggr)  V^H_{11}
 \nonumber\\&&+ \biggr( \bar\zeta U^F_{2a'}  V^H_{5a}- \sqrt{3} \bar\zeta U^F_{4a'}  V^H_{4a} +
  \frac{2i \bar\zeta }{ \sqrt{3} } U^F_{2a'}  V^H_{6a} -\frac{4 \eta }{ \sqrt{3} } U^F_{2a'}  V^H_{3a}-\bar\gamma U^F_{2a'}  V^H_{1a}\biggr)  V^H_{21}
 + \biggr(\frac{4 \eta }{ \sqrt{3} } U^F_{2a'}  V^H_{2a}\nonumber\\&&-\sqrt{3} \zeta U^F_{4a'}  V^H_{4a} +\frac{2i \zeta }{
\sqrt{3} }  U^F_{2a'}  V^H_{6a} + \zeta U^F_{2a'} V^H_{5a} -2
\sqrt{3} \eta i U^F_{1a'} V^H_{4a}+ \gamma U^F_{2a'}
V^H_{1a}\biggr) V^H_{31} \nonumber\\&&
 + \biggr(2  \sqrt{3} \eta i U^F_{1a'}  V^H_{3a} + \frac{i \rho}{\sqrt {3} } U^F_{4a'}  V^H_{6a} -
  i \bar\zeta U^F_{1a'}  V^H_{5a} - \sqrt{3} \bar\zeta U^F_{1a'}  V^H_{6a}+ i \bar\gamma U^F_{1a'}  V^H_{1a}+\sqrt{3} \zeta U^F_{4a'}  V^H_{3a} \nonumber\\&&-\kappa U^F_{4a'}  V^H_{1a}
 +
  \sqrt{3} \bar\zeta U^F_{4a'}  V^H_{2a} \biggr)  V^H_{41}+
\biggr( i \bar\zeta U^F_{1a'}  V^H_{4a}-\bar\zeta U^F_{2a'}
V^H_{2a} - \zeta U^F_{2a'} V^H_{3a}
 +
 \frac{i \rho}{3} U^F_{2a'}  V^H_{6a} \biggr)  V^H_{51}\nonumber\\&&+\biggr( i \kappa U^F_{2a'}  V^H_{1a}- \frac{2 i \bar\zeta }{ \sqrt{3} }  U^F_{2a'}  V^H_{2a}
- \frac{i \rho}{\sqrt {3} } U^F_{4a'}  V^H_{4a} -\frac{2 i \zeta
}{ \sqrt{3} } U^F_{2a'}  V^H_{3a} +\sqrt{3}\bar\zeta U^F_{1a'}
V^H_{4a}\nonumber\\&&-
  \frac{i \rho}{3} U^F_{2a'}  V^H_{5a}\biggr)  V^H_{61}
 | ^2 F_{12} (m^H _a,m^F_{a'}, Q)
\nonumber\\&& -\sum_ {a = 2}^{\mbox{d(H)}}2{g_{10}^2}
\biggr|i\biggr(U^{H*}_{1a}V^H_{11}+U^{H*}_{2a}V^H_{21}+U^{H*}_{3a}V^H_{31}
  +U^{H*}_{5a}V^H_{51}+U^{H*}_{6a}V^H_{61}\biggr)\biggr|^2F_{12} (m^H _a, m_{\lambda_F}, Q)\nonumber\\&&-
2{g_{10}^2}
\biggr|i\biggr(U^{H*}_{11}V^H_{11}+U^{H*}_{21}V^H_{21}+U^{H*}_{31}V^H_{31}
  +U^{H*}_{51}V^H_{51}+U^{H*}_{61}V^H_{61}\biggr)\biggr|^2F_{11} (m_{\lambda_F}, Q)
\eea

\bea K_{G\bar H} &=& \sum_ {a = 2}^{\mbox{d(H)}} \sum_ {a' =
  1}^{\mbox{d(G)}} \biggr|\biggr( \frac{\gamma } {\sqrt{2}} V^G_{3a'} U^H_{3a} -\gamma V^G_{2a'} U^H_{3a} - \sqrt{2} i\gamma
V^G_{4a'} U^H_{4a}  - \bar\gamma V^G_{2a'} U^H_{2a} -
  \frac{ \bar\gamma  } {\sqrt{2}} V^G_{3a'} U^H_{2a}\nonumber\\&&+  \frac{ i \kappa  } {\sqrt{2}} V^G_{3a'}U^H_{6a} +
  \kappa V^G_{1a'} U^H_{5a}\biggr) V^H_{11} + \biggr(
  2  \sqrt{\frac{2}{3}} \eta V^G_{3a'} U^H_{3a} - \frac{4 \eta  } {\sqrt{3}} V^G_{2a'} U^H_{3a} -
  2  \sqrt{6} i \eta V^G_{4a'} U^H_{4a} \nonumber\\&& - \sqrt{\frac{2}{3}} i\bar\zeta V^G_{3a'} U^H_{6a} +
  i \bar\zeta V^G_{1a'}  U^H_{6a} -
  \frac{\bar\zeta } {\sqrt{2}} V^G_{3a'} U^H_{5a}- \bar\gamma V^G_{2a'}  U^H_{1a} + \frac{ \bar\gamma  } {\sqrt{2}} V^G_{3a'}
U^H_{1a} \biggr)  V^H_{21} \nonumber\\&&- \biggr(
\sqrt{\frac{2}{3}} i\zeta V^G_{3a'} U^H_{6a} +
  i \zeta V^G_{1a'} U^H_{6a} +
 \frac{\zeta}{\sqrt{2} } V^G_{3a'}  U^H_{5a} +
  \gamma V^G_{2a'}  U^H_{1a} +
  \frac{\gamma}{\sqrt{2} } V^G_{3a'} U^H_{1a} +
  \frac{4 \eta}{\sqrt{3} } V^G_{2a'} U^H_{2a}\nonumber\\&& +
  2 \sqrt{\frac{2}{3}} \eta V^G_{3a'} U^H_{2a}\biggr)V^H_{31} + \biggr( \sqrt{6}  \lambda V^G_{3a'} U^H_{4a}-2 \sqrt{3} \lambda V^G_{2a'}
U^H_{4a} - \sqrt{2} i \bar\zeta V^G_{5a'} U^H_{5a} -\sqrt{6}
\bar\zeta V^G_{5a'} U^H_{6a} \nonumber\\&&+ \sqrt{2}i \bar\gamma
V^G_{5a'} U^H_{1a} +
  2  \sqrt{6}i  \eta  V^G_{5a'} U^H_{3a}\biggr)  V^H_{41} + \biggr(
  \kappa V^G_{1a'} U^H_{1a}-\frac{i\rho }{3 \sqrt{2}} V^G_{3a'} U^H_{6a} + \frac{\bar\zeta} {\sqrt{2} } V^G_{3a'}
U^H_{2a}
  \nonumber\\&&+ \frac{\zeta}{\sqrt{2} } V^G_{3a'} U^H_{3a} +
 \sqrt{2} i \zeta V^G_{4a'} U^H_{4a} \biggr)  V^H_{51}+ \biggr(\sqrt{\frac{2}{3}} \bar\zeta i V^G_{3a'} U^H_{2a} + i\bar\zeta  V^G_{1a'} U^H_{2a}-\frac{2  \rho}{3 \sqrt{3} } V^G_{2a'} U^H_{6a}
 \nonumber\\&&- \frac{ i\kappa}{\sqrt{2}}  V^G_{3a'} U^H_{1a}
  +
 \sqrt{\frac{2}{3}}i \zeta V^G_{3a'} U^H_{3a} -
  i \zeta V^G_{1a'} U^H_{3a} -
 \sqrt{6} \zeta V^G_{4a'} U^H_{4a} + \frac{i\rho }{3  \sqrt{2} }
  V^G_{3a'}  U^H_{5a}\biggr)  V^H_{61}\biggr| ^ 2 F_{12} (m^H_a, m^G_{a'}, Q) \nonumber\\&&
 -\sum_ {a = 2}^{\mbox{d(H)}} 2{g_{10}^2} \biggr|\frac{i}{\sqrt{5}}\biggr(V^{H*}_{1a}V^H_{11}+V^{H*}_{2a}V^H_{21}+V^{H*}_{3a}V^H_{31}-4 V^{H*}_{4a}V^H_{41}
  +V^{H*}_{5a}V^H_{51}\nonumber\\&&+V^{H*}_{6a}V^H_{61}\biggr)\biggr|^2F_{12} (m^H_a, m_{\lambda_G}, Q)\nonumber\\&&
+\sum_ {a' =
 1}^{\mbox{d(G)}} \biggr|\biggr( \frac{\gamma } {\sqrt{2}} V^G_{3a'} U^H_{31} -\gamma V^G_{2a'} U^H_{31} - \sqrt{2} i\gamma
V^G_{4a'} U^H_{41}  - \bar\gamma V^G_{2a'} U^H_{21} -
  \frac{ \bar\gamma  } {\sqrt{2}} V^G_{3a'} U^H_{21}+  \frac{ i \kappa  } {\sqrt{2}} V^G_{3a'}U^H_{61} \nonumber\\&&+
  \kappa V^G_{1a'} U^H_{51}\biggr) V^H_{11} + \biggr(
  2  \sqrt{\frac{2}{3}} \eta V^G_{3a'} U^H_{31} - \frac{4 \eta  } {\sqrt{3}} V^G_{2a'} U^H_{31} -
  2  \sqrt{6} i \eta V^G_{4a'} U^H_{41} - \sqrt{\frac{2}{3}} i\bar\zeta V^G_{3a'} U^H_{61} +
  i \bar\zeta V^G_{1a'}  U^H_{61} \nonumber\\&&-
  \frac{\bar\zeta } {\sqrt{2}} V^G_{3a'} U^H_{51}- \bar\gamma V^G_{2a'}  U^H_{11} + \frac{ \bar\gamma  } {\sqrt{2}} V^G_{3a'}
U^H_{11} \biggr)  V^H_{21} - \biggr( \sqrt{\frac{2}{3}} i\zeta
V^G_{3a'} U^H_{61} +
  i \zeta V^G_{1a'} U^H_{61} +
 \frac{\zeta}{\sqrt{2} } V^G_{3a'}  U^H_{51}\nonumber\\&& +
  \gamma V^G_{2a'}  U^H_{11} +
  \frac{\gamma}{\sqrt{2} } V^G_{3a'} U^H_{11} +
  \frac{4 \eta}{\sqrt{3} } V^G_{2a'} U^H_{21} +
  2 \sqrt{\frac{2}{3}} \eta V^G_{3a'} U^H_{21}\biggr)V^H_{31} + \biggr( \sqrt{6}  \lambda V^G_{3a'} U^H_{41}-2 \sqrt{3} \lambda V^G_{2a'}
U^H_{41} \nonumber\\&&- \sqrt{2} i \bar\zeta V^G_{5a'} U^H_{51}
-\sqrt{6} \bar\zeta V^G_{5a'} U^H_{61} + \sqrt{2}i \bar\gamma
V^G_{5a'} U^H_{1a} +
  2  \sqrt{6}i  \eta  V^G_{5a'} U^H_{31}\biggr)  V^H_{41} + \biggr(
  \kappa V^G_{1a'} U^H_{11}-\frac{i\rho }{3 \sqrt{2}} V^G_{3a'} U^H_{61} \nonumber\\&& + \frac{\bar\zeta} {\sqrt{2} } V^G_{3a'}
U^H_{21}
  + \frac{\zeta}{\sqrt{2} } V^G_{3a'} U^H_{31} +
 \sqrt{2} i \zeta V^G_{4a'} U^H_{41} \biggr)  V^H_{51}+ \biggr(\sqrt{\frac{2}{3}} \bar\zeta i V^G_{3a'} U^H_{21} + i\bar\zeta  V^G_{1a'} U^H_{21}-\frac{2  \rho}{3 \sqrt{3} } V^G_{2a'} U^H_{61}
 \nonumber\\&&- \frac{ i\kappa}{\sqrt{2}}  V^G_{3a'} U^H_{11}
  +
 \sqrt{\frac{2}{3}}i \zeta V^G_{3a'} U^H_{31} -
  i \zeta V^G_{1a'} U^H_{31} -
 \sqrt{6} \zeta V^G_{4a'} U^H_{41} + \frac{i\rho }{3  \sqrt{2} }
  V^G_{3a'}  U^H_{51}\biggr)  V^H_{61}\biggr| ^ 2 F_{11} (m^G_{a'}, Q) \nonumber\\&&
  -2{g_{10}^2} \biggr|\frac{i}{\sqrt{5}}\biggr(V^{H*}_{11}V^H_{11}+V^{H*}_{21}V^H_{21}+V^{H*}_{31}V^H_{31}-4 V^{H*}_{41}V^H_{41}
  +V^{H*}_{51}V^H_{51}\nonumber\\&&+V^{H*}_{61}V^H_{61}\biggr)\biggr|^2F_{11} ( m_{\lambda_G}, Q)
\eea

For the ${\overline H}[1,2,-1]$ line we have
 \bea (16 \pi^2){\cal
K}_{\overline H} &=&8K_{R C}+
 3K_{D\bar J}+3K_{J\bar E}
 + 9K_{P \bar X} +3K_{T\bar X}
  + 9K_{E\bar P}+
 3K_{E\bar T}
 + 6K_{L\bar Y}
 + K_{\bar V\bar F}
 + 8K_{Z\bar C}\nonumber\\&&
 + 3K_{I\bar D}+
 24K_{Q C}
 + 9K_{U\bar E}
 + 9K_{D\bar U}
 + 6K_{B\bar L}
 + 3K_{X \bar K}
 + 6K_{M\bar B}
 + 18K_{B\bar W}\nonumber\\&&
 + 18K_{W\bar Y}
 + 3K_{\bar V O}
 + 6K_{Y\bar N}
 + K_{ V A}
 + 3K_{\bar H\bar O}
 + 3K_{SH}
 + K_{F\bar H}+ K_{GH}\eea

\bea K_{RC} &=&\sum_ {a = 1}^{\mbox{d(R)}} \sum_ {a' =
  1}^{\mbox{d(C)}} \biggr|\biggr(\gamma V^R_ {1 a}V^C_ {1 a'} +
    \frac{ \gamma }{\sqrt{2}}V^R_ {2 a}V^C_ {1 a'} + \bar \gamma V^
   R_ {1 a}V^C_ {2 a'} -\frac{\bar \gamma }{\sqrt{2}}V^R_ {2 a} V^C_ {2 a'} + \frac{i \kappa}{\sqrt {2}}V^
   R_ {2 a} V^C_ {3 a'}\biggr) U^
   H_ {11} \nonumber \\ && + \biggr( \frac{i\bar\zeta}{\sqrt {6}} V^R_ {2 a} V^
   C_ {3 a'} -\frac{2\eta}{\sqrt {3}} V^R_ {1 a} V^
   C_ {1 a'} -\sqrt {\frac{2}{3}} \eta V^R_ {2 a} V^
   C_ {1 a'}\biggr) U^
   H_ {21} + \biggr(
   \sqrt {\frac{2}{3}}\eta V^R_ {2 a}V^C_ {2 a'} -\frac{2\eta}{\sqrt {3}} V^R_ {1 a}V^C_ {2 a'} \nonumber \\ && + \frac{i\zeta}{\sqrt {6}} V^
   R_ {2 a} V^C_ {3 a'}\biggr) U^
   H_ {31} + \biggr(  \frac{\zeta}{\sqrt{2}}V^R_ {2 a}V^C_ {1 a'} +
    \frac{ \bar\zeta}{\sqrt{2}}V^R_ {2 a}V^C_ {2 a'}-\frac{i \rho}{3\sqrt {2}} V^R_ {2 a} V^
   C_ {3 a'}\biggr) U^
   H_ {51} - \biggr(\frac{\rho}{3\sqrt {3}} V^R_ {1 a} V^
    C_ {3 a'} \nonumber \\ && + \frac{i\zeta}{\sqrt {6}} V^R_ {2 a}V^C_ {1 a'}+
    \frac{i\bar\zeta}{\sqrt {6}} V^R_ {2 a} V^C_ {2 a'}\biggr) U^
   H_ {61}\biggr|^2 F_{12} (m^R_a, m^C_{a'},Q)
\eea

\bea K_{D \bar J} &=&\sum_ {a = 1}^{\mbox{d(D)}} \sum_ {a' =
  1}^{\mbox{d(J)}} \biggr|\biggr( \frac{\bar \gamma}{\sqrt{2}} V^D_ {1 a}U^J_ {3 a'}- \bar \gamma V^D_ {1 a}U^
   J_ {2 a'} - \gamma V^D_ {2 a}U^J_ {2 a'} - \frac{\gamma }{\sqrt{2}}V^D_ {2 a}U^J_ {3a'} - \frac{i \kappa}{\sqrt{2}} V^
   D_ {3 a}U^J_ {3a'} \biggr) U^
   H_ {11} \nonumber \\ &&+ \biggr( \sqrt{\frac{2}{3}}\eta V^
   D_ {2 a}U^J_ {3 a'}- \frac{2\eta}{\sqrt{3}}V^
   D_ {2 a}U^J_ {2 a'} - \frac{2i\bar\zeta}{\sqrt {3}} V^D_ {3 a}U^
   J_ {2 a'} - \frac{i\bar\zeta}{\sqrt {6}}V^D_ {3 a}U^J_ {3 a'}\biggr) U^
   H_ {21} + \biggr(
   \sqrt {6}\eta V^ D_ {1 a} U^J_ {3 a'}\nonumber \\ &&-\frac{2\eta}{\sqrt {3}} V^ D_ {1 a}U^J_ {2 a'}- \frac{2i\zeta}{\sqrt {3}} V^ D_ {3 a} U^J_ {2 a'} + \sqrt {\frac{3}{2}} i\zeta V^ D_ {3 a}U^J_ {3 a'}\biggr) U^
   H_ {31} + \biggr(\frac{i \rho}{3} V^D_ {3 a}U^J_ {5 a'}-4\eta V^D_ {1 a} U^
   J_ {1 a'} -
   2\zeta V^D_ {2 a}U^J_ {5 a'}\nonumber \\ && + 2i \zeta V^
   D_ {3 a}U^J_ {1 a'}\biggr) U^
   H_ {41} + \biggr(\frac{i\rho}{3} V^
   D_ {3 a} - \zeta V^D_ {2 a} - \bar\zeta V^
   D_ {1 a}\biggr) \frac{U^J_ {3 a'}}{\sqrt {2}} U^
   H_ {51} + \biggr( \frac{2i\zeta}{\sqrt {3}} V^D_ {2 a} U^J_ {2 a'}+
    \frac {i\zeta }{\sqrt {6}}V^D_ {2 a}U^J_ {3 a'} \nonumber \\ && -\sqrt {\frac{3}{2}}i\bar\zeta V^ D_ {1 a}U^J_ {3 a'} +\frac{2i\bar\zeta }{\sqrt {3}}
   V^ D_ {1 a} U^J_ {2 a'} -\frac{\rho}{3\sqrt {3}}V^D_ {3 a}U^
   J_ {2 a'} + \frac{\sqrt {2}\rho}{3\sqrt {3}} V^D_ {3 a} U^J_ {3 a'}\biggr) U^
  H_{61}|^2 F_{12} (m^D_a, m^J_{a'},Q)\nonumber \\ &&
  -2{g_{10}^2} \biggr|\frac{2i}{\sqrt{3}}\biggr(U^{D*}_{2a}U^H_{21}+U^{D*}_{1a}U^H_{31}
  +U^{D*}_{3a'}U^H_{61}\biggr)\biggr|^2 F_{12}(m^D_a, m_{\lambda_{J}},Q)
\eea

\bea K_{J \bar E} &=&\sum_ {a = 1}^{\mbox{d(J)}} \sum_ {a' =
  1}^{\mbox{d(E)}} \biggr|\biggr( \frac{\bar \gamma}{\sqrt {2}} V^J_ {3 a}U^E_ {2 a'}- i\sqrt {2}\bar \gamma U^E_ {3 a'} V^J_ {1 a} -\bar \gamma U^E_ {2 a'}V^J_ {2
a} - \gamma V^J_ {2 a} U^E_ {1 a'} - \frac{\gamma}{\sqrt {2}} V^J_
{3 a}U^E_ {1 a'}
    -\frac{i \kappa }{\sqrt {2}}V^ J_ {3 a}U^E_ {6 a'}\nonumber \\ &&
-i \kappa V^J_ {5 a}U^E_ {4 a'}\biggr) U^H_ {11} + \biggr( \sqrt
{\frac{3}{2}}i\bar \zeta V^
 J_ {3 a} U^E_ {6 a'} + \frac{2i\bar \zeta}{\sqrt{3}} V^J_ {2 a} U^E_ {6 a'} +
 \frac{ 2\sqrt {2}i\bar \zeta}{\sqrt{3}}V^J_ {5 a}U^E_ {3 a'} - \frac{i\bar \zeta}{\sqrt{3}}
 V^J_ {5 a}U^E_ {4 a'}-\frac{2\eta}{\sqrt{3}} V^J_ {2 a} U^ E_ {1 a'} \nonumber \\ && - \sqrt {6}\eta
V^J_ {3 a} U^ E_ {1 a'}\biggr) U^H_ {21}+ \biggr( -
\frac{i\zeta}{\sqrt{6}} V^J_ {3 a} U^
 E_ {6 a'} +\frac{2i\zeta}{\sqrt{3}} V^J_ {2 a} U^E_ {6 a'} + \frac{i\zeta}{\sqrt {3}} V^J_ {5 a}U^E_ {4 a'}- \frac{2\eta}{\sqrt {3}} V^J_ {2 a} U^
E_ {2 a'} -\sqrt {\frac{2}{3}}\eta V^J_ {3 a} U^ E_ {2
a'}\nonumber \\ &&- \frac{4i \eta}{\sqrt{3}} V^J_ {1 a}U^E_ {4 a'}
 - \sqrt {\frac{8}{3}}i \eta V^J_ {1 a} U^E_ {3 a'}\biggr) U^
  H_ {31}+ \biggr(2\sqrt {2}i \lambda V^ J_ {2 a}U^E_ {3 a'} - \sqrt{2}i \lambda V^J_ {3 a}U^E_ {4
a'}- 2i \lambda V^J_ {3 a}U^E_ {3 a'}\biggr) U^H_ {41} \nonumber
\\ &&+ \biggr( \frac{i \rho}{3\sqrt {2}}V^J_ {3 a} U^ E_ {6 a'}
-\frac{i\rho}{3} V^J_ {5 a} U^E_ {4 a'} - \frac{\bar
\zeta}{\sqrt{2}} V^J_ {3 a}U^E_ {2 a'} + \sqrt{2}i\bar \zeta U^
 E_ {3 a'} V^J_ {1 a} -\frac{\zeta}{\sqrt{2}} V^J_ {3 a} U^E_ {1 a'}\biggr) U^
 H_ {51} + \biggr(\frac{2}{\sqrt {3}}\bar \zeta V^J_{1 a}U^E_ {4 a'}\nonumber \\ && + \sqrt{\frac{2}{3}}\bar \zeta  V^J_{1 a}U^E_ {3 a'}
 + \frac{ i \bar \zeta}{\sqrt{6}} V^J_ {3 a} U^E_ {2 a'} -
 \frac{2i}{\sqrt {3}} \bar \zeta  V^J_ {2 a} U^E_ {2 a'}- \sqrt{\frac{3}{2}}i\zeta V^J_ {3 a}U^
  E_ {1 a'}-\frac{2i}{\sqrt {3}}\zeta V^J_ {2 a} U^
  E_ {1 a'} - \frac{\rho}{3 \sqrt{3}} V^J_ {5 a}U^E_ {4 a'}\nonumber \\ && -\frac{\sqrt{2}\rho}{3 \sqrt{3}} V^J_ {5 a} U^E_ {3 a'} -
   \frac{\sqrt {2}\rho}{3 \sqrt{3}} V^J_ {3 a} U^E_ {6 a'} -
  \frac{\rho}{3 \sqrt{3}} V^J_ {2 a}U^E_ {6 a'}\biggr) U^H_{61}\biggr|^2 F_{12 }(m^J_a,m^E_{a'},Q)\nonumber \\ &&
   -2{g_{10}^2} \biggr|-\frac{2i}{\sqrt{3}}V^{E*}_{1a'}U^H_{21}-\frac{2i}{\sqrt{3}}
   V^{E*}_{2a'}U^H_{31}-\sqrt{2} V^{E*}_{3a'}U^H_{41}-\frac{2i}{\sqrt{3}}V^{E*}_{6a'}U^H_{61}\biggr|^2
   F_{12}(m_{\lambda_J}, m^E _{a'}, Q) \nonumber \\ &&
   -2{g_{10}^2} \biggr|\frac{2}{\sqrt{3}}U^{J*}_{1a}U^H_{31}-iU^{J*}_{2a}U^H_{41}
   +\frac{i}{\sqrt{2}}U^{J*}_{3a}U^H_{41}-i U^{J*}_{5a}U^H_{51}-\frac{1}{\sqrt{3}}U^{J*}_{5a}U^H_{61}\biggr|^2
   F_{12} (m^J_a,m_{\lambda_E},
   Q)
 \eea

\bea K_{P\bar X} &=&\sum_ {a = 1}^{\mbox{d(P)}} \sum_ {a' =
  1}^{\mbox{d(X)}} \biggr|\biggr(
 \frac{\kappa}{\sqrt{2}}V^P_{2 a} U^X_{2 a'}-\gamma V^P_{1a}U^X_{1a'} \biggr) U^
  H_{11} + \biggr( \frac{2\sqrt{2}\eta}{\sqrt{3}} V^P_{1 a} U^X _{2 a'} -\frac{2\eta}{\sqrt{3}}
  V^P_{1 a} U^X_{1 a'} \nonumber \\ &&-
  \frac{\bar \zeta}{\sqrt{6}} V^P_{2 a} U^X_{2 a'}\biggr) U^H_{21} +
 \frac{\zeta}{ \sqrt{3}} V^
  P_{2 a} \biggr(2 U^X_{1 a'} + \frac{U^X_{2 a'}}{\sqrt {2}}\biggr) U^ H_{31} - \biggr(\frac{\rho}{3\sqrt {2}} V^P_{2 a} U^X _{2 a'} +
 \zeta V^P_{1 a} U^X_{1 a'}\biggr) U^
  H _{51} \nonumber \\ && + \biggr( \frac{i\sqrt{2}\zeta}{\sqrt{3}}V^P_{1 a} U^X_{2 a'}-\frac{i\zeta}{\sqrt{3}}V^P_{1 a} U^X_{1 a'} +
  \frac{i\rho}{3 \sqrt {3}}V^ P_{2 a} U^X _{1 a'}- \frac{i\rho}{3 \sqrt {6}}V^
  P_{2 a}U^X _{2 a'}\biggr ) U^H_{61}\biggr|^ 2 F_{12} (m^P_a, m^X_{a'},
  Q) \nonumber\\&&
  -2{g_{10}^2} \biggr|i \sqrt{\frac{2}{3}} U^{P*}_{1a}U^H_{21}+\frac{U^{P*}_{2a}}{\sqrt{2}}U^H_{51}
  +\frac{i}{\sqrt{6}}U^{P*}_{2a}U^H_{61}\biggr|^2 F_{12} (m^P_a, m_{\lambda_X}, Q) \eea

\bea K_{T\bar X} &=&\sum_ {a' = 1}^{\mbox{d(X)}} \sum_ {a =
1}^{\mbox{d(T)}}\biggr|\biggr(\gamma U^X_ {2 a'} V^T_ {3 a} -
i\bar \gamma U^X_ {1 a'} V^T_ {4 a} + \bar\gamma U^X_ {2 a'} V^T_
{2 a} -\frac{i\kappa }{\sqrt {2}} U^X_ {2 a'} V^T_ {7 a} -\kappa
U^X_ {1 a'} V^T_ {6 a}\biggr) U^H_{11}\nonumber \\ && + \biggr(
\frac{\bar \zeta}{\sqrt {3}} U^X_ {1 a'} V^T_ {6 a} + \frac{2i\bar
\zeta}{\sqrt {3}} U^X_ {1 a'}V^T_ {7 a} - \frac{i\bar\zeta}{\sqrt
{6}} V^T_ {7 a}U^X_ {2 a'} - \sqrt{\frac{2}{3}}\bar\zeta U^X_ {2
a'}V^T_ {6 a}-\frac{\sqrt {2}i\bar \gamma}{\sqrt {3}} U^X_ {1
a'}V^T_ {1 a} -\frac{\bar \gamma}{\sqrt {3}} U^X_ {2 a'} V^T_ {1
a} \nonumber \\ &&-\frac{4\eta}{\sqrt {6}}U^X_ {1 a'} V^T_ {3 a}
\biggr) U^ H_ {21} + \biggr( \frac{\gamma}{\sqrt {3}} U^X_ {2 a'}
V^T_ {1 a}-\frac{i\sqrt {2}\gamma}{\sqrt {3}} U^X_ {1 a'} V^T_ {1
a} + 2\sqrt{\frac{2}{3}}\eta U^X_ {1 a'}V^T_ {2 a} -
2\sqrt{\frac{2}{3}}i \eta U^X_ {2 a'}V^T_ {4 a} \nonumber \\ &&-
\frac{2i\eta}{\sqrt{3}} U^X_ {1 a'}V^T_ {4 a} + \frac{\zeta}{\sqrt
{3}}U^X_ {1 a'} V^T_ {6 a} + \frac{i\zeta}{\sqrt {6}} U^X_ {2 a'}
V^T_ {7 a} + \frac{\sqrt {2}\zeta}{\sqrt {3}}U^X_ {2 a'} V^T_ {6
a}\biggr) U^H_ {31} + 2 i\lambda \biggr(U^X_ {2 a'} + \sqrt {2}
U^X_ {1 a'}\biggr) V^T_ {5 a} U^H_ {41}\nonumber \\ && +
\biggr(\kappa U^X_ {2 a'}V^T_ {1 a} - \frac{i\rho}{3\sqrt {2}} U^
X_ {2 a'} V^T_ {7 a} + i \bar\zeta U^X_ {1 a'} V^T_ {4 a} \biggr)
U^ H_ {51} + \biggr( \frac{i\sqrt {2}\bar\zeta}{\sqrt {3}} U^X_ {1
a'}V^T_ {2 a} + \frac{\bar\zeta}{\sqrt {3}} U^X_ {1 a'}V^T_ {4 a}
+ \frac{\sqrt {2}\bar\zeta}{\sqrt {3}} U^ X_ {2 a'} V^T_ {4
a}\nonumber \\ &&+ \frac{i\bar\zeta}{\sqrt {3}} U^ X_ {2 a'}V^T_
{2 a}-\sqrt{\frac{2}{3}}i \kappa U^X_ {1 a'}V^T_ {1 a} + \frac{i
\zeta}{\sqrt {3}} U^X_ {2 a'}V^ T_ {3 a} - \frac{\sqrt {2}i \zeta
}{\sqrt {3}}
 U^X_ {1 a'}V^ T_ {3 a}
-\frac{\rho}{3\sqrt {3}}U^X_ {1 a'} V^T_ {7 a} -
\frac{\rho}{3\sqrt {6}} U^X_ {2 a'} V^T_ {7 a}\nonumber \\ && +
\frac{i\rho}{3\sqrt {3}} U^X_ {1 a'} V^T_ {6 a} \biggr) U^H_
{61}\biggr|^2 F_{12} (m^X_{a'},m^T_a,Q)
 -2{g_{10}^2} \biggr|U^{T*}_{1a}U^H_{11}-\frac{i}{\sqrt{3}}U^{T*}_{2a}U^H_{21}+\biggr(\sqrt{\frac{2}{3}}U^{T*}_{4a}
 -\frac{i U^{T*}_{3a}}{\sqrt{3}}\biggr)U^H_{31}\nonumber\\&&- iU^{T*}_{5a}U^H_{41}-\frac{i}{\sqrt{2}}U^{T*}_{7a}U^H_{51}
 +\biggr(i\sqrt{\frac{2}{3}}U^{T*}_{6a}-\frac{U^{T*}_{7a}}{\sqrt{6}} \biggr)U^H_{61}\biggr|^2F_{12} (m_{\lambda_X}, m^T_a, Q) \eea

\bea K_{E\bar P} &=&\sum_ {a = 1}^{\mbox{d(E)}} \sum_ {a' =
1}^{\mbox{d(P)}}\biggr|\biggr( \frac{\kappa }{\sqrt{2}} V^ E_ {4
a}
 U^P_ {2 a'}-\bar \gamma V^E_ {3 a}U^P_ {1 a'} \biggr) U^ H_ {11} +
\biggr(\frac{2\bar\zeta}{\sqrt {3}} V^E_ {3 a} U^ P_ {2 a'} +
\frac{\bar \zeta}{\sqrt {6}}V^E_ {4 a} U^ P_ {2 a'}\biggr) U^H_
{21} \nonumber \\ &&+ \biggr( \frac{2\sqrt{2} \eta}{\sqrt {3}}V^E_
{4 a} U^P_ {1 a'} -\frac{2 \eta}{\sqrt {3}}V^E_ {3 a}U^P_ {1 a'} -
\frac{ \zeta}{\sqrt {6}}V^E_ {4 a} U^P_ {2 a'} \biggr) U^ H_ {31}+
\biggr(2\sqrt {2}i \eta V^ E_ {2 a} U^P_ {1 a'} +
\frac{\rho}{3\sqrt {2}}V^ E_ {6 a} U^P_ {2 a'} \nonumber \\ && +
i\sqrt {2} \bar\zeta V^ E_ {1 a} U^P_ {2 a'} +\sqrt {2} \bar\zeta
V^E_ {6 a} U^P_ {1 a'}\biggr) U^ H_ {41} -
\biggr(\frac{\rho}{3\sqrt {2}} V^E_ {4 a} U^P_ {2 a'}+ \bar \zeta
V^E_ {3 a} U^P_ {1 a'}\biggr) U^ H_ {51} + \biggr(\frac{i\bar
\zeta}{\sqrt {3}} V^E_ {3 a}U^P_ {1 a'}\nonumber \\
&&-\frac{i\sqrt{2}\bar \zeta}{\sqrt {3}}V^ E_ {4 a}U^P_ {1 a'}
-\frac{i\rho}{ 3\sqrt{3}}V^E_ {3 a} U^ P_ {2 a'} +\frac{i\rho}{
3\sqrt{6}} V^E_ {4 a} U^ P_ {2 a'}\biggr) U^ H_ {61}\biggr|^2
F_{12} (m^E_a, m^P_{a'}, Q) \nonumber\\&&
  -2{g_{10}^2} \biggr|-i \sqrt{\frac{2}{3}}V^{P*}_{1a'}U^H_{31}+\frac{V^{P*}_{2a'}}{\sqrt{2}}U^H_{51}
  -\frac{i}{\sqrt{6}}V^{P*}_{2a'}U^H_{61}\biggr|^2 F_{12} ( m_{\lambda_E},m^P_{a'}, Q) \eea

\bea
 K_{E\bar T} &=& \sum_ {a'
= 1}^{\mbox{d(T)}}\sum_ {a = 1}^{\mbox{d(E)}}\biggr|\biggr(\gamma
U^T_ {3 a'} V^E_ {4 a} - \gamma U^T_ {5 a'} V^E_ {2 a} - i\gamma
U^T_ {4 a'} V^E_ {3 a} + \bar \gamma U^T_ {2 a'} V^E_ {4 a} +\bar
\gamma U^T_ {5 a'} V^E_ {1 a} -\kappa U^T_ {6 a'} V^E_ {3 a}
\nonumber \\ &&-i\kappa U^T_ {5 a'} V^E_ {6 a}
-\frac{i\kappa}{\sqrt {2}} U^T_ {7 a'} V^E_ {4 a}\biggr) U^ H_
{11} + \biggr(\frac{\bar \gamma}{\sqrt {3}} U^ T_ {1 a'} V^E_ {4
a} -\frac{\sqrt {2}i\bar \gamma}{\sqrt {3}} U^ T_ {1 a'} V^E_ {3a}
+\frac{2\sqrt {2}\eta}{\sqrt {3}} U^T_ {3 a'} V^E_ {3 a} -
\frac{2i\eta}{\sqrt {3}} U^T_ {4 a'} V^E_ {3 a} \nonumber \\
&&+\frac{2\eta}{\sqrt {3}} U^T_ {5 a'} V^E_ {2 a} - \frac{2\sqrt
{2}i\eta}{\sqrt {3}} U^ T_ {4 a'} V^E_ {4 a} + \frac{\bar
\zeta}{\sqrt {3}} U^T_ {6 a'} V^ E_ {3 a} + \frac{\sqrt {2}\bar
\zeta}{\sqrt {3}} U^T_ {6 a'} V^E_ {4 a} + \frac{i\bar\zeta}{\sqrt
{6}} U^T_ {7 a'} V^E_ {4 a} - \frac{i\bar\zeta}{\sqrt{3}} U^T_ {5
a'} V^E_ {6 a}\biggr) U^ H_ {21} \nonumber \\ &&+ \biggr(
2\sqrt{3}\eta U^T_{5 a'} V^E_{1 a} -\frac{2\sqrt {2}\eta}{\sqrt
{3}} U^T_{2 a'} V^E_{3 a} - \frac{ \sqrt {2}i\gamma}{\sqrt {3}}U^
T_ {1 a'} V^E_ {3 a} -\frac{\gamma}{\sqrt {3}}U^T_{1 a'}V^E_ {4 a}
 +
 \frac{ \zeta}{\sqrt {3}} U^T_ {6 a'}V^ E_ {3 a} -\frac{ \sqrt {2} \zeta}{\sqrt {3}} U^T_ {6 a'} V^E_
{4 a}\nonumber \\ &&+i\zeta\sqrt{3} U^T_ {5 a'} V^ E_ {6 a}
+\frac{2i\zeta}{\sqrt{3}} U^T_ {7 a'} V^E_ {3 a} -
\frac{i\zeta}{\sqrt {6}}U^T_ {7 a'} V^E_ {4 a}\biggr) U^H_ {31}+
\biggr( 2 i \eta U^T_ {2 a'} V^E_ {2 a}-2 i \eta U^ T_ {3 a'} V^E_
{1 a} + 2\sqrt {2} \eta
U^T_ {4 a'} V^ E_ {1 a} \nonumber \\
&&- \bar \gamma U^T_ {1 a'}V^E_ {1 a} - \gamma U^T_ {1 a'} V^ E_
{2 a}+ \kappa U^T_ {1 a'}V^E_ {6 a} +
 \frac{\sqrt {2}\rho}{3} U^T_ {6 a'}V^E_ {6 a} - \frac{i\rho}{3\sqrt {2}} U^T_ {7 a'} V^E_ {6 a}
+ \sqrt {2} i\zeta U^T_ {6 a'} V^E_ {2 a} -\zeta U^T_ {3 a'} V^E_
{6 a} \nonumber \\ &&- \sqrt {2} i\zeta U^T_ {4 a'} V^E_ {6 a}+
\sqrt {2}\zeta U^T_ {7 a'} V^E_ {2 a} + \sqrt {2} i\bar\zeta U^T_
{6 a'} V^ E_ {1 a} +\bar\zeta U^T_ {2 a'} V^E_ {6 a}\biggr) U^H_
{41}  + \biggr(\kappa U^T_ {1 a'} V^E_ {4 a} -
\frac{i\rho}{3\sqrt{2}} U^T_ {7 a'}V^E_ {4 a} \nonumber \\ &&
+\frac{i\rho}{3} U^T_ {5 a'} V^E_ {6 a}- \zeta U^T_ {5 a'}V^E_ {2
a}+i\zeta U^T_ {4 a'} V^E_ {3 a} - \bar\zeta U^T_ {5 a'}V^E_ {1
a}\biggr) U^ H_ {51}+ \biggr(i\kappa \sqrt {\frac{2}{3}} U^T_ {1
a'} V^E_ {3 a} + \frac{i\zeta}{\sqrt {3}} U^ T_ {5 a'}V^E_ {2
a}\nonumber \\ && - \frac{\sqrt {2}\zeta}{\sqrt {3}} U^T_ {4 a'}
V^E_ {4 a} - \frac{i\zeta}{\sqrt {3}} U^T_ {3 a'} V^E_ {4 a} -
\frac{\zeta}{\sqrt {3}}U^T_ {4 a'} V^E_ {3 a} - \frac{ \sqrt
{2}i\zeta}{\sqrt {3}} U^T_ {3 a'} V^E_ {3 a} +
 \sqrt {\frac{2}{3}}i\bar \zeta U^ T_ {2 a'} V^E_ {3 a} - \sqrt
{3} i\bar \zeta U^T_ {5 a'}V^ E_ {1 a} \nonumber \\ &&-
\frac{i\bar \zeta }{\sqrt {3}}U^T_ {2 a'} V^E_ {4 a}+
\frac{\rho}{3\sqrt {3}} (U^T_ {7 a'}V^E_ {3 a} + U^T_ {7
a'}\frac{V^E_ {4 a}}{\sqrt {2}} + 2 U^T_ {5 a'}V^E_ {6 a} - i U^T_
{6 a'} V^ E_ {3 a})\biggr) U^H_ {61}\biggr|^2 F_{12} (
m^T_{a'},m^E_a, Q)\nonumber \\ && -2{g_{10}^2}
\biggr|V^{T*}_{1a'}U^H_{11}+\biggr(\frac{i}{\sqrt{3}}V^{T*}_{2a'}-\sqrt{\frac{2}{3}}V^{T*}_{4a'}\biggr)U^H_{21}
+\frac{i}{\sqrt{3}}V^{T*}_{3a'}U^H_{31}-\frac{i}{\sqrt{2}}V^{T*}_{7a'}U^H_{51}+\biggr(\frac{V^{T*}_{7a'}}{\sqrt{6}}\nonumber
\\ &&-i \sqrt{\frac{2}{3}}V^{T*}_{6a'} \biggr)U^H_{61}\biggr|^2
F_{12} ( m^T_{a'},m_{\lambda_E}, Q) \eea

 \bea
  K_{L\bar Y} &=&\sum_ {a = 1}^{\mbox{d(L)}} \biggr|-\biggr(i \bar \gamma V^L_ {1 a} + \kappa V^L_ {2 a}\biggr) U^H_ {11}-
   \frac{\bar \zeta}{\sqrt{3}} V^L_ {2 a}U^H_ {21} + \biggr(2 i \eta V^L_ {1 a} - \zeta V^L_ {2 a}\biggr) \frac{U^
   H_ {31}}{\sqrt {3}} +
    i\bar \zeta V^
   L_ {1 a}U^H_ {51}\nonumber \\ && - \biggr(\frac{i\rho}{3\sqrt {3}} V^L_ {2 a} +
  \frac{\bar \zeta}{\sqrt {3}} V^L_ {1 a}\biggr) U^
  H_ {61}\biggr|^ 2 F_{12}(m^L_a,m_Y,Q)
\eea \bea
 K_{\bar V \bar F} &=&\sum_ {a = 1}^{\mbox{d(F)}}\biggr|-\biggr(i \bar \gamma U^F_ {1 a} + \kappa U^F_ {4
a}\biggr) U^H_ {11} + \sqrt {3}
   \bar\zeta U^F_ {4a}U^H_ {21} + \biggr( \sqrt{3} \zeta U^F_ {4a}-2 \sqrt{3}i\eta U^F_ {1 a} \biggr)
U^H_ {31}\nonumber \\ && +2 \sqrt {3} \lambda U^F_ {2a} U^H_ {41}
+
   i \bar\zeta U^F_ {1a}U^H_ {51} +\biggr(\frac{i \rho}{
  \sqrt {3}} U^F_ {4a} + \sqrt{3} \bar\zeta U^F_ {1a}\biggr) U^H_ {6
 1}\biggr|^2 F_{12}(m^V, m^F_{a}, Q)\nonumber \\ &&-2{g_{10}^2} \biggr|i U^H_{41}
\biggr|^2F_{12}(m^V, m_{\lambda_F}, Q) \quad \quad \eea

 \bea K_{Z \bar C} &=&\sum_ {a = 1}^{\mbox{d(C)}}\biggr| \biggr(i
\kappa U^C_ {3 a} + \gamma U^C_ {2 a} - \bar\gamma U^C_ {1
a}\biggr) U^
   H_ {11} +\biggr( i \bar\zeta U^C_ {3 a}-2 \eta U^
   C_ {2 a} \biggr) \frac{U^H_ {21}}{\sqrt {3}} + \biggr( i\zeta U^C_ {3 a} +
   2 \eta U^C_ {1 a}\biggr) \frac{U^H_ {31}}{\sqrt {3}} \nonumber \\ &&+
   \biggr( \zeta U^C_ {2 a}-\frac{i
   \rho}{
   3} U^C_ {3 a} +
    \bar\zeta U^C_ {1 a}\biggr) U^
   H_ {51} -\frac{i}{\sqrt {3}} \biggr(\zeta U^C_ {2 a} +
  \bar\zeta U^C_ {1 a}\biggr) U^
  H_ {61}\biggr|^ 2 F_{12} (m^Z,m^C_a,Q)
  \eea

  \bea
  K_{I \bar D} &=&\sum_ {a = 1}^{\mbox{d(D)}}\biggr|\biggr( \bar\gamma U^D_ {2 a}-i \kappa U^D_ {3 a} - \gamma U^D_ {1 a}
  \biggr) U^H_ {11} + \sqrt {3}\biggr(i \bar\zeta U^D_ {3 a} -
   2 \eta U^D_ {1 a}\biggr) U^H_ {21} +
   \biggr( - i \zeta U^D_ {3 a}-2 \eta U^D_ {2 a}\biggr) \frac{U^H_ {31}}{\sqrt {3}} \nonumber \\ && + \biggr(\frac{i\rho U^D_ {3 a}}{3} -
   \zeta U^D_ {1 a} - \bar\zeta U^D_ {2 a}\biggr) U^
    H_ {51} - \biggr(\sqrt {3} i\zeta U^D_ {1 a} -\frac{i\bar\zeta U^D_ {2 a}}{\sqrt {3}}
   + \frac{2 \rho U^D_ {3 a}}{3 \sqrt{3}} \biggr) U^
  H_ {61}\biggr|^2 F_{12} (m^I,m^D_a,Q)
 \quad \quad \eea
 \bea
 K_{QC} &=&\sum_ {a = 1}^{\mbox{d(C)}}\biggr|\biggr(i \gamma V^C_ {1 a} + \kappa V^C_ {3 a} - i \bar\gamma V^
   C_ {2 a}\biggr) \frac{U^H_ {11}}{\sqrt {2}} - \biggr(2i \eta V^C_ {1 a} +
   \bar\zeta V^C_ {3 a}\biggr) \frac{U^H_ {21}}{\sqrt {6}} +
    \biggr(\zeta V^C_ {3 a} + 2i \eta V^
   C_ {2 a}\biggr) \frac{U^H_ {31}}{\sqrt {6}} \nonumber \\ &&+ \biggr(\frac{\rho}{3} V^C_ {3 a} - i\zeta V^
   C_ {1 a} - i \bar\zeta V^C_ {2 a}\biggr) \frac{U^
  H_ {51}}{\sqrt {2}}- \biggr(\zeta V^C_ {1 a} +
  \bar\zeta V^C_ {2 a}\biggr) \frac{U^H_ {61}}{\sqrt {6}}\biggr|^2 F_{12}(m^Q,m^C_a,Q)
  \eea

 \bea
  K_{U\bar E} &=&\sum_ {a = 1}^{\mbox{d(E)}}\biggr|\biggr( i\bar\gamma U^E_ {2 a}-i \gamma U^E_ {1 a} - \kappa U^E_ {6 a} \biggr) \frac{U^H_ {11}}{\sqrt{2}} + \biggr( 2i\eta U^
   E_ {1 a}-\bar\zeta U^
   E_ {6 a} \biggr)\frac{U^H_ {21}}{\sqrt {6}} + \biggr( 6i \eta U^
   E_{2 a} +3 \zeta U^E_ {6 a}\biggr) \frac{U^H_ {31}}{\sqrt {6}} \nonumber \\ && +
    \biggr(
   2 \lambda U^E_ {3 a}-\sqrt {2} \lambda U^E_ {4 a} \biggr) U^
   H_ {41} + \biggr(i \zeta U^
   E_ {1 a}- \frac{\rho}{3} U^E_ {6 a} + i \bar\zeta U^E_ {2 a}\biggr) \frac{U^
   H_ {51}}{\sqrt {2}} + \biggr(\zeta U^E_{1 a}-3 \bar\zeta U^
   E_ {2 a} \nonumber \\ &&+\frac{2i\rho}{3} U^
  E_ {6 a}\biggr) \frac{U^H_ {61}}{\sqrt {6}}\biggr|^2 F_{12} (m^U,m^E_a,Q)-2{g_{10}^2} \biggr|\frac{ U^H_{41}}{\sqrt{2}}
\biggr|^2F_{12}(m^U, m_{\lambda_{E}}, Q)
  \eea

  \bea
  K_{D\bar U } & = &\sum_ {a =
  1}^{\mbox {d(D)}}\biggr | \biggr( i\bar\gamma V^D_ {1 a}-i\gamma V^D_ {2 a} - \kappa V^
  D_ {3 a} \biggr)\frac {U^
 H_ {11}} {\sqrt {2}} + \biggr(\sqrt{\frac{3}{2}}\bar\zeta V^D_ {3 a}
 -\sqrt{6} i\eta V^D_ {2 a}\biggr)U^
 H_ {21} + \biggr(-\frac{\zeta}{\sqrt{6}} V^
  D_ {3 a}-\sqrt{\frac{2}{3}}i\eta V^D_ {1 a} \biggr)U^H_ {31}\nonumber \\ && + \biggr( i\zeta V^D_ {2 a}-\frac{\rho}{3} V^D_ {3 a} +
 i\bar\zeta V^D_ {1 a}\biggr)\frac {U^H_ {51}} {\sqrt {2}} + \biggr( \bar\zeta V^
  D_ {1 a} -3\zeta V^
  D_ {2 a} -\frac {2 i\rho} {3} V^
  D_ {3 a}\biggr) \frac {U^H_ {61}} {\sqrt {6}} \biggr|^2 F_ {12} (m^U, m^D_a, Q)
  \eea

\bea
  K_{B\bar L} &=&\sum_ {a = 1}^{\mbox{d(L)}}\biggr|-\biggr(i \gamma U^L_ {1 a} + \kappa U^L_ {2 a}\biggr)U^H_ {11} + \biggr( 2i\eta U^
   L_ {1 a}-\bar\zeta U^
   L_ {2 a}\biggr)\frac{U^H_ {21}}{\sqrt {3}} - \zeta U^L_ {2 a} \frac{U^H_ {31}}{\sqrt {3}} +
   i \zeta U^
   L_ {1 a}U^H_ {51} \nonumber \\ && + \biggr(\zeta U^
   L_ {1 a} + \frac{i\rho}{3} U^
  L_ {2 a}\biggr) \frac{U^
  H_ {61}}{\sqrt{3}}\biggr|^2 F_{12} (m^B,m^L_a,Q)
  \eea

\bea
  K_{X\bar K} &=&\sum_ {a = 1}^{\mbox{d(X)}} \sum_ {a' =
  1}^{\mbox{d(K)}} \biggr| -\biggr(i \gamma \sqrt{2}V^X_ {1 a} U^K_{1 a'} +i
   \kappa V^X_ {2 a}U^K_{2 a'}\biggr) U^
  H_ {11} - \biggr( \frac{2 \sqrt{2}i\eta}{\sqrt{3}} V^X_ {1 a}U^K_{1 a'} +\frac{4i\eta}{\sqrt{3}}
   V^X_ {2 a} U^K_{1 a'} \nonumber \\ && -
  \frac{i\bar \zeta}{\sqrt {3}} V^X_ {2 a}U^K_{2 a'} \biggr) U^
  H_ {21} + \frac{i\zeta}{\sqrt {3}} \biggr(2\sqrt{2} V^X_ {1 a} -
   V^X_ {2 a}\biggr) U^K_{2 a'} U^
  H_ {31} + \biggr(
  \sqrt{2} i \zeta V^X_ {1 a} U^K_{1 a'}-\frac{i\rho}{3} V^X_ {2 a} U^K_{2 a'}\biggr) U^
  H_ {51} \nonumber \\ && + \biggr(\frac{\rho}{3\sqrt {3}}V^X_ {2 a}U^ K_{2 a'} + \frac{\sqrt {2}\rho}{3\sqrt {3}}V^X_ {1 a} U^ K_{2 a'}- \frac{\sqrt{2}\zeta}{\sqrt{3}}V^X_ {1 a} U^
   K_{1 a'} - \frac{2\zeta}{\sqrt{3}} V^X_ {2 a}U^
   K_{1 a'} \biggr) U^
  H_ {61}\biggr|^2 F_{12} (m^X _{a},m^K _{a'}, Q)\nonumber\\&&-2{g_{10}^2} \biggr|\frac{-2}{\sqrt{3}}V^{K*}_{1a'}U^H_{21}
  -iV^{K*}_{2a'}U^H_{51}
  +\frac{V^{K*}_{2a'}}{\sqrt{3}}U^H_{61}\biggr|^2 F_{12} (m_{\lambda_X},m^K _{a'}, Q)
  \eea

\bea
  K_{M\bar B} &=&\biggr|- \sqrt{2}i\bar \gamma U^H_ {11} + 2\sqrt{\frac{2}{3}}i\eta U^H_
  {31}+\sqrt{2} i\bar\zeta U^H_ {51} -\sqrt{\frac{2}{3}}\bar \zeta U^
  H_ {61}\biggr|^2 F_{12} (m^M,m^B,Q)
  \eea

  \bea
  K_{B\bar W} &=&\biggr|-\bar \gamma U^H_ {11} + \frac{2\eta}{\sqrt{3}} U^H_
  {31}-\bar \zeta U^H_ {51} -\frac{i\bar \zeta}{\sqrt{3}} U^
  H_ {61}\biggr|^2 F_{12} (m^B,m^W,Q)
  \eea
\bea
  K_{W\bar Y} &=&\biggr|- \gamma U^H_ {11} + \frac{2\eta}{ \sqrt{3}} U^H_
  {21}- \zeta U^H_ {51} +\frac{i\zeta}{\sqrt{3}} U^
  H_ {61}\biggr|^2 F_{12} (m^W,m^Y,Q)
  \eea

\bea
  K_{\bar V O} &=&\biggr|- \gamma U^H_ {11} -2\sqrt{3}\eta U^H_
  {21}- \zeta U^H_ {51} -\sqrt{3}i\zeta U^
  H_ {61}\biggr|^2 F_{12} (m^V,m^O,Q)
  \eea

  \bea
  K_{Y\bar N} &=&\biggr|- \sqrt{2}i \gamma U^H_ {11} + 2\sqrt{\frac{2}{3}}i\eta U^H_
  {21}+\sqrt{2} i\zeta U^H_ {51} +\sqrt{\frac{2}{3}}\zeta U^
  H_ {61}\biggr|^2 F_{12} (m^M,m^B,Q)
  \eea

  \bea
  K_{VA} &=&\biggr|- \sqrt{2}i\gamma U^H_ {11} - 2\sqrt{6}i\eta U^H_
  {21}+\sqrt{2}i \zeta U^H_ {51} -\sqrt{6} \zeta U^
  H_ {61}\biggr|^2 F_{12} (m^V,m^A,Q)
  \eea
  \bea
  K_{\bar O\bar H} &=&\sum_ {a = 2}^{\mbox{d(H)}} \biggr|\bar \gamma U^H_ {4 a} U^H_ {11} + 2\sqrt {3} \eta U^
   H_{4 a} U^H_{31} +\biggr(2\sqrt {3} \eta U^H_{3 a} +
  \bar\zeta U^H_ {5 a} - \sqrt {3}i\bar\zeta U^H_ {6 a}+\bar\gamma U^H_ {1 a}\biggr) U^
  H_ {41}\nonumber \\ && +
   \bar\zeta U^H_ {4 a} U^H_ {51} - \sqrt {3}i \bar\zeta U^
  H_ {4 a} U^
  H_ {61}\biggr|^2 F_{12}(m^H_a, m^O,Q)\nonumber \\ &&+\biggr|2\bar \gamma U^H_ {41} U^H_ {11} + 4\sqrt {3} \eta U^
   H_{41} U^H_{31}+2
   \bar\zeta \biggr(U^H_ {41} U^H_ {51} - \sqrt {3}i U^
  H_ {41} U^
  H_ {61}\biggr)\biggr|^2 F_{11}( m^O,Q)
  \eea
  \bea
  K_{S H} &=&\sum_ {a = 2}^{\mbox{d(H)}}\biggr |\biggr(i \gamma V^H_ {3 a} - i\bar\gamma V^H_ {2 a} +
   \kappa V^H_ {6 a}\biggr) \frac{U^H_ {11}}{\sqrt {2}} + \biggr(2\sqrt
   {\frac{2}{
   3}}i \eta V^H_ {3 a} - \sqrt {\frac{2}{3}} \bar\zeta V^
   H_ {6 a} + \frac{i\bar\zeta}{\sqrt {2}} V^
   H_ {5 a} +\frac{i\bar\gamma}{\sqrt {2}} V^H_ {1 a}\biggr) U^
   H_ {21} \nonumber \\ &&+ \biggr(
   \frac{i\zeta}{\sqrt {2}} V^
   H_ {5 a}-\sqrt {\frac{2}{3}} \zeta V^H_ {6 a} - \frac{i\gamma}{\sqrt {2}} V^H_ {1 a} -
   2 \sqrt {\frac{2}{3}}i \eta V^H_ {2 a}\biggr) U^
   H_ {31} - \sqrt {6}i \lambda V^H_ {4 a} U^
   H_ {41} + \biggr(\frac{\rho}{3} V^
   H_ {6 a} - i\zeta V^H_ {3 a}\nonumber \\ && -i
   \bar\zeta V^H_ {2 a}\biggr) \frac{U^
   H_ {51}}{\sqrt {2}} + \biggr( \sqrt {\frac{2}{3}} \bar\zeta V^
   H_ {2 a}-\frac{\kappa}{\sqrt {2}}V^
   H_ {1 a} + \sqrt {\frac{2}{3}} \zeta V^H_ {3 a} -
  \frac{\rho}{3\sqrt {2}} V^H_ {5 a}\biggr) U^
  H_ {61}\biggr|^2 F_{12} (m^H_a,m^S,Q)\nonumber\\&&+
   \biggr|\biggr(i \gamma V^H_ {31} - i\bar\gamma V^H_ {21} +
   \kappa V^H_ {61}\biggr) \frac{U^H_ {11}}{\sqrt {2}} + \biggr(2\sqrt
   {\frac{2}{
   3}}i \eta V^H_ {31} - \sqrt {\frac{2}{3}} \bar\zeta V^
   H_ {61} + \frac{i\bar\zeta}{\sqrt {2}} V^
   H_ {51} +\frac{i\bar\gamma}{\sqrt {2}} V^H_ {11}\biggr) U^
   H_ {21} \nonumber \\ &&+ \biggr(
   \frac{i\zeta}{\sqrt {2}} V^
   H_ {51}-\sqrt {\frac{2}{3}} \zeta V^H_ {61} - \frac{i \gamma}{\sqrt {2}} V^H_ {11} -
   2 \sqrt {\frac{2}{3}}i \eta V^H_ {21}\biggr) U^
   H_ {31} - \sqrt {6}i \lambda V^H_ {41} U^
   H_ {41} + \biggr(\frac{\rho}{3} V^
   H_ {61} - i\zeta V^H_ {31}\nonumber \\ &&-i
   \bar\zeta V^H_ {21}\biggr) \frac{U^
   H_ {51}}{\sqrt {2}} + \biggr( \sqrt {\frac{2}{3}} \bar\zeta V^
   H_ {21} -\frac{\kappa}{\sqrt {2}}V^
   H_ {11} + \sqrt {\frac{2}{3}} \zeta V^H_ {31} -
  \frac{\rho}{3\sqrt {2}} V^H_ {51}\biggr) U^
  H_ {61}\biggr|^2 F_{11} (m^S,Q)
  \eea
  \bea
  K_{F \bar H} &=&\sum_ {a = 1}^{\mbox{d(F)}} \sum_ {a'= 2}^{\mbox{d(H)}}
 \biggr|\biggr(i \gamma V^F_ {1 a} U^H_ {4 a'} + \gamma U^H_ {3 a'} V^
   F_ {2 a} - \bar \gamma V^F_ {2 a} U^H_ {2 a'} +
   \kappa V^F_ {4 a} U^H_ {4 a'} + i \kappa V^
   F_ {2 a}U^H_ {6 a'}\biggr) U^H_ {11} + \biggr(\frac {4\eta}{\sqrt{3}} V^F_ {2 a} U^H_ {3 a'} \nonumber \\ &&+2\sqrt {3}i \eta V^F_ {1 a}U^H_ {4 a'} - \sqrt {3}\bar \zeta V^F_ {4 a} U^
   H_ {4 a'} - \frac{2i\bar \zeta}{\sqrt {3}} V^F_ {2 a}U^
   H_ {6 a'} - \bar \zeta V^F_ {2 a}U^H_ {5 a'} + \bar \gamma V^
   F_ {2 a} U^H_ {1 a'}\biggr) U^H_ {21} -\biggr(\frac {4}{\sqrt{3}} \eta V^
   F_ {2 a}U^H_ {2 a'} \nonumber \\ && + \frac{2 i \zeta }{\sqrt {3}}V^F_ {2 a}U^H_ {6 a'} +\zeta
   V^F_ {2 a} U^H_ {5 a'} + \sqrt {3}\zeta V^F_ {4 a} U^
   H_ {4 a'} + \gamma V^F_ {2 a} U^
   H_ {1 a'}\biggr) U^H_ {31} + \biggr( i \zeta V^F_ {1 a}U^H_ {5 a'} -\sqrt {3}\zeta
   V^F_ {1 a}U^H_ {6 a'}\nonumber \\ && -\frac{i \rho}{\sqrt{3}} V^
   F_ {4 a} U^H_ {6 a'} - 2\sqrt {3}i \eta V^
   F_ {1 a} U^H_ {2 a'} - i\gamma V^
   F_ {1 a} U^H_ {1 a'} -
   \kappa V^F_ {4 a} U^H_ {1 a'} + \sqrt {3}\bar\zeta V^
   F_ {4 a} U^
   H_ {2 a'} +\sqrt {3} \zeta V^
   F_ {4 a}U^H_ {3 a'}\biggr) U^H_ {41}\nonumber \\ && + \biggr( \bar\zeta V^F_ {2 a} U^
   H_ {2 a'}-\frac{i\rho}{3} V^F_ {2 a} U^
   H_ {6 a'} - i\zeta V^F_ {1 a} U^
   H_ {4 a'} + \zeta V^F_ {2 a} U^H_ {3 a'}\biggr) U^
   H_ {51} + \biggr(\sqrt {\frac{4}{3}}i \zeta V^
   F_ {2 a} U^H_ {3 a'}+\sqrt {\frac{4}{3}} i \bar \zeta V^
   F_ {2 a}U^H_ {2 a'}\nonumber \\ && + \sqrt {3}\zeta  V^
   F_ {1 a}U^H_ {4 a'} +\frac{i\rho}{\sqrt{3}} V^F_ {4 a} U^
   H_ {4 a'} + \frac{i\rho}{3} V^F_ {2 a} U^
  H_ {5 a'} - i \kappa V^F_ {2 a} U^H_ {1 a'}\biggr) U^
   H_ {61}\biggr|^2 F_{12} (m^F_a, m^H_{a'},Q)\nonumber\\&&
- \sum_ {a'= 2}^{\mbox{d(H)}}2{g_{10}^2}
\biggr|i\biggr(V^{H*}_{1a'}U^H_{11}+V^{H*}_{2a'}U^H_{21}+V^{H*}_{3a'}U^H_{31}
  +V^{H*}_{5a'}U^H_{51}+V^{H*}_{6a'}U^H_{61}\biggr)\biggr|^2F_{12} (m^H_{a'}, m_{\lambda_F}, Q)\nonumber \\ &&-
  2{g_{10}^2} \biggr|i\biggr(V^{H*}_{11}U^H_{11}+V^{H*}_{21}U^H_{21}+V^{H*}_{31}U^H_{31}
  +V^{H*}_{51}U^H_{51}+V^{H*}_{61}U^H_{61}\biggr)\biggr|^2F_{11} ( m_{\lambda_F}, Q)
\eea

  \bea
  K_{G H} &=&\sum_ {a = 1}^{\mbox{d(G)}} \sum_ {a'= 2}^{\mbox{d(H)}}
  \biggr|\biggr(\gamma V^G_ {2 a}V^H_ {3 a'} +\frac{ \gamma}{\sqrt {2}} V^G_ {3 a}V^H_ {3 a'}-
    \sqrt {2}i \bar\gamma V^G_ {5 a}V^H_ {4 a'} +
   \bar\gamma V^G_ {2 a}V^H_ {2 a'} - \frac{\bar\gamma}{\sqrt {2}} V^G_ {3 a}V^H_ {2 a'}
   -\kappa V^G_ {1 a} V^H_ {5 a'}\nonumber \\ && +
   \frac{i\kappa}{\sqrt {2}} V^G_ {3 a} V^H_ {6 a'}\biggr) U^
   H_ {11} + \biggr(2 \sqrt {\frac{2}{3}} \eta V^
   G_ {3 a}V^H_ {3 a'} + \frac{4\eta}{\sqrt{3}} V^G_ {2 a} V^H_ {3 a'}
   -\sqrt {\frac{2}{3}} i \bar\zeta V^G_ {3 a}V^H_ {6 a'} - \frac{\bar\zeta}{\sqrt {2}}V^G_ {3 a}V^H_ {5 a'}
    -
   \bar\zeta i V^G_ {1 a}V^H_ {6 a'} \nonumber \\ && +
   \bar\gamma V^G_ {2 a}V^H_ {1 a'} + \frac{\bar\gamma}{\sqrt {2}} V^G_ {3 a}V^H_ {1 a'}
    \biggr ) U^
   H_ {21} + \biggr( \frac{4\eta}{\sqrt {3}} V^G_ {2 a} V^
   H_ {2 a'}- \frac{4\eta}{\sqrt {6}} V^G_ {3 a} V^
   H_ {2 a'}-2\sqrt {6} \eta i V^G_ {5 a} V^
   H_ {4 a'} -\sqrt {\frac{2}{3}}i \zeta V^G_ {3 a}V^H_ {6 a'}\nonumber \\ && +i \zeta
   V^G_ {1 a} V^H_ {6 a'} - \frac{\zeta}{\sqrt {2}}V^G_ {3 a} V^
   H_ {5 a'} +
   \gamma V^G_ {2 a}V^H_ {1 a'} - \frac{\gamma}{\sqrt {2}} V^G_ {3 a} V^H_ {1 a'}
    \biggr) U^
   H_ {31} + \biggr(2\sqrt {6} i \eta V^
   G_ {4 a}V^H_ {2 a'} -\sqrt {2}i\zeta V^G_ {4 a} V^H_ {5 a'} \nonumber \\ && +
    \sqrt {6}\zeta V^G_ {4 a} V^H_ {6 a'} +
   \sqrt {2}\gamma i V^G_ {4 a}V^H_ {1 a'} +
   2\sqrt {3}\lambda V^G_ {2 a}V^H_ {4 a'} - \sqrt {6}\lambda V^G_ {3 a}V^H_ {4 a'}
   \biggr) U^
   H_ {41} + \biggr( \frac{\zeta}{\sqrt {2}} V^G_ {3 a} V^H_ {3 a'} - \kappa V^G_ {1 a} V^H_ {1 a'}\nonumber \\ &&-\frac{i\rho}{3\sqrt {2}} V^
   G_ {3 a}V^H_ {6 a'} +
   \sqrt {2} i \bar\zeta V^
   G_ {5 a}V^H_ {4 a'} +
    \frac{\bar\zeta}{\sqrt {2}} V^G_ {3 a}V^H_ {2 a'} \biggr) U^
   H_ {51} + \biggr( \sqrt {\frac{2}{3}}i\zeta V^G_ {3 a}V^H_ {3 a'} +i\zeta
   V^G_ {1 a} V^H_ {3 a'} +
   \sqrt {\frac{2}{3}} i \bar\zeta V^G_ {3 a}V^
   H_ {2 a'} \nonumber \\ &&- i \bar\zeta V^G_ {1 a} V^
   H_ {2 a'} + \frac{2 \rho}{3\sqrt {3}} V^H_ {6 a'} V^
   G_ {2 a} + \frac{i\rho}{3\sqrt {2}} V^H_ {5 a'} V^G_ {3 a} -
   \frac{i\kappa}{\sqrt {2}} V^H_ {1 a'} V^
   G_ {3 a} + \sqrt {6} \bar\zeta V^H_ {4 a'} V^G_ {5 a}\biggr) U^
  H_ {61}\biggr|^2 F_{12} (m^G_a, m^H_{a'},Q)\nonumber\\&&
   -\sum_ {a'= 2}^{\mbox{d(H)}}2{g_{10}^2} \biggr|\frac{-i}{\sqrt{5}}\biggr(U^{H*}_{1 a'}U^H_{11}+U^{H*}_{2a'}U^H_{21}+U^{H*}_{3a'}U^H_{31}-4 U^{H*}_{4a'}U^H_{41}
  +U^{H*}_{5a'}U^H_{11}\nonumber\\&&+U^{H*}_{6 a'}U^H_{61}\biggr)\biggr|^2F_{12}(m^H_{a'}, m_{\lambda_G},Q)\nonumber\\&&
  + \sum_ {a = 1}^{\mbox{d(G)}}
  \biggr|\biggr(\gamma V^G_ {2 a}V^H_ {3 1} + \frac{\gamma}{\sqrt {2}} V^G_ {3 a}V^H_ {31}-
    \sqrt {2} i\bar\gamma V^G_ {5 a}V^H_ {41} +
   \bar\gamma V^G_ {2 a} V^H_ {21} - \frac{\bar\gamma}{\sqrt {2}}V^G_ {3 a}V^H_ {21}
   -
   \kappa V^G_ {1 a} V^H_ {51}\nonumber \\ && +
   \frac{i\kappa}{\sqrt {2}} V^G_ {3 a} V^H_ {61}\biggr) U^
   H_ {11} + \biggr(2\sqrt {\frac{2}{3}} \eta V^
   G_ {3 a}V^H_ {31} + \frac{4\eta}{\sqrt{3}} V^G_ {2 a} V^H_ {31}
   -\sqrt {\frac{2}{3}} i \bar\zeta V^G_ {3 a}V^H_ {61} - \frac{\bar\zeta}{\sqrt {2}}V^G_ {3 a}V^H_ {51}
    -
    i \bar\zeta V^G_ {1 a}V^H_ {61} \nonumber \\ && +
   \bar\gamma V^G_ {2 a} V^H_ {11} +
   \frac{\bar\gamma }{\sqrt {2}}V^G_ {3 a}V^H_ {11}\biggr) U^
   H_ {21} + \biggr(i \zeta
   V^G_ {1 a} V^H_ {61} -\sqrt {\frac{2}{3}}i \zeta V^G_ {3 a}V^H_ {61} - \frac{\zeta}{\sqrt {2}}V^G_ {3 a} V^
   H_ {51}-2\sqrt {6}i \eta V^G_ {5 a} V^
   H_ {41} \nonumber \\ && +
   \gamma V^G_ {2 a}V^H_ {11} -
   \frac{\gamma }{\sqrt {2}}V^G_ {3 a}V^H_ {11} + \frac{4\eta}{\sqrt {3}} V^G_ {2 a} V^
   H_ {21} - \frac{4\eta}{\sqrt {6}}V^G_ {3 a} V^
   H_ {2 1}\biggr) U^
   H_ {31} + \biggr(2\sqrt {6} i \eta V^
   G_ {4 a}V^H_ {21} -\sqrt {2}i\zeta V^G_ {4 a}V^H_ {51} \nonumber \\ && +
    \sqrt {6}\zeta  V^G_ {4 a} V^H_ {61}+
   \sqrt {2}i\gamma V^G_ {4 a} V^H_ {11} +
   2\sqrt {3}\lambda V^G_ {2 a}V^H_ {41} - \sqrt {6}\lambda
   V^G_ {3 a}V^H_ {41}\biggr) U^
   H_ {41} + \biggr(-\frac{i\rho}{3\sqrt {2}} V^
   G_ {3 a}V^H_ {61} - \kappa V^G_ {1 a} V^H_ {1 a'}\nonumber \\ && +
   \sqrt {2} i \bar\zeta V^
   G_ {5 a} V^H_ {41} + \frac{\zeta }{\sqrt {2}}V^G_ {3 a}V^H_ {31} +
   \frac{ \bar\zeta}{\sqrt {2}} V^G_ {3 a} V^H_ {21} \biggr) U^
   H_ {51} + \biggr( \sqrt {\frac{2}{3}}i\zeta V^G_ {3 a}V^H_ {31} +i\zeta
   V^G_ {1 a} V^H_ {31} +
    \sqrt {\frac{2}{3}}i \bar\zeta V^G_ {3 a} V^
   H_ {2 1}\nonumber \\ &&- i \bar\zeta V^G_ {1 a} V^
   H_ {2 1} + \frac{2 \rho}{3\sqrt {3}} V^
   G_ {2 a} V^H_ {61} + \frac{i\rho}{3\sqrt {2}} V^G_ {3 a}V^H_ {5 1} -
   \frac{i\kappa}{\sqrt {2}} V^
   G_ {3 a}V^H_ {1 1}+ \sqrt {6} \bar\zeta V^G_ {5 a}V^H_ {41}\biggr) U^
  H_ {61}\biggr|^2 F_{11} (m^G_a,Q)\nonumber\\&&
  -2{g_{10}^2} \biggr|\frac{-i}{\sqrt{5}}\biggr(U^{H*}_{11}U^H_{11}+U^{H*}_{21}U^H_{21}+U^{H*}_{31}U^H_{31}-4 U^{H*}_{41}U^H_{41}
  +U^{H*}_{51}U^H_{51}\nonumber\\&&+U^{H*}_{61}U^H_{61}\biggr)\biggr|^2F_{11}( m_{\lambda_G},Q)
  \eea

\clearpage

\begin{table}
\textbf{Appendix B} $$
 \begin{array}{|c|c|c|c|}
 \hline
 {\rm Parameter }&{\rm Value} &{\rm  Field }&\hspace{10mm} {\rm Masses}\\
 &&{\rm}[SU(3),SU(2),Y]&\hspace{10mm}({\rm Units\,\,of 10^{16} Gev })\\ \hline
       \chi_{X}&  0.3988           &A[1,1,4]&      1.68 \\ \chi_{Z}&
    0.1168
                &B[6,2,{5/3}]&            0.0718\\
           h_{11}/10^{-6}&  3.4611         &C[8,2,1]&{      0.94,      2.41,      5.15 }\\
           h_{22}/10^{-4}&  3.0937    &D[3,2,{7/ 3}]&{      0.08,      3.39,      6.02 }\\
                   h_{33}&  0.0230     &E[3,2,{1/3}]&{      0.09,      0.71,      1.85 }\\
 f_{11}/10^{-6}&
  0.0038+  0.2167i
                      &&{     1.854,      2.65,      5.33 }\\
 f_{12}/10^{-6}&
 -1.0760-  2.0474i
          &F[1,1,2]&      0.29,      0.57
 \\f_{13}/10^{-5}&
  0.0632+  0.1223i
                  &&      0.57,      3.33  \\
 f_{22}/10^{-5}&
  5.0702+  3.6293i
              &G[1,1,0]&{     0.015,      0.14,      0.50 }\\
 f_{23}/10^{-4}&
 -0.3765+  1.7999i
                      &&{     0.498,      0.65,      0.68 }\\
 f_{33}/10^{-3}&
 -0.9059+  0.2815i
              &h[1,2,1]&{     0.291,      2.32,      3.41 }\\
 g_{12}/10^{-4}&
  0.1310+  0.1177i
                 &&{      4.89,     23.26 }\\
 g_{13}/10^{-5}&
 -8.5199+  6.9958i
     &I[3,1,{10/3}]&      0.23\\
 g_{23}/10^{-4}&
 -3.1937-  1.2230i
          &J[3,1,{4/3}]&{     0.201,      0.65,      1.21 }\\
 \lambda/10^{-2}&
 -3.8826+  1.0500i
                 &&{      1.21,      3.83 }\\
 \eta&
 -0.3134+  0.1210i
   &K[3,1, {8/ 3}]&{      1.86,      3.84 }\\
 \rho&
  0.6305-  0.5268i
    &L[6,1,{2/ 3}]&{      1.93,      2.56 }\\
 k&
  0.1926+  0.2311i
     &M[6,1,{8/ 3}]&      2.17\\
 \zeta&
  0.9082+  0.8524i
     &N[6,1,{4/ 3}]&      2.04\\
 \bar\zeta &
  0.2737+  0.6140i
          &O[1,3,2]&      2.77\\
       m/10^{16}GeV&  0.0086    &P[3,3,{2/ 3}]&{      0.64,      3.56 }\\
     m_\Theta/10^{16}GeV&  -2.375e^{-iArg(\lambda)}     &Q[8,3,0]&     0.181\\
             \gamma&  0.3234        &R[8,1, 0]&{      0.08,      0.24 }\\
              \bar\gamma& -3.6166     &S[1,3,0]&    0.2828\\
 x&
  0.78+  0.58i
         &t[3,1,{2/ 3}]&{      0.16,      0.45,      0.90,      2.52  }\\\Delta_X^{tot},\Delta_X^{GUT}&      0.67,      0.74 &&{      4.08,      4.37,     25.68 }\\
                                \Delta_{G}^{tot},\Delta_{G}^{GUT}&-20.46,-23.49           &U[3,3,{4/3}]&     0.238\\
      \Delta\alpha_{3}^{tot}(M_{Z}),\Delta\alpha_{3}^{GUT}(M_{Z})& -0.0126,  0.0020               &V[1,2,3]&     0.187\\
    \{M^{\nu^c}/10^{12}GeV\}&{0.000648,    0.99,   37.28    }&W[6,3,{2/ 3}]&              1.95  \\
 \{M^{\nu}_{ II}/10^{-10}eV\}&  2.41, 3700.98,       138823.42               &X[3,2,{5/ 3}]&     0.063,     2.068,     2.068\\
                  M_\nu(meV)&{1.169109,    7.32,   41.46    }&Y[6,2, {1/3}]&              0.08  \\
  \{\rm{Evals[f]}\}/ 10^{-6}&{0.017143,   26.28,  985.21         }&Z[8,1,2]&              0.24  \\
 \hline\hline
 \mbox{Soft parameters}&{\rm m_{\frac{1}{2}}}=
          -152.899
 &{\rm m_{0}}=
         11400.993
 &{\rm A_{0}}=
         -2.0029 \times 10^{   5}
 \\
 \mbox{at $M_{X}$}&\mu=
          1.5966 \times 10^{   5}
 &{\rm B}=
         -1.7371 \times 10^{  10}
  &{\rm tan{\beta}}=           51.0000\\
 &{\rm M^2_{\bar H}}=
         -2.0655 \times 10^{  10}
 &{\rm M^2_{  H} }=
         -1.7978 \times 10^{  10}
 &
 {\rm R_{\frac{b\tau}{s\mu}}}=
  0.1998
  \\
 Max(|L_{ABCD}|,|R_{ABCD}|)&
          8.1104 \times 10^{ -22}
  {\,\rm{GeV^{-1}}}&& \\
 \hline\hline
 \mbox{Susy contribution to}&&&
 \\
 {\rm \Delta_{X,G,3}}&{\rm \Delta_X^{Susy}}=
            -0.070
 &{\rm \Delta_G^{Susy}}=
             3.04
 &{\rm \Delta\alpha_3^{Susy}}=
            -0.015
 \\
 \hline\end{array}
 $$
 \label{table a} \caption{\small{Solution 1: Values   of the NMSGUT-SUGRY-NUHM  parameters at $M_X$
  derived from an  accurate fit to all 18 fermion data and compatible with RG constraints.
 Unification parameters and spectra of superheavy and superlight fields are  also given.
 }}\end{table}
\eject

 \begin{table}
 $$
 \begin{array}{|c|c|c|c|c|}
 \hline
 &&&&\\
 {\rm  Parameter }&{\rm Target} =\bar O_i & {\rm Uncert.= \delta_i  }  & {\rm Achieved= O_i} & {\rm Pull =(O_i-\bar O_i)/\delta_i}\\
 \hline
    y_u/10^{-6}&  2.062837&  0.788004&  2.066323&  0.004424\\
    y_c/10^{-3}&  1.005548&  0.165915&  1.010599&  0.030440\\
            y_t&  0.369885&  0.014795&  0.369792& -0.006256\\
    y_d/10^{-5}& 11.438266&  6.668509& 12.421488&  0.147443\\
    y_s/10^{-3}&  2.169195&  1.023860&  2.189195&  0.019534\\
            y_b&  0.456797&  0.237078&  0.527664&  0.298917\\
    y_e/10^{-4}&  1.240696&  0.186104&  1.224753& -0.085665\\
  y_\mu/10^{-2}&  2.589364&  0.388405&  2.603313&  0.035911\\
         y_\tau&  0.543441&  0.103254&  0.532427& -0.106669\\
             \sin\theta^q_{12}&    0.2210&  0.001600&    0.2210&           -0.0003\\
     \sin\theta^q_{13}/10^{-4}&   29.1907&  5.000000&   29.0755&           -0.0230\\
     \sin\theta^q_{23}/10^{-3}&   34.3461&  1.300000&   34.3574&            0.0087\\
                      \delta^q&   60.0212& 14.000000&   59.7774&           -0.0174\\
    (m^2_{12})/10^{-5}(eV)^{2}&    5.2115&  0.552419&    5.2189&            0.0133\\
    (m^2_{23})/10^{-3}(eV)^{2}&    1.6647&  0.332930&    1.6650&            0.0011\\
           \sin^2\theta^L_{12}&    0.2935&  0.058706&    0.2926&           -0.0152\\
           \sin^2\theta^L_{23}&    0.4594&  0.137809&    0.4412&           -0.1317\\
           \sin^2\theta^L_{13}&    0.0250&  0.019000&    0.0267&            0.0892\\
 \hline
                  (Z_{\bar u})&   0.957467&   0.957908&   0.957908&\\
                  (Z_{\bar d})&   0.950892&   0.951332&   0.951333&\\
                (Z_{\bar \nu})&   0.925116&   0.925579&   0.925580&\\
                  (Z_{\bar e})&   0.944853&   0.945306&   0.945308&\\
                       (Z_{Q})&   0.968740&   0.969189&   0.969190&\\
                       (Z_{L})&   0.949564&   0.950011&   0.950013&\\
              Z_{\bar H},Z_{H}&        0.000273   &        0.001151    &{}&\\
 \hline
 \alpha_1 &
  0.1609-  0.0000i
 & {\bar \alpha}_1 &
  0.1188-  0.0000i
 &\\
 \alpha_2&
 -0.3140-  0.6026i
 & {\bar \alpha}_2 &
 -0.4802-  0.2961i
 &\\
 \alpha_3 &
 -0.0477-  0.4786i
 & {\bar \alpha}_3 &
 -0.4842-  0.2469i
 &\\
 \alpha_4 &
  0.3903-  0.1942i
 & {\bar \alpha}_4 &
  0.5795+  0.0171i
 &\\
 \alpha_5 &
 -0.0449+  0.0061i
 & {\bar \alpha}_5 &
 -0.0415-  0.1241i
 &\\
 \alpha_6 &
 -0.0071-  0.2982i
 & {\bar \alpha}_6 &
  0.0274-  0.1349i
 &\\
  \hline
 \end{array}
 $$
   \caption{\small{Solution 1: Fit   with $\chi_X=\sqrt{ \sum_{i=1}^{17}
 (O_i-\bar O_i)^2/\delta_i^2}=
    0.3988
 $. Target values,  at $M_X$ of the fermion Yukawa
 couplings and mixing parameters, together with the estimated uncertainties, achieved values and pulls.
 Eigenvalues of the wavefunction renormalization   for fermion   and
  Higgs lines are given with
 Higgs fractions $\alpha_i,{\bar{\alpha_i}}$ which control the MSSM fermion Yukawa couplings.  }}
 \end{table}
 \begin{table}
 $$
 \begin{array}{|c|c|c|c|}
 \hline &&&\\ {\rm  Parameter }&{\rm SM(M_Z) }&{\rm m^{GUT}(M_Z)} & {\rm m^{MSSM}=(m+\Delta m)^{GUT}(M_Z) }\\
 \hline
    m_d/10^{-3}&   2.90000&   1.08183&   3.01515\\
    m_s/10^{-3}&  55.00000&  19.06631&  53.14737\\
            m_b&   2.90000&   3.17508&   3.05602\\
    m_e/10^{-3}&   0.48657&   0.45157&   0.45925\\
         m_\mu &   0.10272&   0.09594&   0.09902\\
         m_\tau&   1.74624&   1.65725&   1.65734\\
    m_u/10^{-3}&   1.27000&   1.10509&   1.27687\\
            m_c&   0.61900&   0.54048&   0.62449\\
            m_t& 172.50000& 145.99987& 170.88573\\
 \hline
 \end{array}
 $$
  \caption{\small{Solution 1: Standard model
 fermion masses in GeV at $M_Z$ compared with masses obtained from
 GUT derived  Yukawa couplings  run down from $M_X^0$ to
 $M_Z$, before and after threshold corrections.
  Fit with $\chi_Z=\sqrt{ \sum_{i=1}^{9} (m_i^{MSSM}- m_i^{SM})^2/ (m_i^{MSSM})^2} =
0.1153$.}}
 \end{table}
 \begin{table}
 $$
 \begin{array}{|c|c|c|c|}
 \hline
 {\rm  Parameter}  &{\rm Value}&  {\rm  Parameter}&{\rm Value} \\
 \hline
                       M_{1}&            210.10&   M_{{\tilde {\bar {u}}_1}}&          14446.81\\
                       M_{2}&            569.81&   M_{{\tilde {\bar {u}}_2}}&          14445.85\\
                       M_{3}&           1000.14&   M_{{\tilde {\bar {u}}_3}}&          24609.79\\
     M_{{\tilde {\bar l}_1}}&           1761.31&               A^{0(l)}_{11}&        -121907.75\\
     M_{{\tilde {\bar l}_2}}&            210.71&               A^{0(l)}_{22}&        -121757.58\\
     M_{{\tilde {\bar l}_3}}&          20777.09&               A^{0(l)}_{33}&         -77289.04\\
        M_{{\tilde {L}_{1}}}&          15308.21&               A^{0(u)}_{11}&        -148456.63\\
        M_{{\tilde {L}_{2}}}&          15258.47&               A^{0(u)}_{22}&        -148455.19\\
        M_{{\tilde {L}_{3}}}&          21320.16&               A^{0(u)}_{33}&         -76985.25\\
     M_{{\tilde {\bar d}_1}}&           8402.95&               A^{0(d)}_{11}&        -122521.00\\
     M_{{\tilde {\bar d}_2}}&           8401.45&               A^{0(d)}_{22}&        -122518.53\\
     M_{{\tilde {\bar d}_3}}&          51842.14&               A^{0(d)}_{33}&         -44046.92\\
          M_{{\tilde {Q}_1}}&          11271.93&                   \tan\beta&             51.00\\
          M_{{\tilde {Q}_2}}&          11270.77&                    \mu(M_Z)&         125591.16\\
          M_{{\tilde {Q}_3}}&          40274.01&                      B(M_Z)&
          2.7861 \times 10^{   9}
 \\
 M_{\bar {H}}^2&
         -1.6336 \times 10^{  10}
 &M_{H}^2&
         -1.7391 \times 10^{  10}
 \\
 \hline
 \end{array}
 $$
  \caption{ \small {Solution 1: Values (in GeV) of soft Susy parameters  at $M_Z$
 (evolved from the soft SUGRY-NUHM parameters at $M_X$)
  determine Susy threshold corrections to fermion Yukawas.
Matching of run down fermion Yukawas in the MSSM to the SM
parameters  determines  soft SUGRY parameters at $M_X$. Note the
heavier third  sgeneration. $\mu(M_Z)$ and   $B(M_Z)=m_A^2 {\sin 2
\beta }/2$ are determined by
   electroweak symmetry breaking conditions. $m_A$ is the
 mass of the CP odd scalar in the doublet Higgs. The sign of
 $\mu$ is assumed positive. }}
 \end{table}
 \begin{table}
 $$
 \begin{array}{|c|c|}
 \hline {\mbox {Field } }&{\rm Mass(GeV)}\\
 \hline
                M_{\tilde{G}}&           1000.14\\
               M_{\chi^{\pm}}&            569.81,         125591.22\\
       M_{\chi^{0}}&            210.10,            569.81,         125591.20    ,         125591.20\\
              M_{\tilde{\nu}}&         15308.069,         15258.322,         21320.059\\
                M_{\tilde{e}}&           1761.89,          15308.29,            211.57   ,          15258.60,          20674.72,          21419.56  \\
                M_{\tilde{u}}&          11271.80,          14446.76,          11270.63   ,          14445.80,          24607.51,          40275.87  \\
                M_{\tilde{d}}&           8402.99,          11272.10,           8401.48   ,          11270.95,          40269.19,          51845.93  \\
                        M_{A}&         377025.29\\
                  M_{H^{\pm}}&         377025.30\\
                    M_{H^{0}}&         377025.28\\
                    M_{h^{0}}&            124.00\\
 \hline
 \end{array}
 $$
  \caption{\small{Solution 1: Spectra of supersymmetric partners  ignoring generation mixing.
   Due to large  values of $\mu,B,A_0$ the LSP and light chargino are  essentially pure Bino and Wino($\tilde W_\pm $).
   The light  gauginos and  light Higgs  $h^0$, are accompanied by a light smuon and  sometimes  selectron.
 The rest of the sfermions have multi-TeV masses. The decoupled and mini-split supersymmetry spectrum and
 large $\mu,A_0$ parameters help avoid problems with FCNC and CCB/UFB instability\cite{kuslangseg}.
Sfermion masses  are ordered by generation not magnitude.
  }}\end{table}

\begin{table}
 $$
 \begin{array}{|c|c|}
 \hline {\mbox {Field } }&{\rm Mass(GeV)}\\
 \hline
                M_{\tilde{G}}&           1000.72\\
               M_{\chi^{\pm}}&            570.11,         125537.00\\
       M_{\chi^{0}}&            210.22,            570.11,         125536.98    ,         125536.98\\
              M_{\tilde{\nu}}&          15257.98,          15307.71,         21350.169\\
                M_{\tilde{e}}&            242.61,           1765.59,          15258.25   ,          15307.93,          20733.03,          21453.81  \\
                M_{\tilde{u}}&          11258.18,          11270.54,          14444.57   ,          14445.53,          24609.90,          40301.29  \\
                M_{\tilde{d}}&           8400.19,           8401.71,          11258.52   ,          11270.84,          40294.63,          51879.28  \\
                        M_{A}&         377430.83\\
                  M_{H^{\pm}}&         377430.84\\
                    M_{H^{0}}&         377430.82\\
                    M_{h^{0}}&            124.13\\
 \hline
 \end{array}
 $$
  \caption{\small{Solution 1: Spectra of supersymmetric partners calculated including  generation mixing effects.
 Inclusion of such effects   changes the spectra only marginally. Due to the large
 values of $\mu,B,A_0$ the LSP and light chargino are  essentially pure Bino and Wino($\tilde W_\pm $).
  Note that the ordering of the eigenvalues in this table follows their magnitudes, comparison
 with the previous table is necessary to identify the sfermions.}}\end{table}

 \begin{table}
 $$
 \begin{array}{|c|c|c|c|}
 \hline
 {\rm Parameter }&{\rm Value} &{\rm  Field }&\hspace{10mm} {\rm Masses}\\
 &&{\rm}[SU(3),SU(2),Y]&\hspace{10mm}( {\rm Units\,\,of 10^{16} Gev} )\\ \hline
       \chi_{X}&  0.1326           &A[1,1,4]&      1.36 \\ \chi_{Z}&
    0.0558
                &B[6,2,{5/3}]&            0.0966\\
           h_{11}/10^{-6}&  3.9601         &C[8,2,1]&{      0.93,      5.17,      7.45 }\\
           h_{22}/10^{-4}&  3.6120    &D[3,2,{7/ 3}]&{      0.29,      6.07,      9.09 }\\
                   h_{33}&  0.0176     &E[3,2,{1/3}]&{      0.11,      0.75,      2.60 }\\
 f_{11}/10^{-6}&
 -0.0130+  0.1591i
                      &&{     2.600,      4.85,      8.23 }\\
 f_{12}/10^{-6}&
 -1.0217-  1.8123i
          &F[1,1,2]&      0.19,      0.65
 \\f_{13}/10^{-5}&
  0.0723+  0.3387i
                  &&      0.65,      4.30  \\
 f_{22}/10^{-5}&
  6.5536+  4.3762i
              &G[1,1,0]&{     0.025,      0.20,      0.76 }\\
 f_{23}/10^{-4}&
 -0.7338+  2.3513i
                      &&{     0.773,      0.77,      0.85 }\\
 f_{33}/10^{-3}&
 -1.2731+  0.5157i
              &h[1,2,1]&{     0.335,      2.67,      5.57 }\\
 g_{12}/10^{-4}&
  0.1284+  0.1895i
                 &&{      7.65,     17.54 }\\
 g_{13}/10^{-5}&
 -9.5431+  2.8232i
     &I[3,1,{10/3}]&      0.36\\
 g_{23}/10^{-4}&
 -1.6403-  0.6279i
          &J[3,1,{4/3}]&{     0.297,      0.39,      1.44 }\\
 \lambda/10^{-2}&
 -4.6906-  0.1490i
                 &&{      1.44,      5.01 }\\
 \eta&
 -0.2495+  0.0683i
   &K[3,1, {8/ 3}]&{      1.73,      5.14 }\\
 \rho&
  1.1753-  0.2967i
    &L[6,1,{2/ 3}]&{      1.79,      2.60 }\\
 k&
 -0.0175+  0.0581i
     &M[6,1,{8/ 3}]&      1.95\\
 \zeta&
  1.2956+  0.9514i
     &N[6,1,{4/ 3}]&      1.88\\
 \bar\zeta &
  0.2238+  0.5885i
          &O[1,3,2]&      3.14\\
       m/10^{16}GeV&  0.0104    &P[3,3,{2/ 3}]&{      0.49,      4.65 }\\
     m_\Theta/10^{16}GeV&  -2.553e^{-iArg(\lambda)}     &Q[8,3,0]&     0.309\\
             \gamma&  0.3925        &R[8,1, 0]&{      0.10,      0.38 }\\
              \bar\gamma& -2.4482     &S[1,3,0]&    0.4403\\
 x&
  0.85+  0.51i
         &t[3,1,{2/ 3}]&{      0.38,      1.16,      1.67,      3.05  }\\\Delta_X^{tot},\Delta_X^{GUT}&      0.80,      0.86 &&{      5.18,      5.70,     20.56 }\\
                                \Delta_{G}^{tot},\Delta_{G}^{GUT}&-20.52,-23.43           &U[3,3,{4/3}]&     0.382\\
      \Delta\alpha_{3}^{tot}(M_{Z}),\Delta\alpha_{3}^{GUT}(M_{Z})& -0.0123, -0.0021               &V[1,2,3]&     0.261\\
    \{M^{\nu^c}/10^{12}GeV\}&{0.000244,    2.33,   81.40    }&W[6,3,{2/ 3}]&              2.50  \\
 \{M^{\nu}_{ II}/10^{-10}eV\}&   0.45, 4292.75,       149682.98               &X[3,2,{5/ 3}]&     0.088,     2.832,     2.832\\
                  M_\nu(meV)&{1.170731,    7.11,   40.21    }&Y[6,2, {1/3}]&              0.11  \\
  \{\rm{Evals[f]}\}/ 10^{-6}&{0.004259,   40.69, 1418.71         }&Z[8,1,2]&              0.38  \\
 \hline\hline
 \mbox{Soft parameters}&{\rm m_{\frac{1}{2}}}=
             0.000
 &{\rm m_{0}}=
         12860.405
 &{\rm A_{0}}=
         -1.9844 \times 10^{   5}
 \\
 \mbox{at $M_{X}$}&\mu=
          1.7240 \times 10^{   5}
 &{\rm B}=
         -1.4927 \times 10^{  10}
  &{\rm tan{\beta}}=           50.0000\\
 &{\rm M^2_{\bar H}}=
         -2.9608 \times 10^{  10}
 &{\rm M^2_{  H} }=
         -2.8920 \times 10^{  10}
 &
 {\rm R_{\frac{b\tau}{s\mu}}}=
  5.6405
  \\
 Max(|L_{ABCD}|,|R_{ABCD}|)&
          7.7373 \times 10^{ -22}
  {\,\rm{GeV^{-1}}}&& \\
 \hline\hline
 \mbox{Susy contribution to}&&&
 \\
 {\rm \Delta_{X,G,3}}&{\rm \Delta_X^{Susy}}=
            -0.053
 &{\rm \Delta_G^{Susy}}=
             2.91
 &{\rm \Delta\alpha_3^{Susy}}=
            -0.010
 \\
 \hline\end{array}
 $$
  \caption{\small{Solution 2: See caption  Table 2}}\end{table}
 \begin{table}
 $$
 \begin{array}{|c|c|c|c|c|}
 \hline
 &&&&\\
 {\rm  Parameter }&{\rm Target} =\bar O_i & {\rm Uncert.= \delta_i  }  & {\rm Achieved= O_i} & {\rm Pull =(O_i-\bar O_i)/\delta_i}\\
 \hline
    y_u/10^{-6}&  2.035847&  0.777694&  2.035834& -0.000017\\
    y_c/10^{-3}&  0.992361&  0.163740&  0.994253&  0.011560\\
            y_t&  0.350010&  0.014000&  0.350076&  0.004715\\
    y_d/10^{-5}& 10.674802&  6.223410& 10.374090& -0.048320\\
    y_s/10^{-3}&  2.024872&  0.955740&  2.118158&  0.097606\\
            y_b&  0.340427&  0.176682&  0.349778&  0.052924\\
    y_e/10^{-4}&  1.121867&  0.168280&  1.122417&  0.003267\\
  y_\mu/10^{-2}&  2.369435&  0.355415&  2.364688& -0.013356\\
         y_\tau&  0.474000&  0.090060&  0.471211& -0.030967\\
             \sin\theta^q_{12}&    0.2210&  0.001600&    0.2210&            0.0009\\
     \sin\theta^q_{13}/10^{-4}&   30.0759&  5.000000&   30.0765&            0.0001\\
     \sin\theta^q_{23}/10^{-3}&   35.3864&  1.300000&   35.3924&            0.0046\\
                      \delta^q&   60.0215& 14.000000&   60.0469&            0.0018\\
    (m^2_{12})/10^{-5}(eV)^{2}&    4.9239&  0.521931&    4.9233&           -0.0012\\
    (m^2_{23})/10^{-3}(eV)^{2}&    1.5660&  0.313209&    1.5664&            0.0011\\
           \sin^2\theta^L_{12}&    0.2944&  0.058878&    0.2931&           -0.0217\\
           \sin^2\theta^L_{23}&    0.4652&  0.139567&    0.4622&           -0.0220\\
           \sin^2\theta^L_{13}&    0.0255&  0.019000&    0.0260&            0.0252\\
 \hline
                  (Z_{\bar u})&   0.972582&   0.972763&   0.972764&\\
                  (Z_{\bar d})&   0.967473&   0.967657&   0.967659&\\
                (Z_{\bar \nu})&   0.946651&   0.946835&   0.946838&\\
                  (Z_{\bar e})&   0.961973&   0.962151&   0.962154&\\
                       (Z_{Q})&   0.983138&   0.983334&   0.983336&\\
                       (Z_{L})&   0.967422&   0.967617&   0.967619&\\
              Z_{\bar H},Z_{H}&        0.000480   &        0.001284    &{}&\\
 \hline
 \alpha_1 &
  0.2016+  0.0000i
 & {\bar \alpha}_1 &
  0.1336-  0.0000i
 &\\
 \alpha_2&
 -0.4805-  0.6320i
 & {\bar \alpha}_2 &
 -0.5177-  0.2850i
 &\\
 \alpha_3 &
  0.0105-  0.3558i
 & {\bar \alpha}_3 &
 -0.3597-  0.2864i
 &\\
 \alpha_4 &
  0.3622-  0.1474i
 & {\bar \alpha}_4 &
  0.4974+  0.3280i
 &\\
 \alpha_5 &
 -0.0159-  0.0451i
 & {\bar \alpha}_5 &
  0.0535-  0.2288i
 &\\
 \alpha_6 &
 -0.0007-  0.2171i
 & {\bar \alpha}_6 &
  0.0189-  0.1050i
 &\\
  \hline
 \end{array}
 $$
  \caption{\small{Solution 2: Fit   with $\chi_X=0.1326 $. See caption  Table 3}}
 \end{table}
 \begin{table}
 $$
 \begin{array}{|c|c|c|c|}
 \hline &&&\\ {\rm  Parameter }&{\rm SM(M_Z)} &{\rm m^{GUT}(M_Z)} &{\rm m^{MSSM}=(m+\Delta m)^{GUT}(M_Z)} \\
 \hline
    m_d/10^{-3}&   2.90000&   1.05215&   2.80332\\
    m_s/10^{-3}&  55.00000&  21.48237&  57.23281\\
            m_b&   2.90000&   2.77488&   2.94586\\
    m_e/10^{-3}&   0.48657&   0.48189&   0.48468\\
         m_\mu &   0.10272&   0.10148&   0.10207\\
         m_\tau&   1.74624&   1.73337&   1.73251\\
    m_u/10^{-3}&   1.27000&   1.09833&   1.27302\\
            m_c&   0.61900&   0.53640&   0.62171\\
            m_t& 172.50000& 146.22372& 172.58158\\
 \hline
 \end{array}
 $$
  \caption{\small{Solution 2: See caption  Table 4.
 Fit with $\chi_Z=0.0557$.}}
 \end{table}
 \begin{table}
 $$
 \begin{array}{|c|c|c|c|}
 \hline
 {\rm  Parameter}  &{\rm Value}&  {\rm  Parameter}&{\rm Value} \\
 \hline
                       M_{1}&            246.41&   M_{{\tilde {\bar {u}}_1}}&          12822.53\\
                       M_{2}&            590.18&   M_{{\tilde {\bar {u}}_2}}&          12822.49\\
                       M_{3}&           1200.01&   M_{{\tilde {\bar {u}}_3}}&          48248.96\\
     M_{{\tilde {\bar l}_1}}&          11957.95&               A^{0(l)}_{11}&        -137311.14\\
     M_{{\tilde {\bar l}_2}}&          11961.97&               A^{0(l)}_{22}&        -137158.88\\
     M_{{\tilde {\bar l}_3}}&          38556.41&               A^{0(l)}_{33}&         -93057.53\\
        M_{{\tilde {L}_{1}}}&          15324.76&               A^{0(u)}_{11}&        -147185.28\\
        M_{{\tilde {L}_{2}}}&          15326.33&               A^{0(u)}_{22}&        -147183.69\\
        M_{{\tilde {L}_{3}}}&          30130.38&               A^{0(u)}_{33}&         -81454.63\\
     M_{{\tilde {\bar d}_1}}&          11245.04&               A^{0(d)}_{11}&        -138168.65\\
     M_{{\tilde {\bar d}_2}}&          11246.12&               A^{0(d)}_{22}&        -138165.70\\
     M_{{\tilde {\bar d}_3}}&          49308.99&               A^{0(d)}_{33}&         -76263.17\\
          M_{{\tilde {Q}_1}}&          13440.51&                   \tan\beta&             50.00\\
          M_{{\tilde {Q}_2}}&          13440.94&                    \mu(M_Z)&         155715.41\\
          M_{{\tilde {Q}_3}}&          48976.61&                      B(M_Z)&
          4.1869 \times 10^{   9}
 \\
 M_{\bar {H}}^2&
         -2.5331 \times 10^{  10}
 &M_{H}^2&
         -2.5545 \times 10^{  10}
 \\
 \hline
 \end{array}
 $$
  \caption{ \small {Solution 2: See caption  Table 5. }}
 \end{table}
 \begin{table}
 $$
 \begin{array}{|c|c|}
 \hline {\mbox {Field } }& {\rm Mass(GeV)}\\
 \hline
                M_{\tilde{G}}&           1200.01\\
               M_{\chi^{\pm}}&            590.18,         155715.46\\
       M_{\chi^{0}}&            246.41,            590.18,         155715.44    ,         155715.44\\
              M_{\tilde{\nu}}&         15324.618,         15326.183,         30130.304\\
                M_{\tilde{e}}&          11958.03,          15324.84,          11961.76   ,          15326.63,          30125.09,          38560.60  \\
                M_{\tilde{u}}&          12822.48,          13440.40,          12822.42   ,          13440.85,          48227.49,          48998.14  \\
                M_{\tilde{d}}&          11245.07,          13440.65,          11246.13   ,          13441.10,          48865.41,          49419.24  \\
                        M_{A}&         457636.54\\
                  M_{H^{\pm}}&         457636.55\\
                    M_{H^{0}}&         457636.54\\
                    M_{h^{0}}&            125.00\\
 \hline
 \end{array}
 $$
 \caption{\small{Solution 2: See caption  Table 6}}\end{table}
\begin{table}
 $$
 \begin{array}{|c|c|}
 \hline {\mbox {Field } }&{\rm Mass(GeV)}\\
 \hline
                M_{\tilde{G}}&           1200.22\\
               M_{\chi^{\pm}}&            590.28,         155704.39\\
       M_{\chi^{0}}&            246.44,            590.28,         155704.37    ,         155704.37\\
              M_{\tilde{\nu}}&          15324.61,          15326.19,         30133.937\\
                M_{\tilde{e}}&          11958.04,          11961.80,          15324.83   ,          15326.64,          30128.73,          38566.31  \\
                M_{\tilde{u}}&          12822.35,          12822.41,          13440.35   ,          13459.00,          48229.64,          48995.86  \\
                M_{\tilde{d}}&          11245.00,          11246.06,          13440.60   ,          13459.25,          48864.06,          49420.94  \\
                        M_{A}&         457783.97\\
                  M_{H^{\pm}}&         457783.98\\
                    M_{H^{0}}&         457783.97\\
                    M_{h^{0}}&            125.02\\
 \hline
 \end{array}
 $$
 \caption{\small{Solution 2: See caption  Table 7}}\end{table}

\begin{table}
 $$
 \begin{array}{|c|c|c|c|c|c|}
 \hline
 {\rm Solution}&\tau_p(M^+\nu) &\Gamma(p\rightarrow \pi^+\nu) & {\rm BR}( p\rightarrow
 \pi^+\nu_{e,\mu,\tau})&\Gamma(p\rightarrow K^+\nu) &{\rm BR}( p\rightarrow K^+\nu_{e,\mu,\tau})\\ \hline
 1
 &
    9.63 \times 10^{ 34}
 &
    4.32 \times 10^{ -37}
 &
 \{
   1.3 \times 10^{ -3}
 ,
   0.34
 ,
   0.66
 \}&
    9.95 \times 10^{ -36}
 &\{
   4.6 \times 10^{ -4}
 ,
   0.15
 ,
   0.85
 \} \\
 2
 &
    3.52 \times 10^{ 34}
 &
    2.14 \times 10^{ -36}
 &
 \{
   1.7 \times 10^{ -3}
 ,
   0.18
 ,
   0.81
 \}&
    2.62 \times 10^{ -35}
 &\{
   1.8 \times 10^{ -3}
 ,
   0.19
 ,
   0.81
 \} \\
 \hline\end{array}
 $$
\caption{\small{$d=5$ operator mediated proton lifetimes
$\tau_p$(yrs), decay rates
 $ \Gamma ( yr^{-1} )$ and Branching ratios in the dominant Meson${}^++\nu$ channels. }} \label{BDEC}\end{table}

\begin{table}
 $$
 \begin{array}{|c|c|c|c|c|c|}
 \hline
 {\rm Solution }&{\rm BR(b\rightarrow s\gamma)} &\Delta a_{\mu}& \Delta \rho&\epsilon/10^{-7}&\delta_{PMNS}  \\ \hline
 1
 &
   3.294 \times 10^{ -4}
 &
   1.06 \times 10^{ -9}
 &
   6.03 \times 10^{ -7}
   &0.12 &6.21^\circ
 \\
 2
 &
   3.289 \times 10^{ -4}
 &
   1.74 \times 10^{ -12}
 &
   1.92 \times 10^{ -7} &0.01&6.27^\circ
 \\
 \hline\end{array}
 $$
 \caption{\small{  Unoptimized  values for  $BR(b\rightarrow s \gamma), \Delta a_\mu$ , $\Delta \rho$,
 $\epsilon^{CP}_{Leptogenesis},\delta_{PMNS}^{CP}$ }}
 \label{LWENRGYCNSTRTS}\end{table}

\end{document}